\documentclass[12pt]{article}
\pdfoutput=1
\usepackage{Max}

\usepackage{empheq}

\newcommand\nc{\newcommand}
\nc\mc\mathcal
\nc\mb\mathbb
\def\Ad{{\rm Ad}}

\newcommand{\hq}{\mathord{/ \!\! / \! \! /_{\xi}}}
\def\Stab{{\rm Stab}}

\def\pt{{\rm pt}}
\def\Rep{{\rm Rep}}

\def\beq{\begin{eqnarray}}
\def\eeq{\end{eqnarray}}

\title{A plethora of K3 metrics}
\author[1]{Arnav Tripathy\thanks{tripathy@math.harvard.edu}}
\author[2]{Max Zimet\thanks{mzimet@fas.harvard.edu}}
\affil[1]{Department of Mathematics,

Harvard University, Cambridge, MA 02138 USA

~}
\affil[2]{Black Hole Initiative and Department of Physics,

Harvard University, Cambridge, MA 02138 USA}

\date{}

\begin{document}

\maketitle

\begin{abstract}
We extend our recent study of K3 metrics near the $T^4/Z_2$ orbifold locus to the other torus orbifold loci. In particular, we provide several new constructions of K3 surfaces as hyper-K\"ahler quotients, which yield new formulae for K3 metrics. We then relate these to the construction of \cite{mz:k3}. As a corollary, we derive infinitely many constraints on the (as yet unknown) BPS spectra of the Minahan-Nemeschansky SCFTs with $E_n$ global symmetry. Specifically, we find linear combinations of $E_n$ characters (evaluated at different points) hiding within K3 metrics and we compute their second order Taylor expansions. We also find novel strong relationships between the BPS spectra of these SCFTs, as well as with that of the $SU(2)$ $N_f = 4$ SCFT. Finally, we provide a new derivation of the class S constructions of these SCFTs and state some experimental observations regarding their BPS spectra.

\end{abstract}

\newpage
\tableofcontents
\hypersetup{linkcolor=blue}

\newpage
\section{Introduction}

In this paper, we generalize the results of \cite{mz:K3HK} by studying the torus orbifold limits of K3 other than $T^4/Z_2$. In a largely unsucessful attempt to prevent this paper from attaining a similar length, we assume that the reader is familiar with the results and notation therein.

In \S\ref{sec:hk}, we review the classification of these limits and then study hyper-K\"ahler quotient constructions of K3 surfaces associated to the cyclic orbifolds: $T^4/Z_q$ for $q=3,4,6$. (We will treat the constructions associated to all of the orbifold limits in a uniform and rigorous manner in \cite{mz:K3HKmath}.) Each of these has 58 parameters and so presumably covers the entire moduli space of Ricci-flat K3 metrics. However, they are particularly useful in the vicinity of their corresponding torus orbifold limit, where the blow-up parameters are small, since in this regime the perturbative approach of \cite{mz:K3HK} allows us to explicitly construct the metrics.

In \S\ref{sec:metrics}, we study this approach in detail for the cyclic orbifolds. In addition to generalizing from $q=2$ to the other values of $q$, we also explicitly carry out to all orders in perturbation theory the iterative procedure that was employed in \cite{mz:K3HK} to obtain the metric and demonstrate that it is equivalent to the recursive solution of an integral equation on the dual torus $\hat T^4$. As in~\cite{mz:K3HK}, we term this the `Higgs branch' construction of K3 metrics.

In addition to being natural generalizations of the $q=2$ case studied in \cite{mz:K3HK}, these orbifolds are of particular interest because when the $T^4$ factors as $T^2\times T^2$ they are naturally regarded as singular elliptically fibered K3 surfaces \cite{dasgupta:constant}. One may therefore probe these surfaces with a D3-brane in F-theory, as in \cite{sen:FOrientifolds,s:fBranes}. In an appropriate scaling limit, the worldvolume of this D3-brane is governed by the $\N=(1,0)$ heterotic little string theory on $T^2$ studied in \cite{mz:k3,mz:K3HK}. The singular fibers present in these orbifolds are all of one of the following types: $I_0^*$, $IV^*$, $III^*$, or $II^*$. Correspondingly, near one of these singular fibers, the theory reduces at low energies, respectively, to either the 4d $SU(2)$ $N_f=4$ SCFT with $\Spin(8)$ global symmetry studied in \cite{sw} or the 4d Minahan-Nemeschansky (MN) SCFTs with $E_6$, $E_7$, or $E_8$ global symmetry studied in \cite{minahan:E6,minahan:En}.\footnote{More precisely, by $E_6$ and $E_7$ we refer to the simply connected forms of these groups, which are respectively the triple and double covers of the centerless groups that are more commonly called $E_6$ and $E_7$. In contrast, $E_8$ is both simply connected and centerless, so there is no ambiguity.} Upon compactification on a circle, the moduli space of this little string theory is the total K3 surface \cite{s:K3,sw:3d,intriligator:compact} and the metric thereon is determined by instanton corrections associated to BPS states of the 4d theory \cite{mz:k3}.

As in \cite{mz:K3HK}, we can read off these BPS state counts by Poisson resumming the metric and comparing with the instanton effects studied in \cite{vafa:spacetimeInsts,seiberg:mirrorT,GMN:walls,mz:k3}. In \S\ref{sec:BPS}, we carry out this procedure, working at leading order in the blow-up parameters. At this order, we are only able to find weak constraints on the BPS spectra of the 4d SCFTs mentioned above (which are consistent with the results of \cite{zwiebach:webs2,zwiebach:webs,neitzke:e6,neitzke:flavorString,neitzke:e7}). Furthermore, as in \cite{mz:K3HK} we will find that there are BPS states in the compactified little string theory which do not contribute at this order, and so one will need to proceed (at least) to the next order in order to study the full spectrum of the little string theory. Nevertheless, it is quite gratifying to find characters of $E_n$ representations and integral BPS state counts hiding within K3 metrics, in addition to the $\Spin(8)$ characters and BPS state counts we found in \cite{mz:K3HK}. This is particularly so because of the tremendous differences between the $SU(2)$ $N_f=4$ and MN spectra: infinitely many $E_n$ \cite{zwiebach:webs} and (conjecturally \cite{neitzke:e6,neitzke:flavorString,neitzke:e7}) supersymmetry representations and charges with arbitrarily large values of $\gcd(p,q)$ are present. And, indeed, our constraints, weak though they may be (at the present order), are infinite in number and involve the BPS index at arbitrarily large values of $\gcd(p,q)$; this is in contrast with the results of \cite{neitzke:e6,neitzke:e7}, which completely determine the BPS index for small values of $\gcd(p,q)$. We are also led to conjecture novel strong relationships between the BPS spectra of the MN theories and of the $SU(2)$ $N_f=4$ SCFT. The central charges of these BPS states, which appear in the metric, again naturally arise from the geometric picture of $(p,q)$-strings winding around the flat F-theory base \cite{bergman:FWebs,sethi:FWebs}.

We recall from \cite{mz:K3HK} that when this procedure is carried out at all orders, one will determine the entire BPS spectrum of the little string theory on $T^2$ (at an orbifold point in its parameter space). One will then obtain the data needed for even more descriptions of smooth K3 metrics, via the `Coulomb branch' construction \cite{mz:k3}. These latter descriptions are valid in an overlapping but different regime from the ones produced by our hyper-K\"ahler quotient perturbation theory.

We conclude this section by discussing, in \S\ref{sec:discuss}, some related ideas; this subsection may be read largely independently from the rest of the paper. First, we relate the above D3-brane pictures of the $SU(2)$ $N_f=4$ and MN theories to the class S ones of \cite{w:MGaugeSol,gaiotto:classS,GMN:classS,tachikawa:e8} which make manifest the connection with Higgs bundle moduli spaces \cite{w:MGaugeSol,kapustin:to3d,kapustin:nonLag,GMN:classS}. The relationship between the Coulomb and Higgs formalisms discussed in \cite{mz:K3HK} then yields isomorphisms between certain Higgs bundle moduli spaces on punctured spheres and equivariant Higgs bundle moduli spaces on $T^2$. Second, we present some experimental observations regarding the BPS spectra of the SCFTs mentioned above, some of which build on those of \cite{neitzke:flavorString}, and these allow us to provide strong evidence against a main conjecture of \cite{neitzke:flavorString}. In particular, we show that the mere knowledge of the existence of the Higgs branch formalism allows us to conjecture many new constraints on the BPS spectra of the MN theories.

\section{Hyper-K\"ahler quotients} \label{sec:hk}

\subsection{Torus orbifolds}

In this subsection, we review the classification of hyper-K\"ahler symmetries of $T^4$, i.e. symmetries which preserve all 3 K\"ahler forms. A related classification problem (of ``special'' symmetries) was solved in \cite{fujiki:t4}, and it was observed in \cite{morrison:triples} that the classification problem of interest has the same solution. Aspects of string compactification on these orbifolds were studied in \cite{morrison:triples,wendland:orb,Volpato:2014zla}.

We follow some of the notation of \cite{fujiki:t4}, splitting these symmetries into case A, consisting of automorphisms by $Z_2, Z_3, Z_4, Z_6$, and case B, with automorphisms by $Q_8\cong 2D_4$ (the order 8 quaternion group / binary dihedral group), $2D_6$ (the order 12 binary dihedral group), and $2T_{12}$ (the order 24 binary tetrahedral group). Generalizing the results of \cite{wati:dBraneT,ho:noncomm,waldram:lol,greene:lol}, for each orbifold we find a corresponding hyper-K\"ahler quotient construction of $K3$s. Namely, by probing the torus orbifold with a D2-brane probe, we expect to find a tri-Hamiltonian action on a flat infinite-dimensional ambient space such that the hyper-K\"ahler quotient with vanishing FI parameters yields the torus orbifold and turning on these parameters yields a full $58$-dimensional family of (almost always) smooth K3s given by resolutions and/or deformations of the original torus orbifold. So, for each of the three new examples in case A and the six examples of case B, we would like to (i) identify these FI parameters and verify that, together with the moduli of the torus, they yield a $58$-dimensional family and (ii) describe the ensuing hyper-K\"ahler quotient. Both parts of this analysis may be conducted in dual formulations; namely, we have the ``D2-brane probe'' picture, which gives a more linear algebraic formulation, and we have the ``D6-brane wrapping'' picture, in which the basic datum is an (equivariant) principal bundle with connection. These two sets of data will be related by Fourier duality. We henceforth neglect the trivial overall volume modulus (of both the $T^4$ and the K3 surface obtained by resolving the orbifold), and so we aim to find 57 parameters in each construction.

We begin with a rather encyclopedic description of the possible hyper-K\"ahler symmetries of $T^4$, paying close attention to the points with nontrivial stabilizers. Unsurprisingly, it is these points (where the quotient geometry will locally resolve to the corresponding ALE geometry) that are crucial for correctly identifying the FI parameters in any frame. For now, we will assume that all of the FI parameters of the standard hyper-K\"ahler quotient construction of each ALE geometry \cite{kronheimer:construct} correspond to parameters in our construction, i.e. that there are no non-local constraints on these blow-up parameters. We recall that in the ALE case, the number of triples of FI parameters is the rank of the Lie algebra associated to the singularity type of an isolated singularity by the McKay correspondence:
\be Z_{n+1} \leftrightarrow A_n \ , \quad 2D_{2(n-2)} \leftrightarrow D_n \ , \quad 2T_{12}\leftrightarrow E_6 \ , \quad 2O_{24} \leftrightarrow E_7 \ , \quad 2I_{60} \leftrightarrow E_8 \ . \ee
This assumption will allow us to verify we have $57$ degrees of freedom, in addition to the overall scale, in all cases; that this assumption is valid will be explained in detail (in both the D2 and D6 frames) in the next subsection and in \S\ref{sec:metrics}.

For fun, we begin with case B. Here, the tori are all isolated -- i.e., for any of these six cases, there exists a unique\footnote{A unique \emph{metric} $T^4$; indeed, our entire discussion is concerned with tori endowed with the structure of a hyper-K\"ahler Riemannian manifold. In particular, we do not make a choice of complex structure, i.e. point on the twistor sphere. Note that this is in contrast to the language and aims of \cite{fujiki:t4}, as the primary concern therein was with the structure of $T^4$ as a complex torus.} $T^4$ admitting such symmetry. We will make extensive use of the quaternions $\mb{H} = \mb{R}\langle 1, i, j, k \rangle$ throughout this discussion. In particular, we write our tori in the form $T^4=\HH/\Lambda$ for $\Lambda$ a lattice in $\HH$.

(B1) $T^4 = \mb{H} / \mb{Z}\langle 1, i, j, k \rangle$ with action by $Q_8 = \{\pm 1, \pm i, \pm j, \pm k\}$ by left-multiplication. Any points with nontrivial stabilizers must in particular be stabilized by $-1$, as every nontrivial element has $-1$ as some power. Hence, to classify all points with nontrivial stabilizers, it suffices to consider the sixteen (two-torsion) points stabilized by $-1$. By explicit computation, we determine
\begin{eqnarray}
\Stab(0) &=& Q_8 \nonumber \\ 
\Stab\Big(\frac{1}{2}\Big) = \Stab\Big(\frac{i}{2}\Big) = \Stab\Big(\frac{j}{2}\Big) = \Stab\Big(\frac{k}{2}\Big) &=& \{\pm 1\} \nonumber \\
\Stab\Big(\frac{1+i}{2}\Big) = \Stab\Big(\frac{j+k}{2}\Big) &=& Z_4 \nonumber \\
\Stab\Big(\frac{1+j}{2}\Big) = \Stab\Big(\frac{i+k}{2}\Big) &=& Z_4 \nonumber \\
\Stab\Big(\frac{1+k}{2}\Big) = \Stab\Big(\frac{i+j}{2}\Big) &=& Z_4 \nonumber \\
\Stab\Big(\frac{1+i+j}{2}\Big) = \Stab\Big(\frac{1+i+k}{2}\Big) = && \nonumber \\
\Stab\Big(\frac{1+j+k}{2}\Big) = \Stab\Big(\frac{i+j+k}{2}\Big) &=& \{\pm 1\} \nonumber \\
\Stab\Big(\frac{1+i+j+k}{2}\Big) &=& Q_8 \ .
\end{eqnarray}
In the above, each series of equalities collates the points that get identified in the quotient, together with the stabilizer of these points. For the stabilizer groups above, the corresponding Dynkin diagrams (under McKay) are $Q_8 \leftrightarrow D_4, Z_4 \leftrightarrow A_3, Z_2 \leftrightarrow A_1$, and so per our running assumptions, we have $2 \cdot 4 + 3 \cdot 3 + 2 \cdot 1 = 19$ triples of FI parameters, as desired. In more detail, the quotient will have $2$ points that are locally given by $\mb{C}^2 / Q_8$, and so the McKay dictionary tells us to expect 4 triples of deformations about each point, as the Dynkin diagram corresponding to $Q_8$ is $D_4$. Here, triples literally means ``triple of real degrees of freedom,'' but we note that once one fixes some choice of complex structure, it is natural to think of each triple as a complex degree of freedom (that deforms the complex structure of the manifold) and a real degree of freedom (that deforms the K\"ahler structure; for complex geometers, the only effect is to (partially or wholly) resolve the manifold). Next, we have two points in the quotient locally isomorphic to $\mb{C}^2 / Z_2$, and so each contributes one triple of parameters under the McKay dictionary. Finally, there are three more points with three triples each, and hence the count above.

(B2) $T^4 = \mb{H} / \mb{Z}\langle 1, i, j, \frac{1+i+j+k}{2}\rangle$ with action by $Q_8 = \{ \pm 1, \pm i, \pm j, \pm k\}$ as before. Note that this lattice is a superlattice, of index $2$, of the prior one. Once again, it suffices to consider the sixteen $2$-torsion points in order to tabulate all points with nontrivial stabilizers. We find the following:
\begin{eqnarray}
\Stab(0) &=& Q_8 \nonumber \\
\Stab\Big(\frac{i}{2}\Big) = \Stab\Big(\frac{j}{2}\Big) = \Stab\Big(\frac{k}{2}\Big) = \Stab\Big(\frac{i+j+k}{2}\Big) &=& \{ \pm 1 \} \nonumber \\
\Stab\Big(\frac{i+j}{2}\Big) &=& Q_8 \nonumber \\
\Stab\Big(\frac{i+k}{2}\Big) &=& Q_8 \nonumber \\
\Stab\Big(\frac{j+k}{2}\Big) &=& Q_8 \nonumber \\
\Stab\Big(\frac{1+i+j+k}{4}\Big) = \Stab\Big(\frac{1-i-j+k}{4}\Big) = && \nonumber \\
\Stab\Big(\frac{1-i+j-k}{4}\Big) = \Stab\Big(\frac{1+i-j-k}{4}\Big) &=& \{\pm 1 \} \nonumber \\
\Stab\Big(\frac{1-i+j+k}{4}\Big) = \Stab\Big(\frac{1+i-j+k}{4}\Big) = && \nonumber \\
\Stab\Big(\frac{1+i+j-k}{4}\Big) = \Stab\Big(\frac{1-i-j-k}{4}\Big) &=& \{ \pm 1 \} \ ,
\end{eqnarray}
for a total of $4 \cdot 4 + 3 \cdot 1 = 19$ triples of FI parameters.

(B3) $T^4$ the same as in (2) but now with action by $2T_{12} = \{\pm 1, \pm i, \pm j, \pm k, \frac{\pm 1 \pm i \pm j \pm k}{2} \}$, where all signs in the last group element are uncorrelated -- i.e., may be picked independently for a total of sixteen possibilities. Note that of these sixteen group elements, the eight with initial sign $+\frac{1}{2}$ have order $6$ while the eight with initial sign $-\frac{1}{2}$ have order $3$. Obtaining all the fixed points is more involved now, as besides the fixed locus of $-1$, we also have order $3$ elements whose fixed loci need not be contained therein. Nevertheless, we find them, and calculate
\begin{eqnarray}
\Stab(0) &=& 2T_{12} \nonumber \\
\Stab\Big(\frac{i}{2}\Big) = \Stab\Big(\frac{j}{2}\Big) = \Stab\Big(\frac{k}{2}\Big) = \Stab\Big(\frac{i+j+k}{2}\Big) = \Stab\Big(\frac{1\pm i \pm j \pm k}{2}\Big) &=& \{ \pm 1 \} \nonumber \\
\Stab\Big(\frac{i+j}{2}\Big) = \Stab\Big(\frac{i+k}{2}\Big) = \Stab\Big(\frac{j+k}{2}\Big) &=& Q_8 \nonumber \\
\Stab\Big(\frac{\widehat{\pm 1 \pm i \pm j \pm k} }{3}\Big) &=& Z_3 \ . \nonumber \\
\end{eqnarray}
The last line needs some explanation: the hat indicates that only three of the four terms are present in the given expressions. Moreover, in contrast to our established notation in prior examples, not every point in the last line is in the same orbit. Rather, these $32$ points assemble into $4$ orbits of size $8$ each. Using now that $2T_{12} \leftrightarrow E_6$ in the McKay dictionary, we count $6 + 1 + 4 + 4 \cdot 2 = 19$ triples of FI parameters.

(B4) $T^4$ the same as in (2) but with action by $Q_8 = \{ \pm 1, \pm i, (\pm j; \frac{1+i}{2}), (\pm k; \frac{1+i}{2}) \}$ where the notation $(\lambda; \mu)$ for $\lambda, \mu \in \mb{H}$ indicates the action $x \mapsto \lambda x + \mu$. It once again suffices as in the first two examples to study the fixed locus of $-1$. In fact, these points are the same as in example $(2)$ above, so we simply check the orbit-stabilizer structure for these points:
\begin{eqnarray}
\Stab(0) = \Stab\Big(\frac{j+k}{2}\Big) &=& Z_4 \nonumber \\
\Stab\Big(\frac{i+j}{2}\Big) = \Stab\Big(\frac{i+k}{2}\Big) &=& Z_4 \nonumber \\
\Stab\Big(\frac{i}{2}\Big) = \Stab\Big(\frac{j}{2}\Big) = \Stab\Big(\frac{k}{2}\Big) = \Stab\Big(\frac{i+j+k}{2}\Big) &=& \{ \pm 1 \} \nonumber \\
\Stab\Big(\frac{1 \pm i \pm j \pm k}{4}\Big) &=& Z_4 \ ,
\end{eqnarray}
where again the notation in the last line is non-standard: the $8$ points organize into $4$ orbits of size $2$, for a total of $6 \cdot 3 + 1 = 19$ triples of FI parameters.

(B5) $T^4$ the same as in (2) but with $2T_{12}$ action by
\be 2T_{12} = \langle Q_8, \Big(\frac{1+i+j+k}{2}; \frac{1+i-j+k}{4}\Big) \rangle \ , \ee
where the $Q_8$-action is as in (4). The points with nontrivial stabilizer will again include the $16$ two-torsion points, but there will be others (stabilized only by $Z_3$s inside $2T_{12}$). We first list the status of the two-torsion points:
\begin{eqnarray}
\Stab(0) = \Stab\Big(\frac{j+k}{2}\Big) = \Stab\Big(\frac{1+i-j+k}{4}\Big) = && \nonumber \\
\Stab\Big(\frac{1-i-j+k}{4}\Big) = \Stab\Big(\frac{1-i+j+k}{4}\Big) = \Stab\Big(\frac{1+i-j-k}{4}\Big) &=& Z_4 \nonumber \\
\Stab\Big(\frac{i+j}{2}\Big) = \Stab\Big(\frac{i+k}{2}\Big) = \Stab\Big(\frac{1+i+j+k}{4}\Big) = && \nonumber \\
\Stab\Big(\frac{1+i+j-k}{4}\Big) = \Stab\Big(\frac{1-i+j-k}{4}\Big) = \Stab\Big(\frac{1-i-j-k}{4}\Big) &=& Z_4 \nonumber \\
\Stab\Big(\frac{i}{2}\Big) = \Stab\Big(\frac{j}{2}\Big) = \Stab\Big(\frac{k}{2}\Big) = \Stab\Big(\frac{i+j+k}{2}\Big) &=& Z_6 \ .
\end{eqnarray}
Note, for example, that $\Stab(\frac{i}{2}) = Z_6$ is generated by $(\frac{1+i+j+k}{2}; \frac{1+i-j+k}{4})$. 

We eschew finding the $Z_3$-fixed points explicitly; we will instead simply count how many there are. Note, however, that we do know that all points with nontrivial stabilizers not yet already found must have stabilizer $Z_3$, as any other nontrivial stabilizer subgroup of $2T_{12}$ would contain $-1$ (and hence have been already found). So, to count these $Z_3$-stabilizer points, note that there are eight order three elements of $2T_{12}$ assembled into four $Z_3$ subgroups. The number of fixed points of any given order three element will be the same as in case $(3)$ above: one way to see this claim is to reflect on the fact that the moduli space of tori with $Z_3$-automorphisms is connected and that the number of fixed points is constant therein. A more abstract approach would be to use the Lefschetz fixed point theorem; this application would be overkill, but it is nonetheless a fun computation which we will detail shortly. 

So, any order $3$ element has $9$ fixed points, either from Lefschetz or from case (3), as in that example the $32$ fixed points with $Z_3$-stabilizer evenly distribute among the four $Z_3$ subgroups to give $8$ for each, and in addition each $Z_3$ stabilizes the trivial fixed point $0$. Returning to this case, the $Z_6$-stabilized points above already contribute $1$ fixed point for any given order $3$ element, so any such element has $8$ more fixed points, for a total of $4 \cdot 8 = 32$ points with stabilizer $Z_3$. Their $2T_{12}$ orbits hence have size $24 / 3 = 8$, so these points assemble into $4$ orbits of size $8$. We may now finish the count of FI parameters: we have $2 \cdot 3 + 5 + 4 \cdot 2 = 19$ triples.

We now explain the Lefschetz calculation somewhat further, as it provides a useful check that we have found all the fixed points in any given case. One may directly show (or read in case A of \cite{fujiki:t4}) that the moduli spaces of tori with $Z_q$-automorphisms are connected, and so we can compute the numbers of fixed points at particularly simple points in moduli space such as $E \times E$ for $E$ the corresponding elliptic curve with extra automorphisms (see our case A discussion below). The Lefschetz calculation for a $Z_q$-action then computes the supertrace of a cyclic generator on $H^*(T^4, \mb{Z}) \simeq \Lambda^* H^1(T^4, \mb{Z})$ as $((1 - e^{2\pi i/q})(1 - e^{-2\pi i/q}))^2$, since $e^{\pm 2 \pi i / q}$ with multiplicities $2$ are the eigenvalues of the cyclic generator on $H^1(T^4; \mb{Z})$. We hence have $16,9,4,1$ fixed points, respectively, for $q = 2,3,4,6$. 

(B6) $T^4 = \mb{H} / \mb{Z}\langle 1, i, h, \ell \rangle$ with $h = \frac{i + \sqrt{3} j}{2}$, $\ell = \frac{1 + \sqrt{3} k}{2}$, and action by \linebreak $2D_6 = \{ \pm 1, \pm i, \pm h, \pm \ell, \pm ih, \pm i \ell \}$. Note that $\{\pm i, \pm h, \pm i \ell\}$ have order $4$ and the powers of $\ell$ form the cyclic group $\{ \ell, ih, -1, -\ell, -ih, 1 \}$. Once again, most of the points with nontrivial stabilizer are among the $16$ fixed points of $-1$, with the rest having $Z_3$ stabilizer. In fact, it is not so difficult to find these either; following the relevant discussion in case A of \cite{fujiki:t4}, it is natural to search among trisectors between lattice points. In any case, we have the following:
\begin{eqnarray}
\Stab(0) &=& 2D_6 \nonumber \\
\Stab\Big(\frac{1}{2}\Big) = \Stab\Big(\frac{i}{2}\Big) = \Stab\Big(\frac{h}{2}\Big) = && \nonumber \\
\Stab\Big(\frac{\ell}{2}\Big) = \Stab\Big(\frac{1+\ell}{2}\Big) = \Stab\Big(\frac{i+h}{2}\Big) &=& \{ \pm 1 \} \nonumber \\
\Stab\Big(\frac{1+i}{2}\Big) = \Stab\Big(\frac{h+\ell}{2}\Big) = \Stab\Big(\frac{1+i+h+\ell}{2}\Big) &=& Z_4 \nonumber \\
\Stab\Big(\frac{1+h}{2}\Big) = \Stab\Big(\frac{1+i+\ell}{2}\Big) = \Stab\Big(\frac{i+h+\ell}{2}\Big) &=& Z_4 \nonumber \\
\Stab\Big(\frac{i+\ell}{2}\Big) = \Stab\Big(\frac{1+i+h}{2}\Big) = \Stab\Big(\frac{1+h+\ell}{2}\Big) &=& Z_4 \nonumber \\
\Stab\Big(\pm \frac{j}{\sqrt{3}}\Big) = \Stab\Big(\pm \frac{k}{\sqrt{3}}\Big) &=& Z_3 \nonumber \\
\Stab\Big(\frac{\pm j \pm k}{\sqrt{3}}\Big) &=& Z_3 \ .
\end{eqnarray}
Using $2D_6 \leftrightarrow D_5$ in the McKay correspondence, we have $5 + 1 + 3 \cdot 3 + 2 \cdot 2 = 19$ triples of FI parameters.


We now turn to case A of \cite{fujiki:t4}, giving a complete description of the cyclic orbifolds (which will be of more import in the following sections of this paper). The tori now have moduli, and so we begin with discussing how to parametrize these tori.

(A0) $T^4$ is arbitrary with $Z_2=\avg{-1}$ action. There are 16 $Z_2$-fixed points, i.e. 16 triplets of FI parameters, and there are 9 moduli in the choice of a unit volume metric on $T^4$, for a total of $16\times 3+9=57$ moduli.

(A1) We now consider tori admitting a $Z_4$ automorphism $\chi$. We may identify the action of $\chi$ on the universal cover $\mb{R}^4 \simeq \mb{C}^2$ as multiplication by $i, -i$ respectively on the two factors. By further identifying the universal cover with $\mb{H}$, we may take $\chi$ to be right-multiplication by $i$. Then the lattice $\Lambda$ by which we quotient may be taken to have basis $\lambda, \chi \lambda, \mu, \chi \mu$. We may eliminate what remains of the $SO(4)$ redundancy in this description of our torus, as well as the overall volume modulus, by putting $\Lambda$ into a standard form via the action of the subgroup of $GO(4)$ that commutes with $\chi$.\footnote{If $g\in GO(4)$ satisfies $g(\chi\lambda)=\chi (g\lambda)$, i.e. $g^{-1}\chi^{-1}g\chi$ stabilizes $\lambda$, then it is easily seen that $g^{-1}\chi^{-1}g\chi$ also stabilizes $\chi\lambda$. The same holds with $\lambda$ replaced by $\mu$. Therefore, if $g$ preserves the form of our basis then $g^{-1}\chi^{-1}g\chi$ stabilizes the whole basis -- i.e., $g$ and $\chi$ commute.} Here $GO(4)$ is the enlargment of the $SO(4)$ rotation group by the $\mb{R}^+$ scaling action. The $SO(4) = (SU(2)_{\ell} \times SU(2)_r)/Z_2$-action may be described by left- and right-multiplication by unit quaternions on $\mb{H}$ (with simultaneous left- and right-multiplication by $-1$ acting trivially), and the scaling action may be included by allowing one side to include any nonzero (rather than only unit-norm) quaternions. The subgroup which commutes with $\chi$ has the left-action arbitrary and the right-action given by multiplication by elements of $\mb{C} \simeq \mb{R}\langle 1, i \rangle$. We hence see that we have five gauge degrees of freedom with which to simplify the lattice $\Lambda$, which {\it a priori} could be parametrized by two quaternionic degrees of freedom $\lambda, \mu$, leaving three real degrees of freedom for the moduli of the $T^4$. In fact, let us use the left-multiplication to set $\lambda = 1$. The remaining gauge degree of freedom is conjugation by the $S^1$ in the $\mb{C}$ above, which we use up by demanding that $\mu$ has no $k$ component and a positive $j$ component. (There is also some discrete freedom in that there are transformations of $\mu$ that give $GO(4)$-equivalent lattices, such as $\mu\mapsto \mu+1$, $\mu\mapsto \mu+i$, and $\mu\mapsto -\mu^{-1}$ (with $GO(4)$ equivalence given by left-multiplication by $-\mu$), but fixing this is not necessary for counting degrees of freedom.)

So, $T^4 = \mb{H} / \mb{Z}\langle 1, i, \mu, \mu i \rangle$ with $Z_4$-action given by right-multiplication by $i$. As before, we only need to understand the $16$ points fixed by $-1$, finding
\begin{eqnarray}
\Stab(0) &=& Z_4 \nonumber \\
\Stab\Big(\frac{1}{2}\Big) = \Stab\Big(\frac{i}{2}\Big) &=& \{ \pm 1 \} \nonumber \\
\Stab\Big(\frac{1+i}{2}\Big) &=& Z_4 \nonumber \\
\Stab\Big(\frac{\mu}{2}\Big) = \Stab\Big(\frac{\mu i}{2}\Big) &=& \{ \pm 1 \} \nonumber \\
\Stab\Big(\frac{\mu + \mu i}{2}\Big) &=& Z_4 \nonumber \\
\Stab\Big(\frac{1+\mu}{2}\Big) = \Stab\Big(\frac{i + \mu i}{2}\Big) &=& \{ \pm 1 \} \nonumber \\
\Stab\Big(\frac{1 + \mu i}{2}\Big) = \Stab\Big(\frac{i + \mu}{2}\Big) &=& \{ \pm 1 \} \nonumber \\
\Stab\Big(\frac{1+i+\mu}{2}\Big) = \Stab\Big(\frac{1+i+\mu i}{2}\Big) &=& \{ \pm 1 \} \nonumber \\
\Stab\Big(\frac{1 + \mu + \mu i}{2}\Big) = \Stab\Big(\frac{i + \mu + \mu i}{2}\Big) &=& \{ \pm 1 \} \nonumber \\
\Stab\Big(\frac{1 + i + \mu + \mu i}{2}\Big) &=& Z_4 \ ,
\end{eqnarray}
for a total of $4 \cdot 3 + 6 = 18$ triples of FI parameters and $3 + 3 \cdot 18 = 57$ total degrees of freedom between the moduli and FI parameters.

(A2) We now consider tori admitting a $Z_3$ automorphism $\chi$. Defining $\kappa_3 = e^{2\pi i/3} = \frac{-1 + \sqrt{3} i}{2}$, we take $\chi$ to be right-multiplication by $\kappa_3$ on $\mb{H}$. Then the discussion of the centralizer of $\chi$ in $GO(4)$ and the simplification of the lattice $\Lambda$ proceeds precisely as before, so our torus is $T^4 = \mb{H} / \mb{Z} \langle 1, \kappa_3, \mu, \mu \kappa_3 \rangle$ with $\mu_k = 0, \mu_j > 0$ as before. The orbit-stabilizer structure of the fixed points in this case is particularly easy, as the stabilizers are all $Z_3$ and the orbits are all single points, namely,
\begin{eqnarray}
0, \frac{2 + \kappa_3}{3}, \frac{1 + 2 \kappa_3}{3}, \nonumber \\
\frac{2 \mu + \mu \kappa_3}{3}, \frac{\mu + 2 \mu \kappa_3}{3}, \frac{2 + \kappa_3 + 2 \mu + \mu \kappa_3}{3}, \nonumber \\
\frac{1 + 2 \kappa_3 + \mu + 2 \mu \kappa_3}{3}, \frac{2 + \kappa_3 + \mu + 2 \mu \kappa_3}{3}, \frac{1 + 2 \kappa_3 + 2 \mu + \mu \kappa_3}{3}
\end{eqnarray}
for a total of $2 \cdot 9 = 18$ triples of FI parameters.

(A3) As above, $T^4 = \mb{H} / \mb{Z}\langle 1, \kappa_3, \mu, \mu \kappa_3 \rangle$ with $Z_6$-automorphism $\chi$ given by right-multiplication by $\kappa_6 = e^{2\pi i/6} = 1 + \kappa_3 = \frac{1 + \sqrt{3} i}{2}$ and
\begin{eqnarray}
\Stab(0) &=& Z_6 \nonumber \\
\Stab\Big(\frac{1}{2}\Big) = \Stab\Big(\frac{1 + \kappa_3}{2}\Big) = \Stab\Big(\frac{\kappa_3}{2}\Big) &=& \{ \pm 1 \} \nonumber \\
\Stab\Big(\frac{\mu}{2}\Big) = \Stab\Big(\frac{\mu \kappa_3}{2}\Big) = \Stab\Big(\frac{\mu + \mu \kappa_3}{2}\Big) &=& \{ \pm 1 \} \nonumber \\
\Stab\Big(\frac{1 + \mu}{2}\Big) = \Stab\Big(\frac{\kappa_3 + \mu \kappa_3}{2}\Big) = \Stab\Big(\frac{1 + \kappa_3 + \mu + \mu \kappa_3}{2}\Big) &=& \{\pm 1 \} \nonumber \\
\Stab\Big(\frac{1 + \mu \kappa_3}{2}\Big) = \Stab\Big(\frac{\kappa_3 + \mu + \mu \kappa_3}{2}\Big) = \Stab\Big(\frac{1 + \kappa_3 + \mu}{2}\Big) &=& \{ \pm 1 \} \nonumber \\
\Stab\Big(\frac{1 + \mu + \mu \kappa_3}{2}\Big) = \Stab\Big(\frac{\kappa_3 + \mu}{2}\Big) = \Stab\Big(\frac{1 + \kappa_3 + \mu \kappa_3}{2}\Big) &=& \{ \pm 1 \} \nonumber \\
\Stab\Big(\frac{2 + \kappa_3}{3}\Big) = \Stab\Big(\frac{1 + 2 \kappa_3}{3}\Big) &=& Z_3 \nonumber \\
\Stab\Big(\frac{2 \mu + \mu \kappa_3}{3}\Big) = \Stab\Big(\frac{\mu + 2 \mu \kappa_3}{3}\Big) &=& Z_3 \nonumber \\
\Stab\Big(\frac{2 + \kappa_3 + 2 \mu + \mu \kappa_3}{3}\Big) = \Stab\Big(\frac{1 + 2 \kappa_3 + \mu + 2 \mu \kappa_3}{3}\Big) &=& Z_3 \nonumber \\
\Stab\Big(\frac{2 + \kappa_3 + \mu + 2 \mu \kappa_3}{3}\Big) = \Stab\Big(\frac{1 + 2 \kappa_3 + 2 \mu + \mu \kappa_3}{3}\Big) &=& Z_3
\end{eqnarray}
for a total of $1 \cdot 5 + 5 \cdot 1 + 4 \cdot 2 = 18$ triples of FI parameters.

\subsection{D6-brane description of hyper-K\"ahler quotients} \label{sec:d6HK}

We now turn to describing the hyper-K\"ahler quotients that result from wrapping D6-branes on the T-duals of the above torus orbifolds. For simplicity, we restrict to the cyclic orbifold (type A) cases here; we will give a complete, general description along with plenty more mathematical rigor in future work. In \S\ref{sec:metrics}, we will T-dualize to the D2-brane probe frame, as that gives us the linear algebra description in which we may most explicitly compute.

So, given a torus cyclic orbifold as above, let $q \in \{2, 3, 4, 6\}$ denote the order of the cyclic automorphism group; the wrapped D6-branes now yield $Z_q$-equivariant $U(q)$-connections on a (trivial) bundle on the dual torus, $\hat T^4=\RR^4/\Lambda^\vee$, with $\Lambda^\vee=\Hom(\Lambda,2\pi\ZZ)$. More precisely, defining $\kappa_q = e^{2\pi i / q}$ and $\sigma_q = \mathrm{diag}(1, \kappa_q, \kappa_q^2, \cdots, \kappa_q^{q-1})$, we denote the infinite-dimensional flat hyper-K\"ahler manifold with which we start and the gauge group, respectively, by
\begin{eqnarray}
\mc{A}' &:=&  \{ \nabla = \partial - i B, B \in \Omega^1(\hat T^4; \mf{u}(q)) \big| \iota^* B = \Ad_{\sigma_q} B \} \nonumber \\
\mc{G}' &:=& \{ g: \hat T^4 \to U(q) \big| g \circ \iota = \Ad_{\sigma_q} g \} \ , \label{eq:d6Proj}
\end{eqnarray}
analogously to the $q = 2$ case in \cite{mz:K3HK}; the equivariance conditions above continue to bear similarity to those imposed by \cite{moore:ALEinst}. Here, we have defined $\iota = (\chi^{-1})^T$. To see that this is a symmetry of the dual lattice $\Lambda^\vee$, we note that if $\hat n\cdot n\in 2\pi\ZZ$ for all $n\in \Lambda$, so that $\hat n\in\Lambda^\vee$, then $(\iota \hat n)\cdot n=\hat n\cdot (\chi^{-1} n)\in 2\pi \ZZ$ for all $n\in \Lambda$, and so $\iota \hat n\in \Lambda^\vee$. We note that the actions of $\chi$ and $\iota$ lift, respectively, to the $\RR^4$ covering spaces of $T^4$ and $\hat T^4$, and if we identify these vector spaces using the metric on $\RR^4$ then these actions coincide. Finally, by $B\in\Omega^1(\hat T^4;\mf{u}(q))$, we mean that $B$ is a smooth 1-form when the FI parameters vanish, and more generally that $B$ has certain prescribed singularities (up to gauge equivalence) at the $Z_q$ fixed points.\footnote{Actually, while this approach of prescribing the singular behavior of $B$ only up to gauge equivalence is formally clean, it comes at the cost of the allowed changes in $B$ not being $L^2$-normalizable, which makes the hyper-K\"ahler structure on $\A$ ill-defined. It is probably therefore best to use the gauge freedom in order to exactly fix the singular behavior of $B$ at the fixed points, and to correspondingly restrict the elements of our gauge group to those $g$ which map the fixed points to the identity. In any case, we will be more careful about such matters in \cite{mz:K3math,mz:K3HKmath}.\label{ft:fixing}} As such, our ambient space in fact depends in this mild way on the choice of $\xi$ and so we denote it by $\mc{A}_{\xi}$. Note in contrast that $g$ is everywhere smooth.

We note that $\G'$ does not act faithfully, since the group of constant maps into the central $U(1)$ acts trivially on $\A'$. Therefore, to be more precise we should take the gauge group to be
\be \tilde\G := \G'/U(1) \ . \ee
Indeed, as in \cite{mz:K3HK}, the $U(1)$ factor entirely decouples, and so for the remainder of this section we take our space $\A$ of equivariant connections to contain only $B\in\Omega^1(\hat T^4;\mf{su}(q))$, the FI parameters to be valued in $\mf{su}(q)$, and our gauge group to be
 \be \G:=\G'/\G_1 \ , \ee
 where
 \be \G_1 := \{ g: \hat T^4 \to U(1) \big| g \circ \iota = g \} \ . \ee
This acts naturally on traceless one-forms $B$: simply pick a representative in $\G'$, act on $B$, and then retain only the traceless part of $B$.

We now claim that (1) $\mc{A}_{\xi}$ has a natural (flat) hyper-K\"ahler structure, (2) the usual action of $\mc{G}$ on $\mc{A}_{\xi}$ is tri-Hamiltonian, and (3) the hyper-K\"ahler quotient $\mc{A}_{\xi} \hq \mc{G}$ with the natural FI parameter deformations of the moment maps (and of the space $\A_{\xi}$, as discussed in footnote \ref{ft:fixing}, although these deformations are quite weak, as they do not change the hyper-K\"ahler structure) yields a family of K3s deforming the $T^4 / Z_q$ orbifold at $\xi = 0$. Indeed, (1) is clear, as the equivariance conditions imposed play well with the hyper-K\"ahler structure on the space of connections on $\hat T^4$, and for (2), we will shortly explicitly write down all three moment maps. That leaves the most interesting claim (3), which we establish at a similar level of rigor as our $q = 2$ analysis \cite{mz:K3HK} (but see forthcoming work \cite{mz:K3math,mz:K3HKmath} to make this fully mathematically rigorous).

Let us first discuss the appearance of the moment map deformation FI parameters in this frame. Recall that they correspond to `cocentral factors' in $(\mathrm{Lie}\,\mc{G})^{\vee}$. As in the case of $q = 2$, these cocentral factors must be supported (as distributions) at the fixed points of $\hat T^4$. Explicitly, the moment map equations take the form
\be F=-*F+\sum_{p} \upsilon_{p} \delta_p \ , \label{eq:D6mu} \ee
where $p$ ranges over the fixed points of $\hat T^4$ and $\upsilon_{p}$ is a harmonic (i.e., constant-coefficient, in standard coordinates obtained from the $\RR^4$ covering space) self-dual 2-form on $\hat T^4$ valued in a subspace of $\mf{su}(q)$. Cocentrality implies the $\mf{su}(q)$ coefficient at a fixed point must lie in the traceless part of the Lie algebra of the center of the centralizer of the image of the stabilizer under the regular representation $\chi\mapsto \sigma_q$ of $Z_q$. Lastly, since the 2-forms $\upsilon_p$ are linear combinations of the K\"ahler forms $\omega_{I,J,K}$ on $\hat T^4$, which are invariant under pullback via $\iota$, while $\iota^*\delta_p=\delta_{\iota^{-1}(p)}$, it follows that the equivariance condition takes the form $\upsilon_{\iota(p)}=\Ad_{\sigma_q}\upsilon_p$. In the following paragraphs, we spell this out more concretely; for ease of notation, we focus on the coefficients in $\upsilon_p$ that multiply one of $\omega_{I,J,K}$, as the analysis is identical for all of these.

We begin with the case $q = 3$ (i.e. (A2) above). Then we claim that the $18$ FI parameters are spanned by $\sigma_3 \delta_p, \sigma_3^2 \delta_p$, as $p$ ranges over the $9$ fixed points discussed above. This is because for each $p$ we obtain FI parameters from the traceless part of the Lie algebra of the $U(1)^3$ centralizer of $\sigma_3$, and this $\mf{u}(1)^2$ Lie algebra is spanned by $\sigma_3$ and $\sigma_3^2$.

The case of $q = 4$ (aka (A1)) is slightly more interesting, as we now have fixed points with nontrivial orbits. Certainly, if $p$ is a $Z_4$-fixed point such as $0$ or $\frac{1+i}{2}$, the story proceeds as above: the cocentral factors supported at such points lie in the traceless part of $\mathrm{Lie}\,Z(C(\rho_{\rm reg}(\mathrm{Stab}(p)))) = \mf{u}(1)^4$, and so they are spanned by $\sigma_4 \delta_p, \sigma_4^2 \delta_p, \sigma_4^3 \delta_p$. Consider, by contrast, a $Z_2$-fixed point $p$ such as $\frac{1}{2}$. Then any cocentral factor supported at such a point $p$ must now be in the traceless part of the Lie algebra of the center of the centralizer of $\rho_{\rm reg}(\mathrm{Stab}(p))$, but $C(\rho_{\rm reg}(\mathrm{Stab}(p)))$ is now $U(2) \times U(2)$ so that $Z(C(\rho_{\rm reg}(\mathrm{Stab}(p)))) = U(1)^2$ and the traceless part of its Lie algebra is spanned by $\sigma_4^2$. So, the FI parameter here is given by $\sigma_4^2 \delta_p$. But, note that the coefficient of $\sigma_4^2 \delta_{\iota(p)}$ is equal to that of $\sigma_4^2 \delta_p$, thanks to the equivariance condition. Hence, we should only take representatives for each orbit. So, we obtain $3 \cdot 4$ triples from the four $Z_4$-fixed points and $1 \cdot 6$ triples from the six orbits of $Z_2$-fixed points for a total of $18$ once again, as claimed. The same match happens in the case $q = 6$, with FI parameters corresponding to orbits of fixed points as found in the prior section once again.

The conclusion of this case-by-case analysis is that the $\upsilon_p$ are valued in the Cartan subalgebra of the Lie algebra associated to the stabilizer of $p$ by the McKay correspondence, regarded as a subalgebra of the $A_{q-1}$ Cartan subalgebra. Furthermore, $\upsilon_{p_1}=\upsilon_{p_2}$ if $p_1$ and $p_2$ are identified by the $Z_q$ action on $\hat T^4$. More generally still, this style of argument yields a corroboration of the earlier claim that FI parameters, appearing here as deformations of the moment maps, are in correspondence with the usual count from the McKay correspondence. Our focus here on the cyclic case is largely for notational consistency with the rest of the paper; see the more abstract \cite{mz:K3HKmath} for a full argument along these lines (i.e., in the D6 frame). That the FI parameters are valued in the Cartan subalgebra of the gauge group of M-theory on $\hat T^4/Z_q$ can alternatively be understood from the fact that the maximal torus of this gauge group is visible in the perturbative type IIA compactification on $\hat T^4/Z_q$ with $q$ D6-branes that defines the Higgs branch picture and the FI parameters are scalars in background 3d $\N=4$ vector multiplets associated to this maximal torus.\footnote{In normal 3d $\N=4$ field theories, these global symmetries to which FI parameters are associated correspond to translations of the dual photons constructed from abelian gauge fields. However, as we discussed in \cite{mz:K3HK}, the FI parameters in this D6-brane theory are not associated to abelian gauge symmetries. Presumably, what is going on here is that this infinite-dimensional gauge theory violates the usual result that instantons in non-abelian gauge theories break these global symmetries. It would be interesting to spell out precisely how these global symmetries act in this theory and why instantons do not break them.}

We finally comment on what to expect from the hyper-K\"ahler quotient for $\xi = 0$. Physically, one expects that the moment map equations imply that $B$ is gauge equivalent to a constant one-form in $\mathrm{Lie}(T^4) \otimes \mf{su}(q) \simeq \mb{C}^2 \otimes_{\mb{C}} \mf{sl}(q)_{\mb{C}}$. As in \cite{mz:K3HK}, we may show this claim directly. We first note that the moment map equations are equivalent to flatness of the connection. (More generally, such moment map equations yield the anti-self-duality equations, but for our topologically trivial bundle, anti-self-duality is equivalent to flatness.) The monodromy representation $\pi_1(T^4) \to SU(q)$ may hence be represented by four commuting matrices, which thus may be simultaneously diagonalized. The commutator contribution to the curvature hence vanishes, so flatness implies $dB = 0$. Note now that the de~Rham differential $d$ is compatible with the equivariant structure; i.e., it commutes with the action of $\text{Ad}_{\sigma_q}^{-1} \circ \iota^*$. Hence the entirety of Hodge theory is compatible with the equivariant structure: the formal adjoint $d^*$ commutes, hence so does $\Delta$, and hence the Hodge splitting respects the equivariant structure. In particular, if we write $B = B_0 + d \lambda$ by Hodge theory, where $B_0$ is harmonic and $d \lambda$ is exact, then $B_0$ and $d \lambda$ are both suitably equivariant. Recall furthermore that harmonic forms on tori are constant, and so the goal is to show that we can gauge away the exact form $d\lambda$. It is not difficult to see that we may furthermore assume $\lambda$ itself is equivariant (for example, by choosing it to vanish at the origin). But then if we take $g = \exp(-i\lambda)$, $g$ is also a satisfactorily equivariant gauge transform with $ig dg^{-1} = - d\lambda$; the $g Bg^{-1}$ contribution to the gauge transformation simply returns $B$ by the common diagonalizability, once again. Hence we may subtract off the $d\lambda$ term and are left with simply the constant, equivariant form $B_0$, as desired. To get from this point to $T^4/Z_q$, we proceed as in \cite{mz:K3HK}; the details are in the next section.

\section{Explicit metrics} \label{sec:metrics}

We now focus on the $T^4/Z_q$ orbifolds and use the perturbation theory of \cite{mz:K3HK} to explicitly describe the metrics. We let $\kappa_q=e^{2\pi i/q}$, $\sigma$ be the diagonal $q\times q$ matrix whose $i$-th diagonal entry is $\kappa_q^{i-1}$ (so the top left entry is 1), and $S_j$ be the length $q$ column vector whose $i$-th entry is $\kappa_q^{(i-1)j}$ (so the top entry is 1).

\subsection{$\CC^2/Z_q$} \label{sec:Zq}

As a warm-up, we study the $\CC^2/Z_q$ orbifold, following \cite{moore:ALEinst}. $U$ and $V$ are $q\times q$ matrices which satisfy
\be U = \kappa_q \sigma U \sigma^\dagger \ , \quad V = \kappa_q^* \sigma V \sigma^\dagger \ . \label{eq:orbConst0} \ee
These constraints are solved by matrices of the form
\be U = \begin{pmatrix} & u_{12} \\ & & u_{23} \\ & & & \ddots \\ & & & & u_{q-1, q} \\ u_{q1} \end{pmatrix} \ , \quad V = \begin{pmatrix} & & & & v_{1q} \\ v_{21} \\ & v_{32} \\ & & \ddots \\ & & & v_{q,q-1} \end{pmatrix} \label{eq:constr0Sol} \ . \ee
Gauge transformations correspond to unitary matrices which satisfy
\be g = \sigma g \sigma^\dagger \ , \label{eq:gaugeConstr0} \ee
i.e. to diagonal unitary matrices. This makes it clear that the gauge group is now $U(1)^q$. By acting on $U$ and $V$ via conjugation by such a $g$, we learn that $(u_{i,i+1},v_{i+1,i}^*)$ (where indices are always defined mod $q$) are the scalars in a hypermultiplet with charge $+1$ under the $i$-th gauge group and $-1$ under the $(i+1)$-th gauge group. That is, we have a circular quiver gauge theory with $q$ nodes corresponding to $U(1)$ gauge groups connected by hypermultiplet links. The diagonal $U(1)$ acts trivially on all of the hypermultiplets.

We now study the orbifold moduli space that manifests when all FI parameters vanish. (With more general FI parameters, one obtains the family of $A_{q-1}$ multi-Eguchi-Hanson ALE manifolds.) The moment maps are
\begin{align}
\mu_+ &= -2[U,V] = -2 \, {\rm diag}(u_{i,i+1} v_{i+1,i} - u_{i-1,i} v_{i,i-1}) \\
\mu_\RR &= [U,U^\dagger] + [V,V^\dagger] = {\rm diag}(|u_{i,i+1}|^2 - |u_{i-1,i}|^2 + |v_{i,i-1}|^2 - |v_{i+1,i}|^2) \ .
\end{align}
Setting $\mu_+=0$ implies that there are $q$ constants $\lambda_i\in \CC$ such that
\be \column{u_{i,i+1}}{v_{i,i-1}} = \lambda_i \column{u_{i-1,i}}{v_{i+1,i}} \ ; \ee
note that composing these equations implies that $\prod_i \lambda_i = 1$, so only $q-1$ of these constants are independent. The real moment map equation $\mu_\RR=0$ now becomes
\be (|\lambda_i|^2-1)(|u_{i-1,i}|^2+|v_{i+1,i}|^2) = 0 \ , \ee
so $|\lambda_i|=1$ for all $i$. We can then use the $i$-th $U(1)$ gauge group to set $\lambda_i=1$. With this gauge choice, we have set all $u_{i,i+1}$ to be equal to a single value, $u$, and similarly $v_{i+1,i}=v$ for all $i$. Lastly, we observe that making a gauge transformation in all $q$ $U(1)$ gauge groups, where the transformation associated to the $i$-th gauge group is given by the phase $\kappa_q^{-i}$, preserves our chosen gauge, and $(u,v)\mapsto (\kappa_q u, \kappa_q^* v)$. So, the moduli space is $\CC^2/Z_q$.

\subsection{K3}

Our starting point is now the $\widehat{U(q)}$ gauge theory with moduli space $\Sym^q(T^4)$ described in \S3.3 of \cite{mz:K3HK}. We first impose the $Z_q$ orbifold projection. Let $\chi$ denote the generator of the $Z_q$ symmetry. Then, the projection is
\begin{align}
X^a_{\chi m;\chi n} &= \chi^a{}_b \sigma X^b_{mn} \sigma^\dagger \nonumber \\
g_{\chi m; \chi n} &= \sigma g_{mn} \sigma^\dagger \ .
\end{align}
Writing
\be X^a = w^a + \sum_n X_n^a e(n) \ , \ee
and noting that
\be w^a_{\chi m; \chi n} = \chi^a{}_b w^b_{mn} \ , \quad e(\ell)_{\chi m; \chi n} = e(\chi^{-1} \ell)_{mn} \ , \ee
the projection takes the form
\begin{align}
X^a_{\chi n} &= \chi^a{}_b \sigma X^b_n \sigma^\dagger \nonumber \\
g_{\chi n} &= \sigma g_n \sigma^\dagger \ . \label{eq:modeProj}
\end{align}
By writing
\be
g(\iota y) = \sum_n g_n e^{i n\cdot (\iota y)} = \sum_n g_n e^{i (\chi^{-1}n)\cdot y} = \sum_n g_{\chi n} e^{i n\cdot y} = \sigma g(y) \sigma^\dagger
\ee
and
\begin{align}
(\iota^* B)^b(y) &= (\chi^{-1})^b{}_a B^a(\iota y)  = (\chi^{-1})^b{}_a \sum_n X_n^a e^{in\cdot (\iota y)} \nonumber \\
&= (\chi^{-1})^b{}_a \sum_n X_n^a e^{i(\chi^{-1}n)\cdot y} = (\chi^{-1})^b{}_a \sum_n X_{\chi n}^a e^{in\cdot y} \nonumber \\
&= (\chi^{-1})^b{}_a \chi^a{}_c \sigma B^c(y) \sigma^\dagger = \sigma B^b(y) \sigma^\dagger \ ,
\end{align}
we find that the projections \eqref{eq:modeProj} coincide with those from the D6-brane picture, \eqref{eq:d6Proj}.

In complex coordinates, we have
\be U_{\chi n} = \kappa_q \sigma U_n \sigma^\dagger \ , \quad V_{\chi n} = \kappa_q^* \sigma V_n \sigma^\dagger \ . \label{eq:orbConst} \ee
Rather than solve these constraints by working with appropriate linear combinations of components of $U_n$'s and $V_n$'s in some basis, as we did in \cite{mz:K3HK}, it will be convenient here to instead work with the redundant variables $U_n$, $V_n$, and keep in mind the constraints \eqref{eq:orbConst}.

The moment maps take the form
\begin{align}
\mu_+ &= -2\sum_n e(n)\brackets{U_n n^v - V_n n^u + \sum_m [U_{n-m}, V_m] - \sum_{i=1}^{q-1} \xi_{n,i,+} \sigma^i } \ , \label{eq:muP} \\
\mu_\RR &= \sum_n e(n)\brackets{ -n^u U^\dagger_{-n} + n^{\bar u} U_n + \sum_m [U_{n+m}, U^\dagger_m] + (U\mapsto V) - \sum_{i=1}^{q-1} \xi_{n,i,\RR} \sigma^i } \ . \label{eq:muR}
\end{align}
We think of $i$ as being defined mod $q$. For later convenience, we also define $\xi_{n,0,+}=\xi_{n,0,\RR}=0$. The FI parameters as written are redundant, as they satisfy a number of constraints. First of all, we consider the constraints from gauge invariance. Following \cite{mz:K3HK}, we compute
\begin{align}
g\sum_n e(n) \xi_{n,i,+} \sigma^i g^\dagger &= g \sum_n e(n) \xi_{n,i,+} \sigma^i \sum_m e(-m) g^\dagger_m \nonumber \\
&= g \sum_n e(n) \xi_{n,i,+} \sum_m e(-m) g^\dagger_{\chi^i m} \sigma^i \nonumber \\
&= g \sum_{n,m} e(n-m) \xi_{n,i,+} g^\dagger_{\chi^i m} \sigma^i \nonumber \\
&= g \sum_{n,m} e(n-m) \xi_{n-m+\chi^{-i} m, i, +} g^\dagger_m \sigma^i \ .
\end{align}
So, we assume that the FI parameter $\xi_{n,i,+}$ only depends on the equivalence class $[n]_i$ of $n$ in $\Lambda/\Lambda_i$, where $\Lambda_i$ is the sublattice of $\Lambda$ consisting of elements of the form $m-\chi^{-i} m$. Then, we have
\begin{align}
g\sum_n e(n) \xi_{n,i,+} \sigma^i g^\dagger &= g \sum_{n,m} e(n-m) \xi_{n,i,+} g^\dagger_m \sigma^i \nonumber \\
&= g g^\dagger \sum_n e(n) \xi_{n,i,+} \sigma^i \nonumber \\
&= \sum_n e(n) \xi_{n,i,+} \sigma^i \ .
\end{align}
We note that $\Lambda_i=\Lambda_{-i}$, since writing any $m\in \Lambda$ as $-\chi^i m'$ gives a relabeling $m-\chi^{-i}m=m'-\chi^i m'$. We also note that both $\chi$ and $-1$ act naturally on equivalence classes:
\begin{align}
\chi(n+m-\chi^i m) &= \chi n + (\chi m)-\chi^i(\chi m) \nonumber \\
-(n+m-\chi^i m) &= -n + (-m) - \chi^i (-m) \ .
\end{align}
When $q=2$, our equivalences classes resided in $\Lambda/2\Lambda$; this agrees with the present equivalence classes, as $\Lambda_1=2\Lambda$ when $q=2$.

Next, we consider the constraint from the orbifold projection on the Lie coalgebra:
\be \sum_i \xi_{\chi^j n, i,+} \sigma^i = \sigma^j \sum_i \xi_{n,i,+}\sigma^i (\sigma^\dagger)^j \ee
implies that $\xi_{\chi^j n,i,+} = \xi_{n,i,+}$ for all $n,i,j$. This is a corollary of the gauge invariance constraint when $j=\pm i$, as follows from $\chi^i n = n - \chi^{-i}(\chi^i n)+(\chi^i n)$ and $\chi^{-i}n = n+\chi^{-i}n-n$, and so it played no role in \cite{mz:K3HK}. However, in general this is a new constraint. We thus find that independent FI parameters $\xi_{n,i,+}$ multiplying $\sigma^i$ correspond to $Z_q$ orbits of equivalence classes in $\Lambda/\Lambda_i$. We will classify these orbits in \S\ref{sec:FI}, but for the present section we will not need the details of this classification. For now, it suffices to note that the counts of these orbits agree with the counts of FI parameters in \S\ref{sec:hk}.

Both of the constraints we have just spelled out hold equally well for $\xi_{n,i,\RR}$. However, there is an extra constraint on the latter, coming from hermiticity of $\mu_\RR$: $\xi^*_{-n,-i,\RR}=\xi_{n,i,\RR}$.

As in \cite{mz:K3HK}, we can relate the moment maps \eqref{eq:muP} and \eqref{eq:muR} to \eqref{eq:D6mu} via
\begin{align}
\sum_{m\in \Lambda_i} e(m+n) \sigma^i &\sim e^{in\cdot y} \sigma^i \sum_{m\in\Lambda_i} e^{im\cdot y} = {\rm vol}(\RR^4/\widehat{\Lambda_i})\,  e^{in\cdot y}\sigma^i \sum_{p\in \widehat{\Lambda_i}} \delta_p(y) \nonumber \\
&= {\rm vol}(\RR^4/\widehat{\Lambda_i})\, \sigma^i \sum_{p\in\widehat{\Lambda_i}} e^{in\cdot p} \delta_p(y) 
= {\rm vol}(\RR^4/\hat \Lambda)\,\frac{1}{|\widehat{\Lambda_i}/\hat\Lambda|} \sigma^i \sum_{p\in\widehat{\Lambda_i}} e^{in\cdot p} \delta_p(y) \ .
\end{align}
Here, $\widehat{\Lambda_i}=\Hom(\Lambda_i,2\pi\ZZ)$ and for any lattice $\Gamma\subset\RR^4$, ${\rm vol}(\RR^4/\Gamma)$ is the volume of a unit cell of $\Gamma$. Summing over all FI parameters, we thus have
\be \sum_{i=1}^{q-1}\sum_{n\in \Lambda/\Lambda_i} \xi_{n,i} \sum_{m\in\Lambda_i} e(m+n)\sigma^i \sim 
{\rm vol}(\RR^4/\hat\Lambda) \sum_{i=1}^{q-1} \frac{1}{|\widehat{\Lambda_i}/\hat\Lambda|} \sum_{n\in\Lambda/\Lambda_i} \xi_{n,i} \sigma^i \sum_{p\in\widehat{\Lambda_i}} e^{in\cdot p} \delta_p(y) \ . \label{eq:fiDist} \ee
(We have omitted the $+/\RR$ subscript, as this identity holds for both real and complex FI parameters.) Noting that $p\in\widehat{\Lambda_i}$ is equivalent to $\iota^i p=p\pmod{\hat\Lambda}$, we see that the natural combinations of FI parameters in the D6-brane language take the form
\be \sum_{i: \iota^i p=p\!\!\!\!\!\pmod{\hat\Lambda}} \frac{1}{|\widehat{\Lambda_i}/\hat\Lambda|} \sum_{n\in\Lambda/\Lambda_i} e^{in\cdot p} \xi_{n,i} \sigma^i \label{eq:TFI} \ee
and are labelled by fixed points $p\in \hat T^4$. We explained in \S\ref{sec:hk} that, after accounting for the constraints on the FI parameters, one finds that they naturally parametrize the (complexified, in the case of the complex FI parameters) Cartan subalgebra of the Lie algebra of the gauge group of M-theory on $\hat T^4/Z_q$.

There is an analogous way to package the FI parameters in the D2-brane picture, where the labelling is now by fixed points $x$ of $T^4$:
\be \sum_{i: \chi^i x=x\!\!\!\!\!\pmod{\Lambda}} \xi_{(\chi^i-1) x,i} \sigma^i \ . \label{eq:D2FI} \ee
Since the FI parameters only depend on $[n]_i\in\Lambda/\Lambda_i$, translating $x$ by an element of $\Lambda$ does not affect this sum. Also, when the FI parameters in this expression are the real FI parameters, the following calculation shows that this expression is Hermitian:
\begin{align}
\parens{\sum_{i: \chi^i x=x\!\!\!\!\!\pmod{\Lambda}} \xi_{(\chi^i-1) x,i,\RR} \sigma^i}^\dagger &= \sum_{i: \chi^i x=x\!\!\!\!\!\pmod{\Lambda}} \xi_{(1-\chi^i) x,-i,\RR} \sigma^{-i} \nonumber \\
&= \sum_{i: \chi^i x=x\!\!\!\!\!\pmod{\Lambda}} \xi_{(1-\chi^{-i}) x,i,\RR} \sigma^{i} \nonumber \\
&= \sum_{i: \chi^i x=x\!\!\!\!\!\pmod{\Lambda}} \xi_{(\chi^i-1) x,i,\RR} \sigma^{i} \ .
\end{align}
Just as the parameters \eqref{eq:TFI} associated to points $p$ related by the $Z_q$ action are equal, the same holds for \eqref{eq:D2FI}.

The reason that \eqref{eq:D2FI} is natural is that, as can be discerned by studying the terms in the perturbative formulae \eqref{eq:jResum} for the K\"ahler forms which are dominant near $x$, it is these FI parameters that should be fed into the hyper-K\"ahler quotient construction of ALE manifolds from \cite{kronheimer:construct,moore:ALEinst} in order to approximate our moduli space near $x$. In particular, this combination of FI parameters naturally parametrizes the (complexified, in the case of the complex FI parameters) Cartan subalgebra of the Lie algebra of the gauge group of M-theory on $T^4/Z_q$. To see this, we first observe that for each $i$, the FI parameter $\xi_{n,i}$ associated to each equivalence class $[n]_i$ appears exactly once. This follows from the fact that $\chi^ix=x\pmod{\Lambda}$ is equivalent to $x\in (\chi^i-1)^{-1}\Lambda=\Lambda(\kappa_q^i-1)^{-1}$ (where we are using the right action of the quaternions on $\RR^4$ in this last equality). So, as $x$ runs over all elements of $\brackets{\Lambda(\kappa_q^i-1)^{-1}}/\Lambda$, $(\chi^i-1)x$ runs over all elements of $\Lambda/\Lambda_i$. Therefore, besides the Hermiticity condition on the real FI parameters, the coefficients that appear in \eqref{eq:D2FI} as $x$ varies over representatives of $Z_q$-orbits of fixed points and $i$ ranges over its allowed values are all independent. Lastly, for a fixed point stabilized by $Z_{q/j}$, there are $q/j-1$ (nonzero) values of $i$ that contribute to \eqref{eq:D2FI}. This is precisely the rank of the $A_{q/j-1}$ Lie algebra associated, via the McKay correspondence, to $Z_{q/j}$.

We note that \eqref{eq:TFI} can equivalently be written as
\be \sum_{i: \iota^ip=p\!\!\!\!\!\pmod{\hat\Lambda}} \frac{1}{|\widehat{\Lambda_i}/\hat\Lambda|} \sum_{x:\chi^i x=x\!\!\!\!\!\pmod{\Lambda}} e^{i ((\chi^i-1)x)\cdot p} \xi_{(\chi^i-1)x,i} \sigma^i \ . \ee
This makes it clear that, while the types of orbifold singularities of $T^4/Z_q$ and $\hat T^4/Z_q$ are the same, and so the Lie algebras of the gauge groups of M-theory compactified on these orbifolds are isomorphic, the map between their Cartan subalgebras given by relating \eqref{eq:D2FI} and \eqref{eq:TFI}, with the parameters $\xi_{n,i}$ serving as intermediaries, does not respect the decomposition of the Lie algebras into their simple factors. That is, physically, the FI parameters associated to a fixed point of $T^4$ in the D2-brane picture affect the FI parameters associated to multiple fixed points of $\hat T^4$ in the D6-brane picture, and vice versa.

The Cartan subalgebras of these isomorphic M-theory gauge algebras have a natural action of the Weyl group. From the point of view of the perturbative type IIA compactification on $T^4/Z_q$, the existence of this action is accidental, and there is no reason that it should leave the moduli space of our D2-brane invariant, since the gauge symmetry in spacetime has been broken by discrete B-flux to its maximal torus. However, the Coulomb branch picture makes it clear that this actually is a redundancy in the space of FI parameters, since in this picture the entire non-abelian symmetry group of M-theory on $T^4/Z_q$ is present. So, the natural action of the Weyl group on the FI parameters, as packaged in \eqref{eq:D2FI}, is actually a gauge symmetry of the parameter space.\footnote{This is only in the sense that the metric on moduli space is invariant: quantities which are not protected by a non-renormalization theorem need not be invariant in the Higgs branch theory.} We stress the lesson from the end of the prior paragraph: this action differs from the natural action of the Weyl group on the FI parameters when packaged as in \eqref{eq:TFI}. That is, the automorphism of the Cartan subalgebra of the gauge algebra of M-theory on $T^4/Z_q$ given by relating \eqref{eq:D2FI} and \eqref{eq:TFI} is not Weyl-equivariant.

One might therefore ask if the Weyl action on \eqref{eq:TFI} gives a second gauge symmetry of the parameter space. As evidence against this, we note that the non-renormalization argument that we used in \cite{mz:K3HK} to relate the Higgs and Coulomb pictures does not apply here, since the infinite coupling limit of our D6-brane configuration is M-theory on the product of the resolution of $\hat T^4/Z_q$ with an ALE $A_{q-1}$ singularity, and there is no way to disentangle the probe from the background in this configuration. There are a number of ways that one could completely determine whether or not this second Weyl action is a gauge symmetry of the parameter space. One is to study the subgroup of $O(3,19,\ZZ)$ that stabilizes points in the $T^4/Z_q$ locus in the $O(3,19)/(O(3)\times O(19))$ Teichm\"uller space of the moduli space of unit-volume Ricci-flat K3 metrics. Alternatively, one could determine whether the Higgs and/or Coulomb branch constructions are insensitive to this Weyl group action. Looking ahead quite a bit, we can report that we took this route and established that this Weyl action is, indeed, not a gauge symmetry, by showing that the functions $F_{n,p,q}$ defined in \eqref{eq:Fdef} are generally not invariant under this action.\footnote{More precisely, we showed this for the Coulomb branches of the compactified 4d $\N=2$ field theories mentioned in the introduction, where the D2-brane probing $T^4$ and D6-branes wrapping $\hat T^4$ are, respectively, replaced by a D2-brane probing $\CC\times T^2$ and D4-branes wrapping $\hat T^2$ in $\CC\times \hat T^2$, but we expect this result to hold for K3 as well. (In fact, even for K3 there are presumably other Weyl actions associated to intermediate configurations where we T-dualize only some of the four 1-cycles of $T^2$.) The failure of the non-renormalization theorem is now clear from the fact that the infinite coupling limit of the D4-branes is a stack of M5-branes, so the low-energy non-linear sigma models to the respective moduli spaces of these configurations do not even exist in the same number of dimensions. From the D4-branes' point of view, the KK momenta associated to this extra dimension are D0-branes, or instanton particles.}

We next turn to the orbifold moduli space, with all FI parameters vanishing. We explained in \S\ref{sec:hk} that the non-zero modes either vanish by the moment map equations or can be gauged away. The remaining moment map equations are just the zero mode equations, which coincide with those of \S\ref{sec:Zq}. Gauge transformations of the form $g=g_0 e(0)$ preserve the condition that $U_n=V_n=0$ for $n\not=0$. So, we can utilize these equations and gauge transformations, as in \S\ref{sec:Zq}, to reduce $U$ and $V$ to be of the form $U=U^{\rm orb}_0 e(0)$, $V=V^{\rm orb}_0 e(0)$, where
\be U^{\rm orb}_0 = u s \ , \quad V_0^{\rm orb} = v s^\dagger \ , \quad s = \begin{pmatrix} & 1 \\ & & 1 \\ & & & \ddots \\ & & & & 1 \\ 1 \end{pmatrix} \ , \ee
and where $u,v$ are well-defined up to $(u,v)\sim (\kappa_q u, \kappa_q^* v)$. We note that $s$ and $s^\dagger$ may be unitarily diagonalized by a discrete Fourier transform, and in this basis it is manifest that the eigenvalues of $U$ and $V$ are, respectively, the $\ZZ^4 \rtimes Z_q$ images of $u$ and $v$. This accords with the construction of the $T^4/Z_q$ orbifold gauge theory in terms of an infinite number of D2-brane probes of $\RR^4$ which are subject to $\ZZ^4 \rtimes Z_q$ orbifold projections.

Lastly, we show that quasi-large gauge transformations identify $(u,v)\sim (u+n^u, v+n^v)$, so that we have $T^4/Z_q$. Specifically, the relevant transformations are
\be \gamma_n = \frac{1}{q} \sum_{i=0}^{q-1} e(\chi^i n)\sum_{j=0}^{q-1} \kappa_q^{-ij} s^j \ . \ee
Note that when $q=2$, these are
\be \gamma_n = \half \brackets{e(n) (I+\sigma_x) + e(-n) (I-\sigma_x)} = \frac{e(n)+e(-n)}{2} I + \frac{e(n)-e(-n)}{2} \sigma_x = e^{in^a y_a \sigma_x} \ , \ee
so they generalize the quasi-large gauge transformations that were important in \cite{mz:K3HK}. Writing them as
\be \gamma_n = \frac{1}{q} \sum_i e(\chi^i n) \sum_j \sigma^i s^j (\sigma^\dagger)^i \ee
makes it clear that they satisfy the orbifold projection, \eqref{eq:modeProj}. Next, we verify that they are unitary:
\begin{align}
\gamma_n^\dagger\gamma_n &= \frac{1}{q^2} \sum_{i,j,i',j'} \kappa_q^{i'j'-ij} e(\chi^in-\chi^{i'} n) s^{j-j'} \nonumber \\
&= \frac{1}{q^2} \sum_{i,j,i',j'} \kappa_q^{j(i'-i)+i'j'} e(\chi^in-\chi^{i'}n) s^{-j'} \label{eq:relabel} \\
&= \frac{1}{q} \sum_{i,j'} \kappa_q^{ij'} e(0) s^{-j'} \nonumber \\
&= e(0) \ .
\end{align}
(\eqref{eq:relabel} obtained by replacing $j'\mapsto j'+j$, the next equality followed from performing the sum over $j$, and the next involved performing the sum over $i$.) So, these are indeed valid gauge transformations. That
\be \gamma_n U_0^{\rm orb} \gamma_n^\dagger = U_0^{\rm orb} \ , \ee
and similarly for $V_0^{\rm orb}$, is clear from the fact that $[s,s]=[s,s^\dagger]=0$. So, the only effect they have is from
\begin{align}
\gamma_n w^a \gamma_n^\dagger &= w^a + [\gamma_n,w^a] \gamma_n^\dagger \nonumber \\
&= w^a + \frac{1}{q} \sum_i [e(\chi^i n),w^a] \sum_j \kappa_q^{-ij} s^j \gamma_n^\dagger \nonumber \\
&= w^a + \frac{1}{q} \sum_i (\chi^i n)^a e(\chi^i n) \sum_j \kappa_q^{-ij} s^j \gamma_n^\dagger \nonumber \\
&= w^a + \frac{1}{q^2} \sum_{i,i',j'} \kappa_q^{j'(i-i')} (\chi^{i'} n)^a e(\chi^i n)\sum_j \kappa_q^{-ij} s^j \gamma_n^\dagger \nonumber \\
&= w^a + \frac{1}{q^2} \sum_{i,i',j,j'} \kappa_q^{-i'j'} (\chi^{i'} n)^a e(\chi^i n) \kappa_q^{-ij} s^{j+j'} \gamma_n^\dagger \nonumber \\
&= w^a + \frac{1}{q} \sum_{i',j'} \kappa_q^{-i' j'} s^{j'} (\chi^{i'} n)^a \gamma_n \gamma_n^\dagger \nonumber \\
&= w^a + \frac{1}{q} \sum_{i',j'} \kappa_q^{-i' j'} s^{j'} (\chi^{i'} n)^a \ .
\end{align}
In complex coordinates, this takes the form
\begin{align}
\gamma_n w^u \gamma_n^\dagger &= w^u + \frac{1}{q}\sum_{i',j'} \kappa_q^{i'(1-j')} s^{j'} n^u \nonumber \\
&= w^u + s n^u \nonumber \\
\gamma_n w^v \gamma_n^\dagger &= w^v + \frac{1}{q}\sum_{i',j'} \kappa_q^{i'(-1-j')} s^{j'} n^v \nonumber \\
&= w^v + s^\dagger n^v \ .
\end{align}
So, as promised, these transformations have the effect of identifying $(u,v)\sim (u+n^u, v+n^v)$.

Now, we study the first-order corrections induced by turning on the FI parameters. At this order, the non-zero mode moment map equations take the form
\begin{align}
\sum_i \xi_{n,i,+} \sigma^i &= U_n n^v - V_n n^u + [U_0^{\rm orb}, V_n] + [U_n, V_0^{\rm orb}] \nonumber \\
\sum_i \xi_{n,i,\RR} \sigma^i &= -n^u U^\dagger_{-n} + n^{\bar u} U_n + [U_0^{\rm orb}, U^\dagger_{-n}] + [U_n, (U^{\rm orb}_0)^\dagger] + (U\mapsto V) \ . \label{eq:approxMu}
\end{align}
The zero mode equations also take this form if we write $U_0=U^{\rm orb}_0 + \Delta U_0$, $V_0=V^{\rm orb}_0 + \Delta V_0$, and replace $U_{\pm n}$ and $V_{\pm n}$ in \eqref{eq:approxMu} with $\Delta U_0$ and $\Delta V_0$, respectively:
\begin{align}
\sum_i \xi_{0,i,+} \sigma^i &= [U^{\rm orb}_0, \Delta V_0] + [\Delta U_0, V^{\rm orb}_0] \nonumber \\
\sum_i \xi_{0,i,\RR} \sigma^i &= [U^{\rm orb}_0, \Delta U_0^\dagger] + [\Delta U_0, (U^{\rm orb}_0)^\dagger] + (U\mapsto V) \ . \label{eq:approxMu0}
\end{align}
In order to make these splittings of $U_0$ and $V_0$ unambiguous, we require
\be \Tr \, (U_0^{\rm orb})^\dagger \Delta U_0 = \Tr \, (V_0^{\rm orb})^\dagger \Delta V_0 =0 \ ; \label{eq:ortho0} \ee
intuitively, any piece of $\Delta U_0$ parallel to $U_0^{\rm orb}$ (in the complexified $\mf{su}(q)$ Lie algebra) is equivalent to a change in $u$, and so we require the former to not exist.\footnote{A priori, one might have thought that \eqref{eq:approxMu0} should include order $\xi^2$ corrections, since, for example, wedging an order $\xi^2$ contribution to $d\Delta U_0$ with $dV^{\rm orb}_0$ in \eqref{eq:flatI} could yield an order $\xi^2$ contribution to $\omega_I$. However, \eqref{eq:ortho0} implies that all corrections of this sort vanish, and so the order $\xi$ approximation \eqref{eq:approxMu0} suffices for determining the leading corrections to the metric.} Lastly, we have the freedom to make order $\xi$ gauge transformations, which take the form
\be \delta U_n = in^u h_n + i[h_n, U^{\rm orb}_0] \ , \quad \delta V_n = in^v h_n + i[h_n, V^{\rm orb}_0] \ , \label{eq:smallGaugeTrans} \ee
where $h_n$ is order $\xi$, $h^\dagger_n=h_{-n}$, and $h_{\chi n} = \sigma h_n \sigma^\dagger$. Even with the FI parameters turned on, the $\mf{u}(1)$ part of the problem decouples, so we can require that $\Tr U_n=\Tr V_n=\Tr h_n=0$ for all $n$.

As in \cite{mz:K3HK}, we find that the infinite-dimensional non-linear problem has reduced to a finite-dimensional linear one. The variables involved are $U_{\pm n}$, $V_{\pm n}$ when $q$ is odd and $U_n$, $V_n$ when $q$ is even (since in the latter case $U_{-n}=-\sigma^{q/2} U_n \sigma^{-q/2}$, and similarly for $V$). After using the orbifold constraints \eqref{eq:orbConst}, one finds that the moment map equations with $n$ replaced by $\chi^j n$ involve these variables and should a priori be included. However, by using $U_0^{\rm orb} = \kappa_q^j \sigma^j U_0^{\rm orb} \sigma^{-j}$ and $V_0^{\rm orb} = \kappa_q^{-j} \sigma^j V_0^{\rm orb} \sigma^{-j}$ (which follow from \eqref{eq:orbConst0}), one finds that these are equivalent to the $j=0$ equations:
\begin{align}
\sum_i \xi_{n,i,+}\sigma^i &= \sigma^j U_n \sigma^{-j} n^v - \sigma^j V_n \sigma^{-j} n^u + [U_0^{\rm orb}, \kappa_q^{-j} \sigma^j V_n \sigma^{-j}] + [\kappa_q^j \sigma^j U_n \sigma^{-j}, V_0^{\rm orb}] \nonumber \\
&= \sigma^j\parens{U_n n^v - V_n n^u + [U_0^{\rm orb}, V_n] + [U_n, V_0^{\rm orb}]} \sigma^{-j} \ , \nonumber \\
\sum_i \xi_{n,i,\RR} \sigma^i &= - \sigma^j U_{-n}^\dagger \sigma^{-j} n^u + \sigma^j U_n \sigma^{-j} n^{\bar u} + [U^{\rm orb}_0, \kappa_q^{-j} \sigma^j U_{-n}^\dagger \sigma^{-j}] + [\kappa_q^j \sigma^j U_n \sigma^{-j}, (U_0^{\rm orb})^\dagger] + (U\mapsto V) \nonumber \\
&= \sigma^j\parens{ -U^\dagger_{-n} n^u + U_n n^{\bar u} + [U_0^{\rm orb}, U^\dagger_{-n}] + [U_n, (U^{\rm orb}_0)^\dagger] + (U\mapsto V)} \sigma^{-j} \ .
\end{align}
When $q$ is odd, it is still important to include the equations obtained by negating $n$. After picking a basis for the $\mf{su}(q)$ Lie algebra,\footnote{Such as the generalized Gell-Mann matrices described at \url{https://en.wikipedia.org/wiki/Generalizations_of_Pauli_matrices\#Construction}.} one thus has $3(q^2-1)$ real equations and $4(q^2-1)$ real unknowns for $q$ even and twice as many of each for $q$ odd. Concretely, these numbers are $(9,12)$ for $q=2$, $(48,64)$ for $q=3$, $(45,60)$ for $q=4$, and $(105,140)$ for $q=6$.\footnote{For $n=0$, the counting is quite different. First of all, the count is the same for both even and odd $q$, since $-0=0$. Second, \eqref{eq:constr0Sol} makes it clear that we have only $4q$ real unknowns, while \eqref{eq:gaugeConstr0} implies that we have only $q-1$ gauge transformations of the form \eqref{eq:smallGaugeTrans}. Analogously, \eqref{eq:approxMu0} encodes only $3(q-1)$ equations. The remaining four are provided by \eqref{eq:ortho0}.} By employing a computer algebra system, one can find the least norm solution, as described in \cite{mz:K3HK}.

However, this solution has an elegant expression if we do not pick a basis. So, we will instead state it and then verify that it solves \eqref{eq:approxMu}. In order to motivate it, we note that the least norm solution from \cite{mz:K3HK} (specifically, equations (3.150)-(3.155), taking into account the factor of $\half$ in (3.159)) may be written as
\begin{align}
U_n &= \frac{1}{4} \sum_{j =0}^1 \frac{2 \xi_{n, +} \bar N^v_j + \xi_{n, \mb{R}} N^u_{j}}{D_{j}} S_{j +1 } S_{ j }^{\dagger} \nonumber \\
V_n &= \frac{1}{4} \sum_{j = 0}^1 \frac{-2 \xi_{n, +}\bar N^u_{j}+ \xi_{n, \mb{R}} N^v_{j}}{D_{j}}  S_{ j +1 } S_{ j}^{\dagger} \ ,
\end{align}
where $N^u_{j} = n^u + 2\kappa_2^{j} u$, $N^v_j = n^v + 2\kappa_2^{-j} v$, and $D_j = |N^u_j|^2 + |N^v_j|^2$. We now claim that a similar result holds more generally:
\begin{align}
U_n &= \frac{1}{2q} \sum_{j=0}^{q-1} \sum_{i = 1}^{q - 1} \frac{2 \xi_{n,i,+} \bar N^v_{i,j} + \xi_{n, i, \mb{R}} N^u_{i,j}}{D_{i,j}} S_{i+j} S_{j}^{\dagger} \nonumber \\
V_n &= \frac{1}{2q} \sum_{j=0}^{q-1} \sum_{i = 1}^{q - 1} \frac{-2 \xi_{n,i,+} \bar N^u_{i,j} + \xi_{n, i, \mb{R}} N^v_{i,j}}{D_{i,j}} S_{i+j} S_{j}^{\dagger} \ , \label{eq:UVsol}
\end{align}
where
\be N^u_{i,j} = n^u + (1 - \kappa_q^{i}) \kappa_q^j u \ , \quad N^v_{i,j} = n^v + (1 - \kappa_q^{-i}) \kappa_q^{-j} v \ , \quad D_{i,j} = |N^u_{i,j}|^2+|N^v_{i,j}|^2 \ . \ee
When we want to make explicit the dependence of these quantities on $n$, we will write expressions such as $N_{n,i,j}^u$ and $D_{n,i,j}$. For later use, we note that
\begin{align}
N^u_{-n,i,j} &= -n^u+(1-\kappa_q^i)\kappa_q^{j}u = -(n^u+(1-\kappa_q^{-i})\kappa_q^{i+j} u) = -N^u_{n,-i,i+j} \nonumber \\
N^v_{-n,i,j} &= -n^v+(1-\kappa_q^{-i})\kappa_q^{-j}v = -(n^v+(1-\kappa_q^{i})\kappa_q^{-i-j} v) = -N^v_{n,-i,i+j} \ , \label{eq:negN}
\end{align}
and so
\begin{align}
U_{-n} &= -\frac{1}{2q}\sum_{j=0}^{q-1}\sum_{i=1}^{q-1} \frac{2\xi_{-n,i,+}\bar N^v_{-i,i+j}+\xi_{-n,i,\RR}N^u_{-i,i+j}}{D_{-i,i+j}} S_{i+j} S^\dagger_j \nonumber \\
&= - \frac{1}{2q} \sum_{j=0}^{q-1}\sum_{i=1}^{q-1} \frac{2\xi_{-n,-i,+}\bar N^v_{i,j} + \xi^*_{n,i,\RR} N^u_{i,j}}{D_{i,j}} S_{j}S^\dagger_{i+j} \nonumber \\
V_{-n} &= - \frac{1}{2q}\sum_{j=0}^{q-1}\sum_{i=1}^{q-1} \frac{-2\xi_{-n,i,+}\bar N^u_{-i,i+j}+\xi_{-n,i,\RR}N^v_{-i,i+j}}{D_{-i,i+j}} S_{i+j} S^\dagger_j  \nonumber \\
&= - \frac{1}{2q}\sum_{j=0}^{q-1}\sum_{i=1}^{q-1} \frac{-2\xi_{-n,-i,+}\bar N^u_{i,j}+\xi^*_{n,i,\RR}N^v_{i,j}}{D_{i,j}} S_j S^\dagger_{i+j} 
\end{align}
and
\begin{align}
U_{-n}^\dagger &= -\frac{1}{2q} \sum_{j=0}^{q-1}\sum_{i=1}^{q-1} \frac{2\xi^*_{-n,-i,+} N^v_{i,j} + \xi_{n,i,\RR} \bar N^u_{i,j}}{D_{i,j}} S_{i+j}S_j^\dagger \nonumber \\
V_{-n}^\dagger &= - \frac{1}{2q}\sum_{j=0}^{q-1}\sum_{i=1}^{q-1} \frac{-2\xi^*_{-n,-i,+} N^u_{i,j} + \xi_{n,i,\RR} \bar N^v_{i,j}}{D_{i,j}} S_{i+j}S^\dagger_j \ .
\end{align}
Of course, we also have $-n=\chi^{q/2}n$ when $q$ is even, but \eqref{eq:negN} is particularly useful (compared to repeated use of \eqref{eq:chiN}) because it works even for $q=3$.

We first verify that these expressions are trivially traceless, since $\Tr S_i S_j^\dagger = q \delta_{i,j}$. Next, we verify that they satisfy the orbifold projections. This relies on the observations
\begin{align}
N_{\chi n,i,j}^u &= \kappa_q n^u + (1-\kappa_q^i)\kappa_q^{j} u = \kappa_q (n^u + (1-\kappa_q^i)\kappa_q^{j-1} u) = \kappa_q N_{n,i,j-1}^u \nonumber \\
N_{\chi n,i,j}^v &= \kappa_q^{-1} n^v + (1-\kappa_q^{-i})\kappa_q^{-j} v = \kappa_q^{-1} (n^v + (1-\kappa_q^{-i})\kappa_q^{-j+1} v) = \kappa_q^{-1} N_{n,i,j-1}^v \ . \label{eq:chiN}
\end{align}
Recalling that $\xi_{\chi n,i,+} = \xi_{n,i,+}$, and similarly for $\xi_{n,i,\RR}$, we thus have
\begin{align}
U_{\chi n} &= \frac{\kappa_q}{2q}\sum_{j=0}^{q-1}\sum_{i=1}^{q-1} \frac{2\xi_{n,i,+} \bar N^v_{i,j-1} + \xi_{n,i,\RR} N^u_{i,j-1}}{D_{i,j-1}} S_{i+j}S^\dagger_j \nonumber \\
&= \frac{\kappa_q}{2q}\sum_{j=0}^{q-1}\sum_{i=1}^{q-1} \frac{2\xi_{n,i,+} \bar N^v_{i,j} + \xi_{n,i,\RR} N^u_{i,j}}{D_{i,j}} S_{i+j+1}S^\dagger_{j+1} \nonumber \\
&= \frac{\kappa_q}{2q}\sum_{j=0}^{q-1}\sum_{i=1}^{q-1} \frac{2\xi_{n,i,+} \bar N^v_{i,j} + \xi_{n,i,\RR} N^u_{i,j}}{D_{i,j}} (\sigma S_{i+j})(\sigma S_{j})^\dagger \nonumber \\
&= \kappa_q \sigma U_n \sigma^\dagger \ ,
\end{align}
and similarly for $V$.

Thirdly, we verify the moment map equations. The key commutation relations that make these work are
\be [S_{i+j} S_j^\dagger,s] = (1-\kappa_q^i)\kappa_q^j S_{i+j} S_j^\dagger \ee
and its adjoint,
\be [S_{i+j} S_j^\dagger, s^\dagger] = (1-\kappa_q^{-i})\kappa_q^{-j} S_{i+j} S_j^\dagger \ . \ee
We also make use of
\be \sum_{j=0}^{q-1} S_{i+j} S_j^\dagger = q\sigma^i \ . \label{eq:Ssum} \ee
We then compute
\begin{align}
U_n n^v &- V_n n^u + [U_0^{\rm orb}, V_n] + [U_n, V_0^{\rm orb}] \nonumber \\
&= \frac{1}{2q}\sum_{j=0}^{q-1}\sum_{i=1}^{q-1} \frac{-N^u_{i,j}(-2\xi_{n,i,+}\bar N^u_{i,j}+\xi_{n,i,\RR} N^v_{i,j})+N^v_{i,j}(2\xi_{n,i,+}\bar N^v_{i,j}+\xi_{n,i,\RR} N^u_{i,j})}{D_{i,j}} S_{i+j} S_j^\dagger \nonumber \\
&= \frac{1}{q} \sum_{i=1}^{q-1} \xi_{n,i,+} \sum_{j=0}^{q-1} S_{i+j}S_j^\dagger 
= \sum_i \xi_{n,i,+} \sigma^i 
\end{align}
and
\begin{align}
-n^u U^\dagger_{-n} &+ n^{\bar u} U_n + [U_0^{\rm orb}, U^\dagger_{-n}] + [U_n, (U^{\rm orb}_0)^\dagger] + (U\mapsto V) \nonumber \\
&= \frac{1}{2q} \sum_{j=0}^{q-1} \sum_{i=1}^{q-1} \brackets{ \frac{\bar N^u_{i,j}(2\xi_{n,i,+}\bar N^v_{i,j}+\xi_{n,i,\RR}N^u_{i,j}) + \bar N^v_{i,j}(-2\xi_{n,i,+}\bar N^u_{i,j}+\xi_{n,i,\RR}N^v_{i,j}) }{D_{i,j}} \right. \nonumber \\
&\left.\quad+ \frac{N^u_{i,j}(2\xi^*_{-n,-i,+}N^v_{i,j}+\xi_{n,i,\RR}\bar N^u_{i,j}) + N^v_{i,j}(-2\xi^*_{-n,-i,+}N^u_{i,j}+\xi_{n,i,\RR}\bar N^v_{i,j})}{D_{i,j}} } S_{i+j}S_j^\dagger  \nonumber \\
&= \frac{1}{q} \sum_{i=1}^{q-1} \xi_{n,i,\RR} \sum_{j=0}^{q-1} S_{i+j}S_j^\dagger = \sum_i \xi_{n,i,\RR} \sigma^i \ .
\end{align}
We should also check the extra zero mode equation \eqref{eq:ortho0}:
\be
\Tr (U_0^{\rm orb})^\dagger \Delta U_0 = \frac{\bar u}{2q} \sum_{j=0}^{q-1}\sum_{i=1}^{q-1} \frac{2\xi_{0,i,+} \bar N^v_{i,j} + \xi_{0,i,\RR} N^u_{i,j}}{D_{0,i,j}} \Tr s^\dagger S_{i+j}S_j^\dagger = 0 \ ,
\ee
since $\Tr s^\dagger S_{i+j}S^\dagger_j = q\kappa_q^{-j} \delta_{i,0}$. We similarly have $\Tr (V_0^{\rm orb})^\dagger \Delta V_0=0$, as $\Tr s S_{i+j}S_j^\dagger=q\delta_{i,0} \kappa_q^j$.

We have now verified that \eqref{eq:UVsol} solves the moment map equations. Although this is not necessary for deriving the metric, we now additionally verify that this solution is the least norm solution with respect to the norm $\|(U_n,V_n,U_{-n},V_{-n})\|^2=\Tr (U_n^\dagger U_n + V_n^\dagger V_n+U_{-n}^\dagger U_{-n} + V_{-n}^\dagger V_{-n})$. To do so, we demonstrate that our solution is orthogonal to generators of gauge transformations, i.e. that\footnote{For $n=0$, we replace $U_n$ and $V_n$ with $\Delta U_0$ and $\Delta V_0$. It does happen to be the case that the zeroth order solution $(U^{\rm orb}_0,V^{\rm orb}_0)$ is orthogonal to first-order gauge transformations, but this is not relevant here, as the norm we a priori should choose to minimize is $\|(\Delta U_0,\Delta V_0)\|^2=\Tr (\Delta U_0^\dagger \Delta U_0 + \Delta V_0^\dagger \Delta V_0)$, not $\Tr(U_0^\dagger U_0+V_0^\dagger V_0)$. For, the former choice is the one that is amenable to the linear-algebraic approach of \cite{mz:K3HK}. In addition, were it not the case that $\Real \Tr\brackets{(U_0^{\rm orb})^\dagger \delta U_0 + (V_0^{\rm orb})^\dagger \delta V_0 }=0$, then minimizing the latter norm would require adding zeroth order quantities to $(\Delta U_0,\Delta V_0)$. Lastly, it would not be well-motivated to minimize this norm over first-order gauge transformations of the form \eqref{eq:smallGaugeTrans}, since the inner product of the first correction to this with the zeroth order expressions $U_0^{\rm orb}$ and $V_0^{\rm orb}$ would contribute at the same order.}
\be \delta \Tr (U_n^\dagger U_n + V_n^\dagger V_n+U_{-n}^\dagger U_{-n} + V_{-n}^\dagger V_{-n}) = 2\Real \Tr (U_n^\dagger \delta U_n + V_n^\dagger \delta V_n+U_{-n}^\dagger \delta U_{-n} + V_{-n}^\dagger \delta V_{-n}) = 0 \ . \label{eq:orthoLN} \ee
Recalling \eqref{eq:smallGaugeTrans}, we have
\begin{align}
\Tr U_n^\dagger \delta U_n &= \Tr\parens{in^u h_n U_n^\dagger + i[h_n,U_0^{\rm orb}] U_n^\dagger } \nonumber \\
&= \Tr \brackets{ih_n\parens{n^u  U_n^\dagger + [U_0^{\rm orb}, U_n^\dagger] }} \ ,
\end{align}
so
\be 2\Real \Tr U_n^\dagger \delta U_n = \Tr\brackets{ih_n(n^u U^\dagger_n+[U_0^{\rm orb},U_n^\dagger]) - ih_{-n} (n^{\bar u}U_n+[U_n,(U_0^{\rm orb})^\dagger])} \ , \ee
where we have used $h_n^\dagger=h_{-n}$. Performing similar manipulations for the other terms in \eqref{eq:orthoLN} and then collecting the terms that involve $h_{-n}$, we find that we want to show that
\be -n^u U_{-n}^\dagger - n^{\bar u} U_{n} + [U_0^{\rm orb},U_{-n}^\dagger] + [(U_0^{\rm orb})^\dagger,U_{n}] + (U\mapsto V) = 0 \ . \label{eq:LNgauge} \ee
(Said another way, this is our gauge fixing condition.) To do so, we compute
\begin{align}
 -n^u U_{-n}^\dagger &- n^{\bar u} U_{n} + [U_0^{\rm orb},U_{-n}^\dagger] + [(U_0^{\rm orb})^\dagger,U_{n}] + (U\mapsto V) \nonumber \\
&= \frac{1}{2q}\sum_{j=0}^{q-1}\sum_{i=1}^{q-1} \brackets{\frac{N^u_{i,j}(2\xi^*_{-n,-i,+}N^v_{i,j}+\xi_{n,i,\RR}\bar N^u_{i,j}) - \bar N^u_{i,j}(2\xi_{n,i,+}\bar N^v_{i,j}+\xi_{n,i,\RR}N^u_{i,j}) }{D_{i,j}} \right. \nonumber \\
&\left.\quad + \frac{N^v_{i,j}(-2\xi^*_{-n,-i,+}N^u_{i,j}+\xi_{n,i,\RR} \bar N^v_{i,j})-\bar N^v_{i,j}(-2\xi_{n,i,+}\bar N^u_{i,j}+\xi_{n,i,\RR}N^v_{i,j})}{D_{i,j}} } S_{i+j}S^\dagger_j \nonumber \\
&= 0 \ . \label{eq:LN}
\end{align}
This completes our verification of \eqref{eq:UVsol}.

We pause at this point to reformulate our gauge fixing condition in the T-dual D6-brane language. The norm we are minimizing is simply the $L^2$ norm $\|\Delta B\|^2=\int_{\hat T^4} \Tr (\Delta B)\wedge *(\Delta B)$ of the difference between the connection 1-form $B^a$ and its orbifold value $(B^{\rm orb})^a$.\footnote{This is a bit dishonest, since the singularities of $\Delta B$ make it non-normalizable. Really, the difference should probably be taken between $B^a$ and a fixed connection with prescribed singularities, while $h$ should be required to vanish at the fixed points. We will ignore such subtleties in this paragraph, as we expect that the ideas contained herein can be made precise along these lines, and we will likely do so in \cite{mz:K3HKmath}. For now, we note that one may write $d_{B^{\rm orb}} * B$ in momentum space and directly verify the conclusion that our gauge choice \eqref{eq:LNgauge} is equivalent to $d_{B^{\rm orb}} * B=0$; the present `derivation' can be viewed as motivation for this result.} Since the infinitesimal gauge transformations \eqref{eq:smallGaugeTrans} take the form
\be \delta B^a = \partial^a h - i[(B^{\rm orb})^a,h] = \nabla^a_{B^{\rm orb}} h \ , \ee
we have
\begin{align}
\delta \|\Delta B\|^2 &= \int_{\hat T^4} \Tr \parens{\nabla_{B^{\rm orb}} h\wedge *\Delta B + \Delta B\wedge *\nabla_{B^{\rm orb}} h} \nonumber \\
&= -2\int_{\hat T^4} \Tr h\, d_{B^{\rm orb}} \!*\!B \ , \label{eq:lnCoulomb}
\end{align}
where $d_{B^{\rm orb}}$ is the exterior covariant derivative. We thus see that our least norm gauge \eqref{eq:LNgauge} coincides with the non-Abelian Coulomb gauge $d_{B^{\rm orb}} *B=0$ with respect to $B^{\rm orb}$.

Now for the K\"ahler forms. Our starting point is the triplet of K\"ahler forms on $\CC^{2\cdot q^2\infty^4}$:\footnote{Really, we have $\CC^{2\cdot q\infty^4}$, thanks to the orbifold projections. But, the K\"ahler forms thereon are obtained by pulling back via the orbifold projections, and so the K\"ahler forms we obtain on K3 are the same regardless of which starting point we take.}\textsuperscript{,}\footnote{Our choice of normalization here differs from that of \cite{mz:K3HK} by a factor of $1/2q$. \label{ft:norm}}
\begin{align}
\omega_I &= \frac{i}{2q} \sum_n \Tr\parens{ -dU_n \wedge dV_{-n} + dU_n^\dagger \wedge dV^\dagger_{-n} } \label{eq:flatI} \\
\omega_J &= -\frac{1}{2q} \sum_n \Tr\parens{ dU_n \wedge dV_{-n} + dU_n^\dagger \wedge dV^\dagger_{-n} } \\
\omega_K &= \frac{i}{2q} \sum_n \Tr\parens{ dU_n \wedge dU_n^\dagger + dV_n \wedge dV_n^\dagger } \ .
\end{align}
We then pull these back via the embeddings that we have just found. At zeroth order, these take the form
\begin{align}
\omega_+^{\rm orb} &= -\frac{i}{q} \Tr\parens{dU_0^{\rm orb}\wedge dV^{\rm orb}_0} = -i \, du\wedge dv  \ , \nonumber \\
\omega_K^{\rm orb} &= \frac{i}{2q} \Tr\parens{dU_0^{\rm orb}\wedge d(U_0^{\rm orb})^\dagger + dV_0^{\rm orb}\wedge d(V_0^{\rm orb})^\dagger } = \frac{i}{2} (du\wedge du^* + dv \wedge dv^*) \ .
\end{align}
At first order, we have
\begin{align}
\varpi(\zeta) &= \varpi^{\rm orb}(\zeta) + \varpi^{\rm pert}(\zeta) \nonumber \\
\varpi^{\rm pert}(\zeta) &= -\frac{i}{2\zeta} \omega_+^{\rm pert} + \omega_K^{\rm pert} - \frac{i\zeta}{2} \omega_-^{\rm pert} \nonumber \\
&= \sum_n \parens{- \frac{i}{2\zeta} \omega'_{n+} + \omega'_{nK} - \frac{i\zeta}{2} \omega'_{n-} } \ .
\end{align}
To compute these corrections, we make use of
\be \Tr S_i S_j^\dagger S_k S_\ell^\dagger = q^2 \delta_{i,\ell} \delta_{j,k} \ . \label{eq:Sortho} \ee
We thus find that
\begin{align}
\omega'_{n+}&= - \frac{i}{q} \Tr dU_n\wedge dV_{-n} \nonumber \\
&= \frac{i}{4q} \sum_{j=0}^{q-1}\sum_{i=1}^{q-1} d\parens{\frac{2\xi_{n,i,+}\bar N^v_{i,j} + \xi_{n,i,\RR} N^u_{i,j}}{D_{i,j}}}\wedge d\parens{\frac{-2\xi_{-n,-i,+}\bar N^u_{i,j} + \xi^*_{n,i,\RR} N^v_{i,j}}{D_{i,j}}} \nonumber \\
\omega'_{nK} &= \frac{i}{2q}\Tr\parens{ dU_n\wedge dU_n^\dagger + dV_n\wedge dV_n^\dagger} \nonumber \\
&= \frac{i}{8q} \sum_{j=0}^{q-1}\sum_{i=1}^{q-1} \brackets{d\parens{\frac{2\xi_{n,i,+}\bar N^v_{i,j}+\xi_{n,i,\RR}N^u_{i,j}}{D_{i,j}}}\wedge d\parens{\frac{2\xi_{n,i,+}^*N^v_{i,j}+\xi_{n,i,\RR}^* \bar N^u_{i,j}}{D_{i,j}}} \right. \nonumber \\
&\left.\quad+ d\parens{\frac{-2\xi_{n,i,+}\bar N^u_{i,j}+\xi_{n,i,\RR}N^v_{i,j}}{D_{i,j}}}\wedge d\parens{\frac{-2\xi^*_{n,i,+}N^u_{i,j}+\xi^*_{n,i,\RR} \bar N^v_{i,j}}{D_{i,j}}} } \ . \label{eq:hkStruct}
\end{align}
Next, we use \eqref{eq:chiN} to write these as
\begin{align}
\omega'_{n+} &= \frac{i}{4q} \sum_{j=0}^{q-1}\sum_{i=1}^{q-1} d\parens{\frac{2\xi_{\chi^{-j} n,i,+} \bar N^v_{\chi^{-j}n,i,0} + \xi_{\chi^{-j}n,i,\RR} N^u_{\chi^{-j}n,i,0}}{D_{\chi^{-j}n,i,0}}} \wedge \nonumber \\
&\qquad\qquad\qquad  d\parens{\frac{-2\xi_{-\chi^{-j}n,-i,+}\bar N^u_{\chi^{-j}n,i,0}+\xi^*_{\chi^{-j}n,i,\RR} N^v_{\chi^{-j}n,i,0}}{D_{\chi^{-j}n,i,0}}} \nonumber \\
\omega'_{nK} &= \frac{i}{8q} \sum_{j=0}^{q-1}\sum_{i=1}^{q-1} \brackets{d\parens{\frac{2\xi_{\chi^{-j}n,i,+}\bar N^v_{\chi^{-j}n,i,0}+\xi_{\chi^{-j}n,i,\RR}N^u_{\chi^{-j}n,i,0}}{D_{\chi^{-j}n,i,0}}} \wedge \right. \nonumber \\
&\left.\qquad\qquad\qquad d\parens{\frac{2\xi_{\chi^{-j}n,i,+}^*N^v_{\chi^{-j}n,i,0}+\xi_{\chi^{-j}n,i,\RR}^* \bar N^u_{\chi^{-j}n,i,0}}{D_{\chi^{-j}n,i,0}}} \right. \nonumber \\
&\left.\qquad\quad\qquad+ d\parens{\frac{-2\xi_{\chi^{-j}n,i,+}\bar N^u_{\chi^{-j}n,i,0}+\xi_{\chi^{-j}n,i,\RR}N^v_{\chi^{-j}n,i,0}}{D_{\chi^{-j}n,i,0}}} \wedge \right. \nonumber \\
&\left.\qquad\qquad\qquad d\parens{\frac{-2\xi^*_{\chi^{-j}n,i,+}N^u_{\chi^{-j}n,i,0}+\xi^*_{\chi^{-j}n,i,\RR} \bar N^v_{\chi^{-j}n,i,0}}{D_{\chi^{-j}n,i,0}}} } \ .
\end{align}
This allows us to eliminate the sums over $j$ by exchanging terms associated to various values of $n$. We thus find
\begin{align}
\varpi^{\rm pert}(\zeta) &= -\frac{i}{2\zeta} \omega_+^{\rm pert} + \omega_K^{\rm pert} - \frac{i\zeta}{2} \omega_-^{\rm pert} \nonumber \\
&= \sum_n\sum_{i=1}^{q-1} \parens{- \frac{i}{2\zeta} \omega'_{ni+} + \omega'_{niK} - \frac{i\zeta}{2} \omega'_{ni-} } \nonumber \\
\omega'_{ni+} &= \frac{i}{4} d\parens{\frac{2\xi_{n,i,+} \bar N^v_{n,i,0} + \xi_{n,i,\RR} N^u_{n,i,0}}{D_{n,i,0}}} \wedge d\parens{\frac{-2\xi_{-n,-i,+}\bar N^u_{n,i,0} + \xi^*_{n,i,\RR} N^v_{n,i,0}}{D_{n,i,0}}} \nonumber \\
\omega'_{niK} &= \frac{i}{8} \brackets{d\parens{\frac{2\xi_{n,i,+}\bar N^v_{n,i,0}+\xi_{n,i,\RR}N^u_{n,i,0}}{D_{n,i,0}}}\wedge d\parens{\frac{2\xi_{n,i,+}^*N^v_{n,i,0}+\xi_{n,i,\RR}^* \bar N^u_{n,i,0}}{D_{n,i,0}}} \right. \nonumber \\
&\left.\quad+ d\parens{\frac{-2\xi_{n,i,+}\bar N^u_{n,i,0}+\xi_{n,i,\RR}N^v_{n,i,0}}{D_{n,i,0}}}\wedge d\parens{\frac{-2\xi^*_{n,i,+}N^u_{n,i,0}+\xi^*_{n,i,\RR} \bar N^v_{n,i,0}}{D_{n,i,0}}} } \ . \label{eq:jResum}
\end{align}
Similarly, when $i>\floor{q/2}$, we can use \eqref{eq:negN} to replace $(n,i,0)$ by $(-n,-i,i)$ and then again use \eqref{eq:chiN} to set $j=0$. So, we obtain
\begin{align}
\varpi^{\rm pert}(\zeta) &= -\frac{i}{2\zeta} \omega_+^{\rm pert} + \omega_K^{\rm pert} - \frac{i\zeta}{2} \omega_-^{\rm pert} \nonumber \\
&= \sum_n\sum_{i=1}^{\floor{q/2}} f_i\sum_{t=\pm 1} \parens{- \frac{i}{2\zeta} \omega'_{nti+} + \omega'_{ntiK} - \frac{i\zeta}{2} \omega'_{nti-} } \nonumber \\
\omega'_{nti+} &= \frac{i}{4} d\parens{\frac{2\xi_{tn,ti,+} \bar N^v_{n,i,0} + \xi_{tn,ti,\RR} N^u_{n,i,0}}{D_{n,i,0}}} \wedge d\parens{\frac{-2\xi_{-tn,-ti,+}\bar N^u_{n,i,0} + \xi^*_{tn,ti,\RR} N^v_{n,i,0}}{D_{n,i,0}}} \nonumber \\
\omega'_{ntiK} &= \frac{i}{8} \brackets{d\parens{\frac{2\xi_{tn,ti,+}\bar N^v_{n,i,0}+\xi_{tn,ti,\RR}N^u_{n,i,0}}{D_{n,i,0}}}\wedge d\parens{\frac{2\xi_{tn,ti,+}^*N^v_{n,i,0}+\xi_{tn,ti,\RR}^* \bar N^u_{n,i,0}}{D_{n,i,0}}} \right. \nonumber \\
&\left.\quad+ d\parens{\frac{-2\xi_{tn,ti,+}\bar N^u_{n,i,0}+\xi_{tn,ti,\RR}N^v_{n,i,0}}{D_{n,i,0}}}\wedge d\parens{\frac{-2\xi^*_{tn,ti,+}N^u_{n,i,0}+\xi^*_{tn,ti,\RR} \bar N^v_{n,i,0}}{D_{n,i,0}}} } \ , \label{eq:halfSum}
\end{align}
where $f_i=\piecewise{\half}{i=q/2}{1}{\rm else}$. Since $j$ will no longer play a role, we now abbreviate $N^u_{n,i}\equiv N^u_{n,i,0}$, and similarly for $N^v$ and $D$; we will furthermore often leave $n$ implicit. We also define
\be \xi_{nti+} \equiv \xi_{tn,ti,+} \ , \quad \xi_{nti\RR} \equiv \xi_{tn,ti,\RR} \ . \ee
 Note that the reality condition on $\xi_{nti\RR}$ is simply $\xi_{n(-t)i\RR}=\xi^*_{nti\RR}$; in particular, for $q=3$, conjugation no longer relates FI parameters associated to different values of $n$, and for $q$ even and $i=q/2$, this takes the form $\xi_{n(-t)(q/2)\RR}=\xi_{nt(q/2)\RR}=\xi_{nt(q/2)\RR}^*$, as $-\frac{q}{2} \equiv \frac{q}{2} \pmod{q}$ and $2n\in\Lambda_{q/2}$.

Expanding these expressions yields
\begin{align}
\omega'_{nti+\, u\bar u} &= -\frac{i |1-\kappa_q^i|^2}{4} \frac{|\xi_{nti\RR}|^2 N^u_i N^v_i + 2\xi_{nti\RR} \xi_{n(-t)i+} (|N^v_i|^2 - |N^u_i|^2) - 4 \xi_{nti+} \xi_{n(-t)i+} \bar N^u_i \bar N^v_i}{D_i^3} \nonumber \\
\omega'_{nti+\, uv} &= \frac{i |1-\kappa_q^i|^2}{2} \frac{\bar N^u_i \bar N^v_i (\xi_{nti\RR} \xi_{n(-t)i+} - \xi_{nti+} \xi^*_{nti\RR})}{D_i^3} \nonumber \\
\omega'_{nti+\, u\bar v} &= -\frac{i (1-\kappa_q^i)^2}{4} \frac{ (2\xi_{nti+}\bar N^u_i - \xi_{nti\RR} N^v_i) (2 \xi_{n(-t)i+} \bar N^u_i - \xi^*_{nti\RR} N^v_i) }{D_i^3} \nonumber \\
\omega'_{nti+\, \bar u v} &= -\frac{i(1-\kappa_q^{-i})^2}{4} \frac{ (2\xi_{nti+} \bar N^v_i + \xi_{nti\RR} N^u_i) (2\xi_{n(-t)i+} \bar N^v_i + \xi^*_{nti\RR} N^u_i)}{D_i^3}  \nonumber \\
\omega'_{nti+\, \bar u \bar v} &= -\frac{i |1-\kappa_q^i|^2}{2} \frac{N^u_i N^v_i (\xi_{nti\RR} \xi_{n(-t)i+} - \xi_{nti+} \xi^*_{nti\RR})}{D_i^3} \nonumber \\
\omega'_{nti+\, v\bar v} &= -\omega'_{n(-t)i+\, u\bar u} \nonumber \\
\omega'_{ntiK\, u\bar u} &= -\frac{i |1-\kappa_q^i|^2}{8} \frac{ (4|\xi_{nti+}|^2 - |\xi_{nti\RR}|^2) (|N^v_i|^2 - |N^u_i|^2) + 4\xi_{nti\RR}\xi^*_{nti+} N^u_i N^v_i + 4\xi_{nti+}\xi^*_{nti\RR} \bar N^u_i \bar N^v_i }{D_i^3} \nonumber \\
\omega'_{ntiK\, uv} &= 0 \nonumber \\
\omega'_{ntiK\, u\bar v} &= \frac{i(1-\kappa_q^i)^2}{4} \frac{\parens{2\xi_{nti+} \bar N^u_i - \xi_{nti\RR} N^v_i}\parens{2\xi_{nti+}^* N^v_i + \xi^*_{nti\RR} \bar N^u_i}}{D_i^3} \nonumber \\
\omega'_{ntiK\, \bar u v} &= \frac{i(1-\kappa_q^{-i})^2}{4} \frac{ (2\xi_{nti+} \bar N^v_i + \xi_{nti\RR} N^u_i)(-2\xi^*_{nti+} N^u_i + \xi^*_{nti\RR} \bar N^v_i)}{D_i^3} \nonumber \\
\omega'_{ntiK\, \bar u \bar v} &= 0 \nonumber \\
\omega'_{ntiK\, v\bar v} &= - \omega'_{ntiK\, u\bar u} \ .
\end{align}
Finally, rearranging terms between $t=\pm 1$ yields
\begin{align}
\varpi^{\rm pert}(\zeta) &= -\frac{i}{2\zeta} \omega_+^{\rm pert} + \omega_K^{\rm pert} - \frac{i\zeta}{2} \omega_-^{\rm pert} \nonumber \\
&= \sum_n\sum_{i=1}^{\floor{q/2}} f_i\sum_{t=\pm 1} \parens{- \frac{i}{2\zeta} \omega_{nti+} + \omega_{ntiK} - \frac{i\zeta}{2} \omega_{nti-} } \nonumber \\
\omega_{nti+\, u\bar u} &= \frac{i |1-\kappa_q^i|^2}{4} \frac{(2\xi_{nti+}\bar N^v_i + \xi_{nti\RR} N^u_i)(2\xi_{n(-t)i+}\bar N^u_i - \xi^*_{nti\RR} N^v_i)}{D_i^3} \nonumber \\
\omega_{nti+\, uv} &= 0 \nonumber \\
\omega_{nti+\, u\bar v} &= \omega'_{nti+\, u\bar v} \nonumber \\
\omega_{nti+\, \bar u v} &= \omega'_{nti+\, \bar u v} \nonumber \\
\omega_{nti+\, \bar u \bar v} &= 0 \nonumber \\
\omega_{nti+\, v\bar v} &= -\omega_{nti+\, u\bar u} \nonumber \\
\omega_{ntiK} &= \omega'_{ntiK} \ .
\end{align}
As in \cite{mz:K3HK}, one may easily obtain the metric and complex structures via matrix multiplication and verify that this is a valid hyper-K\"ahler structure; in particular, the metric is Ricci-flat.

We conclude this subsection by noting that the procedure for generalizing this to higher orders in $\xi$ that was described in \cite{mz:K3HK} works here as well. Indeed, in light of the above results we may carry out this procedure quite explicitly. Suppose that we have solved for $U_n$ and $V_n$ to order $\xi^{\nu-1}$, where $\nu\ge 1$. We now improve this approximation to order $\xi^\nu$. We write $U_n^{(\nu)}=U_n^{(\nu-1)}+\delta U_n^{(\nu)}$, and similarly for $V_n$, where $\delta U_n^{(\nu)}$ and $\delta V_n^{(\nu)}$ will contain terms only of order $\xi^{\nu}$. Then, at order $\nu$ the moment map equations (including those for $n=0$) take the form
\begin{align}
\xi_{n,+}^{(\nu)} &= \delta U_n^{(\nu)} n^v - \delta V_n^{(\nu)} n^u + [U_0^{\rm orb}, \delta V^{(\nu)}_n] + [\delta U_n^{(\nu)}, V_0^{\rm orb}] \nonumber \\
\xi_{n,\RR}^{(\nu)} &= -n^u (\delta U^{(\nu)}_{-n})^\dagger + n^{\bar u} \delta U^{(\nu)}_n + [U_0^{\rm orb}, (\delta U^{(\nu)}_{-n})^\dagger] + [\delta U^{(\nu)}_n, (U^{\rm orb}_0)^\dagger] + (U\mapsto V) \ , \label{eq:approxMuNu}
\end{align}
which are of the same form as \eqref{eq:approxMu}, with $\xi_{n,+}^{(1)}\equiv \sum_i \xi_{n,i,+}\sigma^i$ and $\xi_{n,\RR}^{(1)}\equiv \sum_i \xi_{n,i,\RR}\sigma^i$ replaced by\footnote{Recall that $U_0^{\rm orb}$, $V_0^{\rm orb}$, and their adjoints all commute.}
\begin{align}
\xi_{n,+}^{(\nu)} &= \xi_{n,+}^{(1)} - U_n^{(\nu-1)} n^v + V_n^{(\nu-1)} n^u - [U_0^{\rm orb},V_n^{(\nu-1)}] - [U_n^{(\nu-1)},V_0^{\rm orb}] - \sum_m \sum_{\substack{\nu',\nu''=1,\ldots,\nu-1 \\ \nu'+\nu''\le \nu}} [\delta U_{n-m}^{(\nu')}, \delta V_m^{(\nu'')}] \nonumber \\
\xi_{n,\RR}^{(\nu)} &= \xi_{n,\RR}^{(1)} + \brackets{ (U_{-n}^{(\nu-1)})^\dagger n^u - U_n^{(\nu-1)} n^{\bar u} - [U_0^{\rm orb},(U_n^{(\nu-1)})^\dagger] - [U_n^{(\nu-1)},(U_0^{\rm orb})^\dagger] \vphantom{\sum_{\substack{\nu',\nu''=1,\ldots,\nu-1 \\ \nu'+\nu''\le \nu}}} \right. \nonumber \\
&\left.\qquad - \sum_m \sum_{\substack{\nu',\nu''=1,\ldots,\nu-1 \\ \nu'+\nu''\le \nu}} [\delta U^{(\nu')}_{n+m},(\delta U^{(\nu'')}_{m})^\dagger] + (U\mapsto V) }
\end{align}
when $\nu\ge 2$. Using \eqref{eq:approxMuNu}, we find that for $\nu\ge 2$,
\begin{align}
\xi_{n,+}^{(\nu)} &= \xi_{n,+}^{(\nu-1)} - \delta U_n^{(\nu-1)} n^v + \delta V_n^{(\nu-1)} n^u - [U_0^{\rm orb},\delta V_n^{(\nu-1)}] - [\delta U_n^{(\nu-1)},V_0^{\rm orb}] - \sum_m \sum_{\nu'=1}^{\nu-1} [\delta U_{n-m}^{(\nu')}, \delta V_m^{(\nu-\nu')}] \nonumber \\
&= - \sum_m \sum_{\nu'=1}^{\nu-1} [\delta U_{n-m}^{(\nu')}, \delta V_m^{(\nu-\nu')}] \nonumber \\
\xi_{n,\RR}^{(\nu)} &= \xi_{n,\RR}^{(\nu-1)} + \brackets{ (\delta U_{-n}^{(\nu-1)})^\dagger n^u - \delta U_n^{(\nu-1)} n^{\bar u} - [U_0^{\rm orb},(\delta U_n^{(\nu-1)})^\dagger] - [\delta U_n^{(\nu-1)},(U_0^{\rm orb})^\dagger] \vphantom{\sum_{\nu'=1}^{\nu-1}} \right. \nonumber \\
&\left.\qquad - \sum_m \sum_{\nu'=1}^{\nu-1} [\delta U^{(\nu')}_{n+m},(\delta U^{(\nu-\nu')}_{m})^\dagger] + (U\mapsto V) } \nonumber \\
&= - \sum_m \sum_{\nu'=1}^{\nu-1} [\delta U^{(\nu')}_{n+m},(\delta U^{(\nu-\nu')}_{m})^\dagger] + (U\mapsto V) \ . \label{eq:xiNu}
\end{align}
In particular, we see that $\xi^{(\nu)}_{n,+}$ and $\xi^{(\nu)}_{n,\RR}$ contain only terms of order $\xi^\nu$. Next, at each order we continue to impose the least norm gauge fixing condition,
\be -n^u (\delta U_{-n}^{(\nu)})^\dagger - n^{\bar u} \delta U_{n}^{(\nu)} + [U_0^{\rm orb},(\delta U_{-n}^{(\nu)})^\dagger] + [(U_0^{\rm orb})^\dagger,\delta U_{n}^{(\nu)}] + (U\mapsto V) = 0 \ . \label{eq:LNgaugeNu} \ee
(We note that, in the D6-brane language, summing the equations $d_{B^{\rm orb}} *\delta B^{(\nu)}=0$ over all $\nu$ yields $d_{B^{\rm orb}} *B=0$.) Finally, we impose the orthogonality constraint
\be \Tr \, (U_0^{\rm orb})^\dagger \delta U^{(\nu)}_0 = \Tr \, (V_0^{\rm orb})^\dagger \delta V^{(\nu)}_0 =0 \ . \label{eq:ortho0Nu} \ee
This yields a system of equations which is nearly identical to the one we have just solved; the only change is in the upgrade of $\xi^{(1)}$ to $\xi^{(\nu)}$.

Because of this upgrade, the solution is slightly more complicated. The reason is that $\xi^{(\nu)}$ need not be a linear combination of powers of $\sigma$. That said, the solution is still quite nice:
\begin{align}
\delta U_n^{(\nu)} &= \frac{1}{2q} \sum_{j=0}^{q-1} \sum_{i = \delta_{n,0}}^{q - 1} \frac{2 \tilde\xi_{n,i,j,+}^{(\nu)} \bar N^v_{i,j} + \tilde\xi_{n, i, j, \mb{R}}^{(\nu)} N^u_{i,j}}{D_{i,j}} S_{i+j} S_{j}^{\dagger} \nonumber \\
\delta V_n^{(\nu)} &= \frac{1}{2q} \sum_{j=0}^{q-1} \sum_{i = \delta_{n,0}}^{q - 1} \frac{-2 \tilde\xi_{n,i,j,+}^{(\nu)} \bar N^u_{i,j} + \tilde\xi_{n, i, j, \mb{R}}^{(\nu)} N^v_{i,j}}{D_{i,j}} S_{i+j} S_{j}^{\dagger} \ , \label{eq:solNu}
\end{align}
where $\tilde\xi_{n,i,j,+}^{(\nu)}$ and $\tilde\xi_{n,i,j,\RR}^{(\nu)}$ are defined by
\be \xi_{n,+}^{(\nu)} = \frac{1}{q} \sum_{i,j=0}^{q-1} \tilde \xi^{(\nu)}_{n,i,j,+} S_{i+j} S^\dagger_j \ , \quad \xi_{n,\RR}^{(\nu)} = \frac{1}{q} \sum_{i,j=0}^{q-1} \tilde \xi^{(\nu)}_{n,i,j,\RR} S_{i+j} S^\dagger_j \ . \label{eq:tXi} \ee
This change in the range of summation for $n=0$ is made because the $n=i=0$ terms would be singular, as $N^u_{0,0,j}=N^v_{0,0,j}=0$. That this solution generalizes our $\nu=1$ solution (with $\tilde\xi^{(1)}_{n,i,j,+}=\xi_{n,i,+}$ and $\tilde\xi^{(1)}_{n,i,j,\RR}=\xi_{n,i,\RR}$ for all $j$) follows from \eqref{eq:Ssum}. Explicitly, using \eqref{eq:Sortho} we have
\be \tilde \xi^{(\nu)}_{n,i,j,+} = \frac{1}{q} \Tr S_j S_{i+j}^\dagger \xi_{n,+}^{(\nu)} \ , \quad \tilde \xi^{(\nu)}_{n,i,j,\RR} = \frac{1}{q} \Tr S_j S_{i+j}^\dagger \xi_{n,\RR}^{(\nu)} \ . \label{eq:tXiDef} \ee
The verification of \eqref{eq:solNu} proceeds as in the $\nu=1$ case. In particular, one inductively verifies that $\xi_{\chi n,+}^{(\nu)}=\sigma \xi_{n,+}^{(\nu)}\sigma^\dagger$, $\xi_{\chi n,\RR}^{(\nu)}=\sigma \xi_{n,\RR}^{(\nu)}\sigma^\dagger$, $\tilde\xi_{\chi n,i,j,+}^{(\nu)}=\tilde\xi_{n,i,j-1,+}$, $\tilde\xi_{\chi n,i,j,\RR}^{(\nu)}=\tilde\xi_{n,i,j-1,\RR}$, $U_{\chi n}^{(\nu)}=\kappa_q \sigma U_n^{(\nu)} \sigma^\dagger$, and $V_{\chi n}^{(\nu)}=\kappa_q^* \sigma V_n^{(\nu)} \sigma^\dagger$. For $n=0$, this implies that $\tilde\xi^{(\nu)}_{0,i,j,+}$ and $\tilde\xi^{(\nu)}_{0,i,j,\RR}$ are independent of $j$; further specializing to $i=0$, we find that $\tilde\xi^{(\nu)}_{0,0,j,+}=\frac{1}{q}\sum_{j'} \tilde\xi^{(\nu)}_{0,0,j',+}=\frac{1}{q}\Tr\xi_{0,+}^{(\nu)}=0$, and similarly $\tilde\xi^{(\nu)}_{0,0,j,\RR}=0$, and so we may omit the $i=0$ terms in \eqref{eq:tXi}. The reality condition $(\xi_{-n,\RR}^{(\nu)})^\dagger=\xi_{n,\RR}^{(\nu)}$ also still holds; this translates to $(\tilde\xi^{(\nu)}_{-n,-i,i+j,\RR})^*=\tilde\xi^{(\nu)}_{n,i,j,\RR}$ and allows us to write
\begin{align}
(\delta U_{-n}^{(\nu)})^\dagger &= - \frac{1}{2q} \sum_{j=0}^{q-1} \sum_{i = \delta_{n,0}}^{q - 1} \frac{2 (\tilde\xi_{-n,-i,i+j,+}^{(\nu)})^* N^v_{i,j} + \tilde\xi_{n, i, j, \mb{R}}^{(\nu)} \bar N^u_{i,j}}{D_{i,j}} S_{i+j} S_{j}^{\dagger} \nonumber \\
(\delta V_{-n}^{(\nu)})^\dagger &= - \frac{1}{2q} \sum_{j=0}^{q-1} \sum_{i = \delta_{n,0}}^{q - 1} \frac{-2 (\tilde\xi_{-n,-i,i+j,+}^{(\nu)})^* N^u_{i,j} + \tilde\xi_{n, i, j, \mb{R}}^{(\nu)} \bar N^v_{i,j}}{D_{i,j}} S_{i+j} S_{j}^{\dagger} \ . \label{eq:solNuMDagger}
\end{align}
The only step that is a bit different is verifying tracelessness of $U_n^{(\nu)}$ and $V_n^{(\nu)}$, as there is now an $i=0$ term when $n\not=0$. For this term, one uses the fact that $N^u_{0,j}=n^u$ and $N^v_{0,j}=n^v$, so that all the $j$-dependence is in the $\tilde \xi^{(\nu)}$ coefficients (after we evaluate the trace, which eliminates the $j$-dependence in $S_j S_j^\dagger$). One then uses $\sum_j \tilde\xi_{n,0,j,+}^{(\nu)}=\Tr \xi_{n,+}^{(\nu)}=0$ and $\sum_j \tilde\xi_{n,0,j,\RR}^{(\nu)}=\Tr \xi_{n,\RR}^{(\nu)}=0$.

From here, proceeding to a closed-form solution for $U_n\equiv U_n^{(\infty)}$ and $V_n\equiv V_n^{(\infty)}$ is not too difficult. For, when $\nu\ge 2$ plugging \eqref{eq:xiNu} into \eqref{eq:solNu} yields
\begin{align}
\delta U_n^{(\nu)} &= -\frac{1}{2q^2}\sum_{j=0}^{q-1}\sum_{i=\delta_{n,0}}^{q-1}\sum_m \sum_{\nu'=1}^{\nu-1} \frac{S_{i+j} S_j^\dagger}{D_{i,j}} \times \nonumber \\
&\qquad \Tr\brackets{S_j S^\dagger_{i+j} \parens{2[\delta U_{n-m}^{(\nu')}, \delta V_m^{(\nu-\nu')}] \bar N^v_{i,j} + \parens{[\delta U_{n+m}^{(\nu')}, (\delta U_m^{(\nu-\nu')})^\dagger]+[\delta V_{n+m}^{(\nu')}, (\delta V_m^{(\nu-\nu')})^\dagger]} N^u_{i,j}}} \nonumber \\
\delta V_n^{(\nu)} &= -\frac{1}{2q^2}\sum_{j=0}^{q-1}\sum_{i=\delta_{n,0}}^{q-1}\sum_m \sum_{\nu'=1}^{\nu-1} \frac{S_{i+j} S_j^\dagger}{D_{i,j}} \times \nonumber \\
&\qquad \Tr\brackets{S_j S^\dagger_{i+j} \parens{-2[\delta U_{n-m}^{(\nu')}, \delta V_m^{(\nu-\nu')}] \bar N^u_{i,j} + \parens{[\delta U_{n+m}^{(\nu')}, (\delta U_m^{(\nu-\nu')})^\dagger]+[\delta V_{n+m}^{(\nu')}, (\delta V_m^{(\nu-\nu')})^\dagger]} N^v_{i,j}}} \ ,
\end{align}
which allows us to recursively compute $\Delta U_n=U_n-U^{\rm orb}_n$ and $\Delta V_n=V_n-V^{\rm orb}_n$. Summing over all $\nu\ge 1$ yields the final answer
\begin{align}
\Delta U_n &= \frac{1}{2q}\sum_{j=0}^{q-1}\sum_{i=\delta_{n,0}}^{q-1} \frac{S_{i+j} S_j^\dagger}{D_{i,j}} \brackets{ \vphantom{\sum_m} \parens{2\xi_{n,i,+} \bar N^v_{i,j} + \xi_{n,i,\RR} N^u_{i,j}} \right. \nonumber \\
&\left. - \frac{1}{q} \sum_m \sum_{\nu',\nu''=1}^{\infty} \Tr\brackets{S_j S^\dagger_{i+j} \parens{2[\delta U_{n-m}^{(\nu')}, \delta V_m^{(\nu'')}] \bar N^v_{i,j} + \parens{[\delta U_{n+m}^{(\nu')}, (\delta U_m^{(\nu'')})^\dagger]+[\delta V_{n+m}^{(\nu')}, (\delta V_m^{(\nu'')})^\dagger]} N^u_{i,j}}} } \nonumber \\
\Delta V_n &= \frac{1}{2q}\sum_{j=0}^{q-1}\sum_{i=\delta_{n,0}}^{q-1} \frac{S_{i+j} S_j^\dagger}{D_{i,j}} \brackets{ \vphantom{\sum_m} \parens{-2\xi_{n,i,+} \bar N^u_{i,j} + \xi_{n,i,\RR} N^v_{i,j}} \right. \nonumber \\
&\left. \!\!\!\!\! - \frac{1}{q} \sum_m \sum_{\nu',\nu''=1}^{\infty} \Tr\brackets{S_j S^\dagger_{i+j} \parens{-2[\delta U_{n-m}^{(\nu')}, \delta V_m^{(\nu'')}] \bar N^u_{i,j} + \parens{[\delta U_{n+m}^{(\nu')}, (\delta U_m^{(\nu'')})^\dagger]+[\delta V_{n+m}^{(\nu')}, (\delta V_m^{(\nu'')})^\dagger]} N^v_{i,j}}} } \ .
\end{align}
At the cost of obscuring the algorithmic nature of this formula, we may rewrite it as
\begin{align}
\Delta U_n &= \frac{1}{2q}\sum_{j=0}^{q-1}\sum_{i=\delta_{n,0}}^{q-1} \frac{S_{i+j} S_j^\dagger}{D_{i,j}} \brackets{ \vphantom{\sum_m} \parens{2\xi_{n,i,+} \bar N^v_{i,j} + \xi_{n,i,\RR} N^u_{i,j}} \right. \nonumber \\
&\left.\qquad - \frac{1}{q} \sum_m \Tr\brackets{S_j S^\dagger_{i+j} \parens{2[\Delta U_{n-m}, \Delta V_m] \bar N^v_{i,j} + \parens{[\Delta U_{n+m}, \Delta U_m^\dagger]+[\Delta V_{n+m}, \Delta V_m^\dagger]} N^u_{i,j}}} } \nonumber \\
\Delta V_n &= \frac{1}{2q}\sum_{j=0}^{q-1}\sum_{i=\delta_{n,0}}^{q-1} \frac{S_{i+j} S_j^\dagger}{D_{i,j}} \brackets{ \vphantom{\sum_m} \parens{-2\xi_{n,i,+} \bar N^u_{i,j} + \xi_{n,i,\RR} N^v_{i,j}} \right. \nonumber \\
&\left.\qquad - \frac{1}{q} \sum_m \Tr\brackets{S_j S^\dagger_{i+j} \parens{-2[\Delta U_{n-m}, \Delta V_m] \bar N^u_{i,j} + \parens{[\Delta U_{n+m}, \Delta U_m^\dagger]+[\Delta V_{n+m}, \Delta V_m^\dagger]} N^v_{i,j}}} } \ .
\end{align}
Further defining $\Delta U=\sum_n \Delta U_n e(n)$ and $\Delta V=\sum_n \Delta V_n e(n)$, we obtain
\begin{align}
\Delta U &= \frac{1}{2q}\sum_n \sum_{j=0}^{q-1}\sum_{i=\delta_{n,0}}^{q-1} \frac{S_{i+j} S_j^\dagger}{D_{i,j}} \brackets{ \vphantom{\sum_m} \parens{2\xi_{n,i,+}e(n) \bar N^v_{i,j} + \xi_{n,i,\RR}e(n) N^u_{i,j}} \right. \nonumber \\
&\left.\qquad - \frac{1}{q} \sum_m \Tr\brackets{S_j S^\dagger_{i+j} \parens{2[\Delta U_{n-m}e(n-m), \Delta V_m e(m)] \bar N^v_{i,j} \right.\right.\right. \nonumber \\
&\left.\left.\left.\qquad + \parens{[\Delta U_{n+m} e(n+m), \Delta U_m^\dagger e(-m)]+[\Delta V_{n+m} e(n+m), \Delta V_m^\dagger e(-m)]} N^u_{i,j}} \vphantom{S^\dagger_{i+j}} } \vphantom{\frac{S_{i+j}S_j^\dagger}{D_{i,j}}} } \nonumber \\
\Delta V &= \frac{1}{2q}\sum_n \sum_{j=0}^{q-1}\sum_{i=\delta_{n,0}}^{q-1} \frac{S_{i+j} S_j^\dagger}{D_{i,j}} \brackets{ \vphantom{\sum_m} \parens{-2\xi_{n,i,+}e(n) \bar N^u_{i,j} + \xi_{n,i,\RR}e(n) N^v_{i,j}} \right. \nonumber \\
&\left.\qquad - \frac{1}{q} \sum_m \Tr\brackets{S_j S^\dagger_{i+j} \parens{-2[\Delta U_{n-m}e(n-m), \Delta V_m e(m)] \bar N^u_{i,j} \right.\right.\right. \nonumber \\
&\left.\left.\left.\qquad + \parens{[\Delta U_{n+m} e(n+m), \Delta U_m^\dagger e(-m)]+[\Delta V_{n+m} e(n+m), \Delta V_m^\dagger e(-m)]} N^v_{i,j}} \vphantom{S^\dagger_{i+j}} } \vphantom{\frac{S_{i+j}S_j^\dagger}{D_{i,j}}} } \ , \label{eq:intMom}
\end{align}
which is the momentum-space expression of an integral equation involving the distributions $\Delta U$, $\Delta V$, $\xi_{+} = \sum_{n,i} \xi_{n,i,+} \sigma^i e(n)$, and $\xi_{\RR}=\sum_{n,i} \xi_{n,i,\RR} \sigma^i e(n)$ on $\hat T^4$.

To make this explicit, we first recall the complex coordinates on $\hat T^4$ from \cite{mz:k3}:
\be \psi_{1'} = \frac{y_1 - i y_2}{2} \ , \quad \psi_{2'} = \frac{y_3 - i y_4}{2} \ . \ee
We then define
\begin{align}
G^{1'}_{+}(y)[\Oo] &= \frac{1}{q^2 {\rm vol}(\RR^4/\hat\Lambda)} \sum_{n} \sum_{i=\delta_{n,0}}^{q-1} \sum_{j=0}^{q-1} \frac{\bar N^v_{i,j}}{D_{i,j}} S_{i+j} S_j^\dagger \Tr\brackets{S_j S_{i+j}^\dagger \Oo} e^{in\cdot y} \nonumber \\
G^{1'}_{\RR}(y)[\Oo] &= \frac{-1}{2q^2 {\rm vol}(\RR^4/\hat\Lambda)} \sum_{n} \sum_{i=\delta_{n,0}}^{q-1} \sum_{j=0}^{q-1} \frac{N^u_{i,j}}{D_{i,j}} S_{i+j} S_j^\dagger \Tr\brackets{S_j S_{i+j}^\dagger \Oo} e^{in\cdot y} \nonumber \\
G^{2'}_{+}(y)[\Oo] &= \frac{1}{q^2 {\rm vol}(\RR^4/\hat\Lambda)} \sum_{n} \sum_{i=\delta_{n,0}}^{q-1} \sum_{j=0}^{q-1} \frac{-\bar N^u_{i,j}}{D_{i,j}} S_{i+j} S_j^\dagger \Tr\brackets{S_j S_{i+j}^\dagger \Oo} e^{in\cdot y} \nonumber \\
G^{2'}_{\RR}(y)[\Oo] &= \frac{-1}{2q^2 {\rm vol}(\RR^4/\hat\Lambda)} \sum_{n} \sum_{i=\delta_{n,0}}^{q-1} \sum_{j=0}^{q-1} \frac{N^v_{i,j}}{D_{i,j}} S_{i+j} S_j^\dagger \Tr\brackets{S_j S_{i+j}^\dagger \Oo} e^{in\cdot y} \nonumber \\
G^{\bar 1'}_{-}(y)[\Oo] &= \frac{-1}{q^2 {\rm vol}(\RR^4/\hat\Lambda)} \sum_{n} \sum_{i=\delta_{n,0}}^{q-1} \sum_{j=0}^{q-1} \frac{N^v_{i,j}}{D_{i,j}} S_{i+j} S_j^\dagger \Tr\brackets{S_j S_{i+j}^\dagger \Oo} e^{in\cdot y} \nonumber \\
G^{\bar 1'}_{\RR}(y)[\Oo] &= \frac{1}{2q^2 {\rm vol}(\RR^4/\hat\Lambda)} \sum_{n} \sum_{i=\delta_{n,0}}^{q-1} \sum_{j=0}^{q-1} \frac{\bar N^u_{i,j}}{D_{i,j}} S_{i+j} S_j^\dagger \Tr\brackets{S_j S_{i+j}^\dagger \Oo} e^{in\cdot y} \nonumber \\
G^{\bar 2'}_{-}(y)[\Oo] &= \frac{-1}{q^2 {\rm vol}(\RR^4/\hat\Lambda)} \sum_{n} \sum_{i=\delta_{n,0}}^{q-1} \sum_{j=0}^{q-1} \frac{-N^u_{i,j}}{D_{i,j}} S_{i+j} S_j^\dagger \Tr\brackets{S_j S_{i+j}^\dagger \Oo} e^{in\cdot y} \nonumber \\
G^{\bar 2'}_{\RR}(y)[\Oo] &= \frac{1}{2q^2 {\rm vol}(\RR^4/\hat\Lambda)} \sum_{n} \sum_{i=\delta_{n,0}}^{q-1} \sum_{j=0}^{q-1} \frac{\bar N^v_{i,j}}{D_{i,j}} S_{i+j} S_j^\dagger \Tr\brackets{S_j S_{i+j}^\dagger \Oo} e^{in\cdot y} \nonumber \\
G^{\bar 1'}_+(y)[\Oo] &= G^{\bar 2'}_+(y)[\Oo] = G^{1'}_-(y)[\Oo] = G^{2'}_-(y)[\Oo] = 0 \ , \label{eq:green}
\end{align}
where $\Oo$ is valued in the (complexified) adjoint representation of $\mf{su}(q)$. Straightforward calculations demonstrate that these are the components of 1-forms $G_+$, $G_-$, and $G_\RR$ on $\hat T^4$ valued in $\mf{su}(q)\otimes \mf{su}(q)^\vee$ which satisfy
\be d_{B^{\rm orb}}*G_\pm[\Oo] = d_{B^{\rm orb}}*G_\RR[\Oo] = 0 \ee
and
\be
d^+_{B^{\rm orb}} G_\pm[\Oo] = \tilde\omega_\pm (\delta_0 \Oo - P[\Oo]) \ , \quad
d^+_{B^{\rm orb}} G_\RR[\Oo] = \tilde\omega_K (\delta_0 \Oo - P[\Oo]) \ , \label{eq:green} \ee
where $d^+_{B^{\rm orb}}=\frac{1+*}{2} d_{B^{\rm orb}}$, $P[\Oo] = \frac{1}{q^2 {\rm vol}(\RR^4/\hat\Lambda)} \sum_j S_j S_j^\dagger \Tr[S_j S_j^\dagger \Oo]$, and $\tilde\omega_+=-id\psi_{1'}\wedge d\psi_{2'}$ and $\tilde\omega_K=\frac{i}{2}(d\psi_{1'}\wedge d\psi_{1'}^*+d\psi_{2'}\wedge d\psi_{2'}^*)$ are respectively the holomorphic 2-form (whose conjugate is $\tilde\omega_-$) and K\"ahler form of $\hat T^4$ in complex structure $K$. (By $\delta_0$, we really mean $\sum_{p\in \hat\Lambda} \delta_p$; that is, we are implicitly assuming that $y$ is in the unit cell of $\hat\Lambda$ containing 0.) We will comment further on these identities in \cite{mz:K3math}; hopefully, our notation suggests that these 1-forms $G_\pm$ and $G_\RR$ may be assembled into a Green's function. (The reason for the $P[\Oo]$ terms in \eqref{eq:green} is that $d_{B^{\rm orb}}^+$ is not surjective.) In the meantime, we note that \eqref{eq:intMom} may be re-expressed as
\begin{align}
\Delta U(y) &= \int_{\hat T^4} d^4y' \, \parens{G^{1'}_{+}(y-y')\brackets{\xi_+(y') - [\Delta U(y'),\Delta V(y')]} \right. \nonumber \\
&\left. \qquad - G^{1'}_{\RR}(y-y')\brackets{ \xi_\RR(y') - [\Delta U(y'), \Delta U(y')^\dagger] - [\Delta V(y'), \Delta V(y')^\dagger] } } \nonumber \\
\Delta V(y) &= \int_{\hat T^4} d^4y' \, \parens{G^{2'}_{+}(y-y')\brackets{\xi_+(y') - [\Delta U(y'),\Delta V(y')] } \right. \nonumber \\
&\left. \qquad - G^{2'}_{\RR}(y-y')\brackets{ \xi_\RR(y') - [\Delta U(y'), \Delta U(y')^\dagger] - [\Delta V(y'), \Delta V(y')^\dagger] } } \nonumber \\
\Delta U(y)^\dagger &= \int d^4y'\, \parens{G^{\bar 1'}_-(y-y')[\xi_+(y')^\dagger + [\Delta U(y')^\dagger, \Delta V(y')^\dagger] ] \right. \nonumber \\
&\left. \qquad - G^{\bar 1'}_\RR(y-y')[\xi_\RR(y') - [\Delta U(y'), \Delta U(y')^\dagger] - [\Delta V(y'), \Delta V(y')^\dagger ] ] } \nonumber \\
\Delta V(y)^\dagger &= \int d^4y'\, \parens{G^{\bar 2'}_-(y-y')[\xi_+(y')^\dagger + [\Delta U(y')^\dagger, \Delta V(y')^\dagger] ] \right. \nonumber \\
&\left.\qquad - G^{\bar 2'}_{\RR}(y-y)[\xi_\RR(y') - [\Delta U(y'), \Delta U(y')^\dagger] - [\Delta V(y'), \Delta V(y')^\dagger ] ] } \ , \label{eq:higgsInt}
\end{align}
as may be verified using \eqref{eq:Ssum} and \eqref{eq:Sortho}. Finally, we state an intermediate form of these equations, where the first order contributions on the right hand side, which are independent of $\Delta U$ and $\Delta V$, are written as in \eqref{eq:intMom}, while the higher order terms are as in \eqref{eq:higgsInt}:
\begin{align}
\Delta &U(y) = \frac{1}{2q} \sum_n \sum_{i=1}^{q-1} \sum_{j=0}^{q-1} \frac{S_{i+j} S_j^\dagger}{D_{i,j}} e^{in\cdot y} \brackets{ 2 \xi_{n,i,+} \bar N^v_{i,j} + \xi_{n,i,\RR} N^u_{i,j} } \nonumber \\
& - \int_{\hat T^4} d^4y'\, \parens{ G^{1'}_+(y-y')[ [\Delta U(y'), \Delta V(y')] ] - G^{1'}_\RR(y-y')[ [\Delta U(y'), \Delta U(y')^\dagger] + [\Delta V(y'), \Delta V(y')^\dagger] ] } \nonumber \\
\Delta &V(y) = \frac{1}{2q} \sum_n \sum_{i=1}^{q-1} \sum_{j=0}^{q-1} \frac{S_{i+j}S_j^\dagger}{D_{i,j}} e^{in\cdot y} \brackets{ -2 \xi_{n,i,+} \bar N^u_{i,j} + \xi_{n,i,\RR} N^v_{i,j} } \nonumber \\
& - \int_{\hat T^4} d^4y'\, \parens{ G^{2'}_+(y-y')[ [\Delta U(y'), \Delta V(y')] ] - G^{2'}_\RR(y-y')[ [\Delta U(y'), \Delta U(y')^\dagger] + [\Delta V(y'), \Delta V(y')^\dagger] ] } \ . \label{eq:higgsInt2}
\end{align}
\eqref{eq:higgsInt} is conceptually illuminating, as we will discuss in \cite{mz:K3math}, but \eqref{eq:higgsInt2} is more useful for explicitly computing hyper-K\"ahler structures of K3 surfaces; these expressions are related via \eqref{eq:fiDist}. We expect \eqref{eq:higgsInt2} to also play a starring role in \cite{mz:K3math}, as applying the Banach contraction principle to it should yield a rigorous proof of our results.

This integral equation adds to the similarities between this Higgs branch approach and the Coulomb branch approach of \cite{GMN:walls,mz:k3}. However, here we have a 4-dimensional integral on $\hat T^4$, as opposed to integrals over infinitely many rays in $\CC^\times$. Furthermore, the ALF, ALG, and ALH versions of these results (for those $q$ for which they make sense -- see the discussion in \S\ref{sec:BPS}) involve integrals, respectively, over the 1-, 2-, or 3-dimensional dual torus (corresponding to the fact that in these cases, the relevant D-branes in the `wrapping picture' are D3-, D4-, or D5-branes), whereas the Coulomb branch integrals are still over rays in $\CC^\times$ in the ALG and ALH cases, and the Coulomb branch formalism does not apply to the ALF case. Finally, our approach constructs not only the metric on the moduli space, but also the solutions $U,V$ themselves. In particular, in the ALG case, our integral equation produces not only the metric on the moduli space of equivariant Higgs bundles, but the Higgs bundles themselves. (Similarly, in the ALF, ALH, and K3 cases we produce equivariant solutions, respectively, of Nahm's equations on a circle, the Bogomolny equations on $\hat T^3$, and the anti-self-duality equations on $\hat T^4$, and not just the metrics on the moduli spaces of these solutions.) In this sense, it is perhaps more analogous to the integral equation of \cite{GMN:2d}. This suggests that one might be able to extract 2d-4d (and, more generally, 3d-5d and 4d-6d) BPS degeneracies from our integral equation. We hope to improve our understanding of the relationships between these approaches in future work.

\subsection{FI parameters} \label{sec:FI}

In this section, we classify the $Z_q$ orbits of equivalence classes in $\Lambda/\Lambda_i$ (i.e., equivalence classes in $(\Lambda/\Lambda_i)/Z_q$), for all $i$ and $q$. As we explained in the previous section, this roughly corresponds to the classification of triplets of FI parameters. (The `roughly' caveat is due to the reality condition on the real FI parameters.) We write $\Lambda \simeq L_B \oplus L_F$, where $L_B=\avg{1,i}$, $L_F=\avg{\mu,\mu i}$ when $q=4$ and $L_B=\avg{1,\kappa_3}$, $L_F=\avg{\mu,\mu\kappa_3}$ when $q=3,6$; of course, this decomposition need only hold when $\Lambda$ is regarded as a finitely-generated free abelian group -- that is, it need not respect the inner product. The action of right multiplication by $\kappa_q$ respects this decomposition. We write $L_{B,i}\subset L_B$ for the sublattice $L(\kappa_q^i-1)$, or equivalently $L(\kappa_q^{-i}-1)$, and similarly define $L_{F,i}$, so that $\Lambda_i\cong L_{B,i}\oplus L_{F,i}$. We now note that $L_{B,i}$ and $L_{F,i}$ are canonically isomorphic (as finitely-generated free abelian groups with $Z_q$ automorphisms), and so we can drop the $B$ and $F$ subscripts. A similar observation holds for $L_B$ and $L_F$.

We begin the orbit classification now for $q = 3$. A priori, $L \simeq \mb{Z}[\kappa_3]$, but it is convenient to take the isomorphic presentation $\mb{Z}[\kappa_3] \simeq \mb{Z}^3 / \langle (1, 1, 1) \rangle$, with the isomorphism given by $a + b \kappa_3 + c \kappa_3^2 \leftrightarrow (a, b, c)$. We let $L'$ denote the index $3$ sublattice of $L$ given by cosets with a representative $(a, b, c)$ with $a+b+c=0$ (or, equivalently, for which any representative $(a,b,c)$ satisfies $a+b+c\equiv 0$ (mod 3)), and we claim this sublattice $L'$ coincides with $L_1$ (and thus with $L_2$); indeed, the cyclic generator acts on $L$ by $(a, b, c) \mapsto (c,a,b)$, rendering the claim a straightforward computation. The three cosets in $L/L'$ are labelled by the value of $a+b+c$ (mod 3), for any representative $(a,b,c)$ of a representative in $L$. The $Z_3$ action on $L/L'$ is trivial, and so $Z_3$ orbits of $\Lambda / \Lambda_i$ are classified by $(L/L_i)^{\oplus 2}$ for each $i$, giving us $2 \cdot 3^2 = 18$ triplets of FI parameters, as promised.

Moving on to $q = 4$, we now have $L = \mb{Z}[\kappa_4]$, i.e. $L = \mb{Z}[i]$, whose elements we write as $a + bi$. Then $L_1 = L_3 = \{ (a, b) | a + b \equiv 0 \pmod{2} \}$ are index $2$ sublattices of $L$, while $L_2 = 2L$ is index $4$. For $i = 1, 3$, the $Z_4$ action on $L/L_i$ is once again trivial, so we may neglect the orbit structure and find $\Lambda / \Lambda_i \simeq (L/L_i)^{\oplus 2}$ -- a contribution of $4$ triplets from each of $i = 1, 3$. But, for $i = 2$, the 16 elements of $\Lambda / 2 \Lambda$ do have some interesting orbit structure. Indeed, writing equivalence classes of $L/2L$ as $(a, b)$ with $a, b \in \{0, 1\}$, we have that the orbit structure of $Z_4$ on these cosets is $(0, 0) ; (0, 1) \sim (1, 0); (1, 1)$. Taking a product with another copy of $L/2L$ essentially gives products of orbits, except that the `middle' choices in both yield two orbits, namely $(0, 1; 0, 1) \sim (1, 0; 1, 0)$ and $(0, 1; 1, 0) \sim (1, 0; 0, 1)$, where we write equivalence classes of $\Lambda/2\Lambda$ as $(a, b; c, d)$ (with $a, b, c, d \in \{0, 1\}$). Doing the straightforward combinatorics yields $10$ orbits in addition to the $2 \cdot 4$ from before, for a total of $18$ triplets of FI parameters once again.

The case of $q = 6$ returns us to the same lattice $L = \mb{Z}[\kappa_3]$ as before, which we again write as $\mb{Z}^3 / \avg{(1, 1, 1)}$, with the standard cyclic generator now acting by $(a, b, c) \mapsto (-b, -c, -a)$. We may compute that $L_1 = L_5 = L, L_2 = L_4 = L', L_3 = 2L$, so that $i = 1, 5$ immediately yield a single orbit each. For $i = 2$ or $4$, we first note the orbit structure of $L/L'$, where we take $(0, 0, a)$ as representatives for $a \in \{0, 1, 2\}$: $a = 0$ stays fixed while $a = 1, 2$ interchange. In $(L/L')^{\oplus 2}$, we hence have four products of orbits, except that once again multiplying the larger orbits in fact yields the two orbits $(0, 0, 1; 0, 0, 1) \sim (0, 0, 2; 0, 0, 2), (0, 0, 1; 0, 0, 2) \sim (0, 0, 2; 0, 0, 1)$, for a total of five orbits from each of $i = 2, 4$. Finally, for $i = 3$, we take as equivalence class representatives for $L/2L$ the choices $(a, b, c)$ for $a, b, c \in \{0, 1\}$ and at most one of them equal to $1$. Then the orbit structure is $(0, 0, 0) ; (1, 0, 0) \sim (0, 1, 0) \sim (0, 0, 1)$ for a paltry two orbits. In $(L/2L)^{\oplus 2}$, however, we obtain the obvious three orbits involving $(0,0,0)$, but multiplying the `large' orbits together yields an additional three, with representatives given by $(0, 0, 1 ; a, b, c)$ for $(a, b, c)$ any of the three equivalence classes with two $0$s and one $1$. Hence we have six orbits from $i = 3$ for a total of $2 \cdot 1 + 2 \cdot 5 + 6 = 18$ triplets of FI parameters once again.

\section{BPS spectra} \label{sec:BPS}

We now relate these results to the Coulomb branch construction of \cite{mz:k3}. The counts of parameters differ slightly from that of \cite{mz:K3HK}. On the Higgs side, there are 18 triplets of FI parameters, two real degrees of freedom for deforming $T^2_F\times T^2_B$ to a more general torus, and a final degree of freedom determining the ratio of the volumes of $T^2_F$ and $T^2_B$. On the Coulomb side, these map to 18 triplets of mass parameters associated to a non-abelian global symmetry, 2 real masses associated to a $U(1)^2$ symmetry, and $R$. (In contrast, in the $q=2$ case studied in \cite{mz:K3HK} these numbers on both sides were 16, 4, and 1, and in addition we had the freedom to vary $\tau_F$ and $\tau_B$.) We henceforth take the complex FI parameters to vanish, in order to focus on the BPS spectra at the orbifold points. At the order to which we work, the real FI parameters then only appear in the combination $|\xi_{nti\RR}|^2$, which is independent of $t$, and so we henceforth drop the $t$ subscript by defining $\xi_{ni\RR}=\xi_{n1i\RR}$. As in \cite{mz:K3HK}, the real mass parameters that deform the semi-flat limit from being elliptically fibered to being genus 1 fibered are not necessary for reading off the fully flavored BPS spectrum, and so for simplicity we take them to vanish. Since we could easily turn on all 20 real masses if we desired, the BPS spectra at these orbifold points determine (for each $q=3,4,6$) a 21-dimensional family of unit volume K3 metrics. (For $q=2$, the BPS spectra at the orbifold locus determine a 25-dimensional family of unit volume K3 metrics, as one has the extra freedom to vary $\tau_F$ and $\tau_B$.)

We note another interesting difference from the $q=2$ case studied in \cite{mz:K3HK}. Namely, for $q\not=2$ the orbifolds $(\RR\times T^3)/Z_q$ and $(\RR^3\times S^1)/Z_q$ do not make sense. Therefore, our Higgs branch results do not yield metrics on Coulomb branches of 5d field theories on $T^2$ or of 3d field theories at finite coupling. However, the orbifolds $\CC^2/Z_q$, $(\CC\times T^2)/Z_q$, and $T^4/Z_q$ do make sense, and as in \cite{mz:K3HK} they correspond, respectively, to 3d field theories (specifically, $U(1)$ with $N_f=q$) at infinite coupling, 4d SCFTs (the MN theories) on $S^1$, and 6d little string theories on $T^3$.

We begin with the semi-flat orbifold geometry:
\be \omega_+ = da\wedge d\tilde z \ , \quad \omega_K = \frac{i}{2}\parens{R\tau_{F,2} \, da\wedge d\bar a + \frac{1}{R\tau_{F,2}} d\tilde z\wedge d\overline{\tilde z}} \ , \label{eq:CoulombSF} \ee
where
\be \tilde z \sim \tilde z+1 \sim \tilde z+\tau_F \ , \quad a \sim a+1\sim a+\tau_B \ , \quad (a,\tilde z)\sim (\kappa_q a, \kappa_q^* \tilde z) \ . \ee
Thanks to the choice of normalization mentioned in footnote \ref{ft:norm}, we now have
\be u = i\rho a\ , \quad v = \frac{\tilde z}{\rho} \ , \label{eq:change0} \ee
where
\be \rho = \sqrt{R\tau_{F,2}} \ . \ee
We parametrize $n^u$ and $n^v$ by
\be n^u = i\rho(\tilde n^1 + \tau_B \tilde n^2) = i\rho n_B \ , \quad n^v = \frac{1}{\rho}(\tilde n^3 + \tau_F \tilde n^4) = \frac{1}{\rho} n_F \ , \label{eq:NuNv0} \ee
where $\tilde n^a\in \ZZ$. We let $\tau_F=\tau_B=\kappa_3$ for $q=3,6$ and $\tau_F=\tau_B=i$ for $q=4$.

The tilde in $\tilde z$ is due to a novel complication in the $q=3,6$ cases. Namely, in these cases it turns out that the good coordinate on the fibers differs from the elliptic coordinate $\tilde z$ by a translation:
\be z = \tilde z - \frac{1+\tau_F}{2} \ . \label{eq:zZp} \ee
For the sake of uniformity, and to make clear that there is no problem with $\Spin(8)$ singular fibers coexisting with $E_6$ or $E_8$ singular fibers, we make this translation even when $q=2,4$, as in these cases the translation is by a $Z_q$ fixed point, and so it simply permutes the mass parameters.

To understand the necessity of \eqref{eq:zZp} when $q=3,6$, note that the action
\be (a,\tilde z)\mapsto (\kappa_q a, \kappa_q^* \tilde z) \ee
induced by monodromy about the singular fiber at $a=0$ is accompanied by a transformation of the gauge charges $(p,q)$ so that $(p\tau_F+q)a$ is invariant. It should similarly be the case that $e^{i\theta_{p\gamma_m+q\gamma_e}}=(-1)^{pq} e^{i(p\theta_m+q\theta_e)}$ is monodromy-invariant. For $q=3$, we find that the gauge charges transform as $(p,q)\mapsto (-q,p-q)$. So, accounting only for the changes in the gauge charges, but not the coordinates, we find that $e^{i\theta_m}$ maps to $e^{i\theta_e}$, but $e^{i\theta_e}$ maps to $e^{-i(\theta_m+\theta_e+\pi)}$; to compensate for this, we must have $(\theta_m,\theta_e)\mapsto (-\theta_m-\theta_e-\pi,\theta_m+2\pi)$ (where the $2\pi$ shift in the second factor is inserted for our immediate convenience). Using the usual relationship
\be z = \frac{\theta_m-\tau_F\, \theta_e}{2\pi} \ , \quad p\theta_m+q\theta_e = \frac{i\pi}{\tau_{F,2}}\parens{(p\bar\tau_F+q)z-(p\tau_F+q)\bar z} \ , \label{eq:zTheta} \ee
we find that the desired monodromy takes the form
\be z\mapsto \kappa_3^* z - \frac{1+2\kappa_3}{2} \ , \ee
and this coincides with the action $\tilde z\mapsto \kappa_3^* \tilde z$ when written in terms of the $z$ coordinate. Another way of saying this is that the quadratic refinement $\sigma(p\gamma_m+q\gamma_e)=(-1)^{pq}$, which satisfies $\sigma(\gamma_m)=\sigma(\gamma_e)=1$ and so locally identifies $z$ and $\tilde z$ (as described in footnote 29 of \cite{mz:k3}), is not monodromy-invariant, whereas the quadratic refinement $\sigma(p\gamma_m+q\gamma_e)=(-1)^{p-pq+q}=(-1)^{p^2-pq+q^2}=(-1)^{|p\tau_F+q|^2}$ is monodromy-invariant, but it satisfies $\sigma(\gamma_m)=\sigma(\gamma_e)=-1$ and so introduces the shift in \eqref{eq:zZp}. So, if we want to have a global identification between $z$ and $\tilde z$ then we must do so via the latter quadratic refinement. Identical reasoning in the $q=2,4,6$ cases, and considering monodromies about all singular fibers, demonstrates that \eqref{eq:zZp} always induces the correct monodromy actions on $z$.

Note that, in every case, the point $z=0$ is invariant under $\tilde z\mapsto -\tilde z$; this must be the case, as the point $\theta_e=\theta_m=0$ must be invariant under the charge conjugation element of the S-duality group, which acts as $-1\in SL(2,\ZZ)$ (together with an outer automorphism of the Lie algebras associated to any $E_6$ singular fibers).\footnote{The outer automorphisms of the Lie algebras corresponding to any $D_4$ singular fibers associated to $-1\in SL(2,\ZZ)$ are trivial, as $-1$ coincides with the identity when regarded as an element of $SL(2,Z_2)\cong S_3$.} As a check of \eqref{eq:zZp}, we note that for each of $q=3,4,6$, if we were to modify the displacement from $\tilde z$ to $z$ by a $Z_2$ fixed point which is not also a $Z_q$ fixed point then $z$ would no longer have the correct monodromy.

We now review some of the physics encoded in these orbifolds. We begin by studying the $Z_q$ fixed points of the base and their associated singular fibers \cite{dasgupta:constant}. Via the duality between a little string theory on $T^2$ and a D3-brane probe of F-theory on K3, it is clear that the singular fibers yield global symmetries of the little string theory, as they yield gauge symmetries in F-theory; by decompactifying the base, the same is found to hold for the 4d SCFTs. In the latter cases, the only fixed point is the origin of $\CC$, which is stabilized by the entirety of $Z_q$. The corresponding singular fiber yields, respectively, $E_6$, $E_7$, or $E_8$ global symmetry for $q=3,4,6$. In the full little string theory, there are additional fixed points of $T^2_B$. For $q=3$, these are (in $a$ coordinates) at $\frac{\kappa_{12}}{\sqrt{3}}$ and $\frac{i}{\sqrt{3}}$; again, each point has a $Z_3$ stabilizer and an associated $E_6$ global symmetry. For $q=4$, we have the point $\frac{1+i}{2}$ with a $Z_4$ stabilizer and $E_7$ symmetry and the orbit $\frac{1}{2}\sim \frac{i}{2}$, each element of which has a $Z_2$ stabilizer, and whose corresponding singular fiber yields a $\Spin(8)$ global symmetry. Lastly, for $q=6$ we have the orbits $\frac{\kappa_{12}}{\sqrt{3}}\sim \frac{i}{\sqrt{3}}$ and $\half\sim \frac{\kappa_3}{2}\sim \frac{\kappa_6}{2}$, whose elements are stabilized, respectively, by $Z_3$ and $Z_2$ subgroups of $Z_6$, and whose corresponding singular fibers yield, respectively, $E_6$ and $\Spin(8)$ global symmetries.

Of course, the classification of singularities within the singular fibers is identical. This is useful for studying the moduli space of the theory compactified on a circle. The 4d global symmetry groups are now broken down to the products of ADE groups associated to the singularities of the moduli spaces. (Actually, while this is true at the level of Lie algebras, we will find that there are interesting non-local correlations between the symmetry groups associated to the various singularities.) This is clear from the duality between a little string theory on $T^3$ and an M2-brane probe of K3. The result is as follows: each $E_6$ Lie algebra breaks to $A_2^3$, each $E_7$ breaks to $A_1\oplus A_3^2$, each $E_8$ breaks to $A_1\oplus A_2\oplus A_5$, and each $D_4$ breaks to $A_1^4$. From the point of view of the probe theory, this global symmetry breaking is implemented by real mass parameters. 

Let us focus for the moment on just a single singular fiber, i.e. on the $(\CC\times T^2)/Z_q$ orbifold, and its associated field theory. The discussion of the above paragraph raises the following problem of immediate interest: given the 4d flavor group $G_{4d}$, if we are given the Lie algebra of the group to which it breaks upon turning on some real mass parameters, can we find the values of those real mass parameters? In this case (essentially because $G_{3d}$ is semisimple and maximal rank), the answer is yes, and we will now do so following the paradigm of Borel-de Siebenthal~\cite{borelDeSieb}. First, let us identify the space of inequivalent real mass parameters (which we recall arise from a flavor Wilson line wound about the $S^1_R$ circle of $\mb{R}^3 \times S^1_R$). Most intrinsically, this space is simply that of conjugacy classes of $G_{4d}$, i.e. the quotient of $G_{4d}$ by its conjugation action on itself. Somewhat more usefully, we may as usual conjugate into a maximal torus $T$; now, the redundancies in the space of mass parameters are given by the Weyl group $W = N(T)/T$, where $N(T)$ is the normalizer of the maximal torus. Then, we may more efficiently write the space of inequivalent real mass parameters as $T/W$. In particular, note that Weyl group-equivalent masses yield the same moduli spaces.

To parametrize $T$, we note that the flavor lattice is the weight lattice of $G_{4d}$ and employ the fundamental weights $\omega_i$ as a basis. Then, as a homomorphism from the flavor lattice to $\RR/2\pi\ZZ$, $\theta$ (restricted to the flavor lattice) takes the form $\sum_i \theta_i \alpha^\vee_i$, where $\alpha^\vee_i\in\mf{h}$ are the simple coroots, which satisfy $\alpha^\vee_i(\omega_j)=\delta_{ij}$. In words, $\theta_i$ is the real mass parameter associated to $\omega_i$. Here, $\mf{h}$ is the Cartan subalgebra of $G_{4d}$. But, this means that we can naturally identify $\theta$ as an element of $\mf{h}$, and $e^{i\theta}$ parametrizes $T$. We denote the orbifold values of these parameters by $\theta^{(0)}=\sum_i \theta^{(0)}_i \alpha_i^\vee$. The 3d flavor group at the orbifold point, $G_{3d}$, is the subgroup of $G_{4d}$ which stabilizes $e^{i\theta^{(0)}}$ (again, under the adjoint action of $G_{4d}$ on itself), a.k.a. the centralizer of $e^{i\theta^{(0)}}$. Although it is not important for the determination of $e^{i\theta^{(0)}}$, it is enjoyable to note that this point is always a $q$-torsion point: one straightforward way to see this claim is to note that $e^{i\theta^{(0)}}$ must be central in $G_{3d}$ (if $G_{3d}$ is to be its centralizer), but the center of $G_{3d}$ consists only of $q$-torsion points. (At this point it is only the Lie algebra of $G_{3d}$ that is clear, but as the center of the simply-connected cover $\tilde G_{3d}$ consists only of $q$-torsion points, the same is certainly true for any central quotient thereof.)

We now note that since $G_{4d}$ is simple, a fundamental domain in $\mf{h}$ for the quotient $T/W\cong \mf{h}/W^{\rm aff}$, where $W^{\rm aff}$ is the affine Weyl group, is given by the simplex with vertices at 0 and $\frac{2\pi}{m_i}\omega_i^\vee$, a.k.a. ($2\pi$ times) the Weyl alcove. Here, $\omega_i^\vee$ is a fundamental coweight and $m_i$ is the positive integral coefficient in the expression $\alpha_0=\sum_i m_i \alpha_i$ of the highest root $\alpha_0$ in the basis of simple roots. These vertices naturally correspond to the nodes of the extended Dynkin diagram associated to $G_{4d}$ (with 0 corresponding to the extra (affine) node). We will choose $\theta^{(0)}$ to lie in this fundamental domain. Our first input from \cite{borelDeSieb} is the fact that since $G_{3d}$ is a semi-simple connected closed subgroup of $G_{4d}$ of maximal rank, $\theta^{(0)}$ must be one of these vertices. Furthermore, the Dynkin diagram of $G_{3d}$ is obtained by simply deleting the corresponding node from the extended Dynkin diagram of $G_{4d}$. By inspecting the affine $D_4$ and $E_n$ Dynkin diagrams, one may verify that in order to obtain the 3d flavor Lie algebras that we found above, this node should be the unique node of valence 3 for $q\not=2$ and the unique node of valence 4 for $q=2$. This node always has $m_i=q$, so we reaffirm that $e^{i\theta^{(0)}}$ is a $q$-torsion point. Lastly, in order to determine the coefficients $\theta_i^{(0)}$ we should convert from the basis of fundamental coweights to the basis of simple coroots. If we denote the Cartan matrix by $A$, then the final result is $\theta^{(0)}_i = \frac{2\pi}{q} (A^{-1})_{ij}$, where $j$ corresponds to the deleted node of the extended Dynkin diagram. Explicitly, using the conventions of the Mathematica package LieART \cite{lieART} for numbering fundamental weights, we have
\begin{align}
q=2&: \qquad \theta_i^{(0)}=\pi (1,2,1,1) \nonumber \\
q=3&: \qquad \theta_i^{(0)}=\frac{2\pi}{3} (2,4,6,4,2,3) \nonumber \\
q=4&: \qquad \theta_i^{(0)}=\frac{\pi}{2} (4,8,12,9,6,3,6) \nonumber \\
q=6&: \qquad \theta_i^{(0)}=\frac{\pi}{3} (10,20,30,24,18,12,6,15) \ . \label{eq:orbPt}
\end{align}
It is also amusing to work out how the flavor symmetry $G_{4d}$ breaks at multiples of the orbifold point $\theta^{(0)}$ in the cases when $q$ is not prime, as we find pretty patterns of partial symmetry breaking of $G_{4d}$ down to $G_{3d}$. We work these symmetry-breaking patterns out explicitly at the Lie algebra level: for $q=4$, this takes the form $E_7\supset A_1\oplus D_6\supset A_1\oplus A_3^2$, while for $q=6$, we have the two chains $E_8\supset A_1\oplus E_7\supset A_1\oplus A_2\oplus A_5$ and $E_8\supset A_2\oplus E_6\supset A_1\oplus A_2\oplus A_5$. For $q=4$, the intermediate subgroup is the centralizer of $2\theta^{(0)}$ in $G_{4d}$, while for $q=6$ the intermediate subgroups are, respectively, the centralizers of $3\theta^{(0)}$ and $2\theta^{(0)}$.


Now that we have the orbifold points $\theta^{(0)}$, we can actually determine the groups $G_{3d}=C_{G_{4d}}(e^{i \theta^{(0)}})$. (We could also do this via heterotic/M-theory duality.) We compute the following:
\begin{align}
q=2&: \qquad G_{3d} = SU(2)^4 \nonumber \\
q=3&: \qquad G_{3d} = \frac{SU(3)^3}{Z_3} \nonumber \\
q=4&: \qquad G_{3d} = \frac{SU(4)^2\times SU(2)}{Z_4} \nonumber \\
q=6&: \qquad G_{3d} = \frac{SU(6)\times SU(3)\times SU(2)}{Z_6} \ . \label{eq:G3d}
\end{align}
We thus learn that the gauge groups associated to the various singularities of a K3 surface in M-theory are not completely independent. Of course, from this M-theory perspective there is no difference between singularities that lie within a single singular fiber and singularities that lie in different singular fibers, and so we learn that there must be relationships between the gauge symmetries associated to different singular fibers. The same must then also hold of the gauge symmetries associated to singular fibers in F-theory on K3. This will not be visible in the BPS spectra of the SCFTs associated to the singular fibers, but it will be visible in the flavor representations that appear in the full little string theory spectrum.

Before proceeding, we comment further on the $q=2$ orbifold point. The basis for the flavor lattice given by the fundamental weights that we employed above is algebraically convenient, but it obscures the geometric meaning of the real mass parameters as the positions of D6-branes on the circle in a type IIA $(\CC\times S^1)/Z_2$ orientifold. The latter is clearer if, as in \cite{mz:K3HK}, we instead employ a `basis' given by linearly independent weights of the $\mathbf{8_v}$ representation; one such basis is called the orthogonal basis in LieART. The cost of this is that the coefficients that specify weights in this basis are generally valued in $\half\ZZ$, rather than $\ZZ$. An upside is that one can employ the intuition described in \cite{mz:K3HK} to guess the orbifold point: one can obtain an $A_1^4$ singularity by starting with two $O6+2\times D6$ stacks and accounting for the non-perturbative physics of the 3d $\N=4$ $SU(2)$ $N_f=2$ gauge theory studied in \cite{sw:3d}. We denote the weights comprising the orthogonal basis by $\tilde\omega_a$. If we write $\theta=\sum_i\theta_i\alpha_i^\vee=\sum_a \tilde\theta_a \tilde\alpha_a^\vee$, where $\tilde\alpha_a^\vee\in \mf{h}$ satisfy $\tilde\alpha_a^\vee(\tilde\omega_b)=\delta_{ab}$, then the change of basis $\omega_i=\sum_a C_{ia} \tilde\omega_a$ induces the relations $\tilde\alpha_a^\vee=\sum_i C_{ia} \alpha_i^\vee$ and $\theta_i=\sum_a C_{ia} \tilde\theta_a$. We note that $\tilde\alpha_a^\vee$ are not coroots, since $C$ has half-integral entries in each column:
\be C_{ia} = \begin{pmatrix} 1&0&0&0\\ 1&1&0&0 \\ \half&\half&\half&-\half\\ \half & \half & \half & \half \end{pmatrix} \ . \ee
Similarly, the periodicities of $\tilde\theta_a$ are a bit funny: we can freely translate $\tilde\theta_1,\tilde\theta_2,\tilde\theta_4$ by integral multiples of $2\pi$, but these also shift $\tilde\theta_3$ by integral multiples of $2\pi$, and the fourth translation freedom only allows us to change $\tilde\theta_3$ by integral multiples of $4\pi$. Accounting for this, it turns out that by permuting and translating by $2\pi$ the entries of $\tilde\theta_a^{(0)}=\pi(0,0,1,1)$ one obtains 12 different points of $T$, and these precisely comprise the Weyl orbit of orbifold points. So, this basis makes it clear that all 12 points describe two $O6+2\times D6$ stacks.

We now consider perturbations away from the orbifold point. We write $\theta=\theta^{(0)}+\delta\theta$, and similarly $\theta_i=\theta^{(0)}_i+\delta\theta_i$ for the numerical coordinates. In order to relate the coordinates $\delta\theta_i$ on $T_{\theta^{(0)}}\mf{h}$ to the FI parameters of the Higgs branch formalism, it proves useful to first relate them to an intermediate basis associated to weights of $\tilde G_{3d}$, the simply connected cover of $G_{3d}$. Specifically, we write $\delta\theta=\sum_\alpha \eta_\alpha \beta_\alpha^\vee$, where $\beta_\alpha^\vee$ are $\tilde G_{3d}$ simple coroots, which we can also think of as $G_{4d}$ coroots. To determine the relationship between $\eta_\alpha$ and $\delta\theta_i$, we invoke one more result from Borel-de Siebenthal theory: a choice of simple roots for $\tilde G_{3d}$ is given by the simple roots of $G_{4d}$ other than the one corresponding to the deleted node, plus the lowest root $-\alpha_0$. However, this clearly means that the chosen positive roots of $\tilde G_{3d}$ are not all positive roots of $G_{4d}$, so we instead prefer to make a different choice of the final simple root of $\tilde G_{3d}$. To do so, we consider the simple factor $\tilde G'_{3d}$ of $\tilde G_{3d}$ of which $\pm\alpha_0$ are roots; since $\alpha_0$ is the highest root of $G_{4d}$, it will also be the highest root of $\tilde G'_{3d}$ if we choose our positive roots of $\tilde G_{3d}$ to be positive roots of $G_{4d}$. But, the simple factors of the Lie algebra of $\tilde G_{3d}$ are all of type $A_n$, and the highest root of an $A_n$ Lie algebra is given by the sum of the simple roots. Therefore, we take the final simple root of $\tilde G_{3d}$ to be $\alpha_0-\sum_{i\in I} \alpha_i$, where $I$ indexes the simple roots of $G_{4d}$ which are also roots of $\tilde G'_{3d}$. Equivalently, $I$ consists of the indices of nodes in the extended Dynkin diagram of $G_{4d}$ which, after we delete a node in order to obtain the Dynkin diagram of $\tilde G_{3d}$, are in the same connected component as (but differ from) the extra node of the extended Dynkin diagram of $G_{4d}$. Explicitly, the simple roots of $\tilde G_{3d}$ and $G_{4d}$ are related as follows:
\begin{align}
q=2&:\qquad \beta_\alpha=\alpha_\alpha\,\, (\alpha\not=2)\ , \quad \beta_2=\alpha_1+2\alpha_2+\alpha_3+\alpha_4 \nonumber \\
q=3&:\qquad \beta_\alpha=\alpha_\alpha\,\, (\alpha\not=3)\ , \quad \beta_3=\alpha_1+2\alpha_2+3\alpha_3+2\alpha_4+\alpha_5+\alpha_6 \nonumber \\
q=4&:\qquad \beta_\alpha=\alpha_\alpha\,\, (\alpha\not=3)\ , \quad \beta_3=\alpha_1+2\alpha_2+4\alpha_3+3\alpha_4+2\alpha_5+\alpha_6+2\alpha_7 \nonumber \\
q=6&:\qquad \beta_\alpha=\alpha_\alpha\,\, (\alpha\not=3)\ , \quad \beta_3=2\alpha_1+4\alpha_2+6\alpha_3+4\alpha_4+3\alpha_5+2\alpha_6+\alpha_7+3\alpha_8 \ . \label{eq:rootHiggs}
\end{align}
For $q=2$, each $\beta_\alpha$ corresponds to a different $A_1$ factor; for $q=3$, the simple roots of the different simple factors in $A_2^3$ are given by $\{\beta_1,\beta_2\},\{\beta_4,\beta_5\},\{\beta_3,\beta_6\}$; for $q=4$, the simple roots of $A_3^2\oplus A_1$ are given by $\{\beta_1,\beta_2,\beta_3\},\{\beta_4,\beta_5,\beta_6\},\{\beta_7\}$; and for $q=6$ the simple roots of $A_5\oplus A_2\oplus A_1$ are given by $\{\beta_3,\beta_4,\beta_5,\beta_6,\beta_7\},\{\beta_1,\beta_2\},\{\beta_8\}$. Because all of our Lie algebras are simply-laced, root lengths are all equal and we can convert \eqref{eq:rootHiggs} into a relationship between the simple coroots of $\tilde G_{3d}$ and the simple coroots of $G_{4d}$ by putting a ${}^\vee$ superscript on all $\beta$s and $\alpha$s. This relationship, combined with $\delta\theta=\sum_\alpha \eta_\alpha \beta^\vee_\alpha = \sum_i \delta\theta_i \alpha^\vee_i$, yields
\begin{align}
q=2&:\qquad \delta\theta_i=\eta_i+\eta_2\,\,(i\not=2)\ , \quad \delta\theta_2 = 2\eta_2 \nonumber \\
q=3&:\qquad \delta\theta_i=\eta_i+\eta_3\,\,(i\in\{1,5,6\})\ ,\quad \delta\theta_i=\eta_i+2\eta_3\,\, (i\in\{2,4\})\ , \quad \delta\theta_3 = 3\eta_3 \nonumber \\
q=4&:\qquad \delta\theta_i=\eta_i+\eta_3\,\,(i\in\{1,6\})\ , \quad \delta\theta_i=\eta_i+2\eta_3\,\,(i\in\{2,5,7\})\ , \nonumber\\
&\qquad\qquad \delta\theta_4=\eta_4+3\eta_3 \ ,\quad \delta\theta_3 = 4\eta_3 \nonumber \\
q=6&:\qquad \delta\theta_7=\eta_7+\eta_3\ ,\quad \delta\theta_i=\eta_i+2\eta_3\,\,(i\in\{1,6\})\ , \quad \delta\theta_i=\eta_i+3\eta_3\,\,(i\in\{5,8\}) \ , \nonumber \\
&\qquad\qquad \delta\theta_i=\eta_i+4\eta_3\,\,(i\in\{2,4\}) \ , \quad \delta\theta_3=6\eta_3 \ . \label{eq:etaDTheta}
\end{align}
As with $\tilde\theta_a$, the periodicities of the $\eta_\alpha$ coordinates are slightly complicated. This correlates with the fact that the fundamental weights of $\tilde G_{3d}$ dual to the simple coroots $\beta_\alpha^\vee$ are not an honest basis for the $G_{4d}$ weight lattice, as their integral span includes $\tilde G_{3d}$ weights which are not $G_{3d}$ or $G_{4d}$ weights. (Of course, as $G_{3d} \subset G_{4d}$ is of maximal rank, these groups share the same weight lattice.)

Next, we sketch how we will relate the $\eta$s to the FI parameters; we will actually carry this out, on a case by case basis, in the following subsections. The discussion so far has been focused on the Higgsing of $G_{4d}$ to $G_{3d}$ within a single singular fiber of the F-theory compactification, but these same results hold for each simple factor of the non-abelian part of the global symmetry group of the little string theory. We therefore now switch our focus to the little string theory, in order to talk about the FI parameters corresponding to all of the singular fibers. So, $G_{3d}$ will now refer to the non-abelian global symmetry at the orbifold point $\theta^{(0)}$ and $\eta$s will have subscripts indicating their associated singular fiber. The purpose of introducing the $\eta$ basis as an intermediary between the $\delta\theta$s and the FI parameters is that the $\eta$s naturally parametrize the Cartan subalgebra of $G_{3d}$, and so do the $\xi$s when packaged as in \eqref{eq:D2FI}. (This is the correct choice, as opposed to \eqref{eq:TFI}, since the $\eta$ parameters are associated to singularities of $T^4/Z_q$, not of $\hat T^4/Z_q$.) So, as is familiar from the examples of 3d mirror symmetry studied in \cite{s:3dmirror}, there is a natural identification between them.

Strangely, however, we will actually find it prudent to slightly modify \eqref{eq:D2FI} before matching with the $\eta$s. Otherwise, the match between the Higgs and Coulomb formulae at second order in the FI parameters does not work for $q=4,6$. Specifically, we replace \eqref{eq:D2FI} with
\be \sum_{i:\chi^i x=x\!\!\!\!\!\pmod{\Lambda}} \frac{\xi_{(\chi^i-1)x,i}}{1-\kappa_q^i} \sigma^i \ . \label{eq:D2mod} \ee
It is not clear why we should perform such a modification; it would certainly be nice to understand this (assuming that this is the right fix for the mismatch between the Higgs and Coulomb formulae). Furthermore, since the symplectic forms are only sensitive to the magnitudes of the real FI parameters at the present order, there is some phase freedom in our modifications in \eqref{eq:D2mod}. However, \eqref{eq:D2mod} does at least lead to a change of variables from $\eta$s to $\xi$s with the property that all of the data from \cite{neitzke:e6,neitzke:e7} agrees with our Poisson-resummed Higgs branch results, and it also simplifies some formulae. Rather than obsess (further) over this, we defer its resolution to \cite{mz:K3HK2,mz:K3math}, since higher order corrections to the symplectic forms should be sensitive to the phases of the FI parameters, and will therefore hopefully clarify this point, and the perturbation theory about ALE metrics that describes K3 metrics near the fixed points may also be helpful.

The only remaining problem is, when there are singularities with isomorphic corresponding factors in $G_{3d}$ within a singular fiber, to figure out which singularity, with its associated FI parameters as in \eqref{eq:D2FI}, should be paired with which simple factor of $G_{3d}$ and its associated $\eta$s. To do so, we will combine the results of our Poisson resummation with a general result of \cite{zwiebach:webs2} that constrains the flavor representations that can appear in the BPS spectrum with given values of the gauge charges, as well as minimal information from \cite{neitzke:e7} when $q=4$.

\bigskip

We now work out the first order (order $\xi^2$) corrections to \eqref{eq:change0}:
\be u = u^{\rm sf} + \sum_{n}\sum_{i=1}^{\floor{q/2}} f_i u_{ni} \ , \quad v = v^{\rm sf} + \sum_{n}\sum_{i=1}^{\floor{q/2}}f_i v_{ni} \ . \label{eq:newVars} \ee
(Henceforth, the summation range of $i$ will be as in \eqref{eq:newVars}.) As in \cite{mz:K3HK}, we do so using our a priori knowledge that $\omega_{+\, z\bar a} = \omega_{+\, z\bar z} = 0$. We first compute
\begin{align}
0 &= \omega_{+\, z\bar a} = \partial_z v^{\rm sf} \partial_{\bar a} \bar u^{\rm sf} \sum_{n,t,i} f_i \omega_{nti+\, v\bar u} + \omega_{+\, vu}^{\rm sf} \partial_z v^{\rm sf} \sum_{n,i} f_i \partial_{\bar a} u_{ni} \\
&= \frac{1}{\rho}(-i\rho) \sum_{n,t,i} f_i \omega_{nti+\, v\bar u} + i\frac{1}{\rho} \sum_{n,i} f_i \partial_{\bar a} u_{ni} \ ,
\end{align}
which suggests the differential equation
\be
\partial_{\bar a} u_{ni} = \rho \sum_t \omega_{nti+\, v\bar u} = \frac{i(1-\kappa_q^{-i})^2 \rho |\xi_{ni\RR}|^2 (N^u_i)^2}{2D_i^3} \ee
(where we use \eqref{eq:change0} on the right hand side) whose solution is\footnote{As in \cite{mz:K3HK}, we will find that the `constant' of integration -- a function of $a,z,\bar z$ -- should be taken to vanish. This is quite sensible, since the denominator of such a function would presumably involve $D_i$, but the latter depends on $\bar a$. Finally, a genuine constant -- independent of $a,z,\bar z$ -- would have no effect at the order to which we work, since at this order the correction \eqref{eq:newVars} to the change of variables only matters via its contribution to the Jacobian. But, even at higher orders, where such a constant could be detected, we expect it to vanish, since the identifications $(\kappa_q u,\kappa_q^* v)\sim (u,v)$ correspond to the analogous identifications for $(a,\tilde z)$.}
\be
u_{ni} = \frac{(1-\kappa_q^{-i}) |\xi_{ni\RR}|^2 N^u_i}{4 D_i^2} \label{eq:uChange}
\ . \ee
As a check, we compute
\begin{align}
\omega_{+ \, z\bar z} &= \partial_z v^{\rm sf} \partial_{\bar z} \bar v^{\rm sf} \sum_{n,t,i} f_i \omega_{nti+\, v\bar v} + \omega^{\rm sf}_{+\, vu} \partial_z v^{\rm sf} \sum_{n,i} f_i \partial_{\bar z} u_{ni} \\
&= \frac{1}{\rho^2} \sum_{n,t,i} f_i \omega_{nti+\, v\bar v} + \frac{i}{\rho} \sum_{n,i} f_i \partial_{\bar z} u_{ni}
\end{align}
and then note that
\be \frac{1}{\rho^2} \sum_{t} \omega_{nti+\, v\bar v} + \frac{i}{\rho} \partial_{\bar z} u_{ni} = 0 \ . \ee
Next, in light of \eqref{eq:uChange}, we make the guess
\be v_{nti} = -\frac{(1-\kappa_q^i) |\xi_{ni\RR}|^2 N^v_i}{4 D_i^2} \ . \label{eq:vChange} \ee
As a check of this, we study $\omega_{+\, az}$:
\begin{align}
\omega_{+\, az} &- \omega^{\rm sf}_{+\, az} = \partial_a u^{\rm sf}\partial_z v^{\rm sf} \sum_{n,t,i} f_i \omega_{nti+\, uv} + \partial_a u^{\rm sf} \omega^{\rm sf}_{+\, uv} \sum_{n,i} f_i \partial_z v_{ni} + \partial_z v^{\rm sf} \omega^{\rm sf}_{+\, uv} \sum_{n,i} f_i \partial_a u_{ni} \nonumber \\
&= \rho \sum_{n,i} f_i \partial_z v_{ni} - \frac{i}{\rho} \sum_{n,i} f_i \partial_a u_{ni} \ .
\end{align}
In \cite{mz:K3HK}, the contributions from these two terms were equal to each other, and this continues to hold here:
\be \rho \partial_z v_{ni} = -\frac{i}{\rho} \partial_a u_{ni} = \frac{|1-\kappa_q^i|^2 |\xi_{ni\RR}|^2 (|N^v_i|^2 - |N^u_i|^2)}{4 D_i^3} \ . \ee
Of course, the ultimate check of \eqref{eq:vChange} will be that our expressions take the form predicted in \cite{mz:k3}, but this seems to be a useful heuristic.

Fourier transformation of the K\"ahler forms now proceeds as in \cite{mz:K3HK}. Specifically, for each $n_B$, $i=1,\ldots,\floor{q/2}$, and $\lambda\in L/L_i$ (where $L$ refers to the fiber copy of $L$ in $\Lambda\cong L\oplus L$ parametrized by $n_F$), we Poisson resum over the representatives of $\lambda$, as the FI parameters are the same for them all. It will also prove convenient to perform a $q$- and $i$-dependent rescaling of $n_B$:
\be n_B' = \frac{1}{\kappa_q^i-1} n_B \ .  \ee
This introduces the complication that terms associated to a given $i$ only contribute when $n_B'\in \frac{1}{\kappa_q^i-1} L\equiv \tilde L_i=\tilde L_{-i}$. However, having made this rescaling we can express gauge and total central charges in the form
\be Z_{\gamma_g'} = (P\tau_F+Q) a \ , \quad Z_{\gamma'} = (P\tau_F+Q)(a-n_B') \ , \ee
where $\gamma'_g=(P,Q)$ will be defined slightly differently depending on $q$ and $i$. This makes it clear that the terms with a given value of $n_B'$ arise from strings in the F-theory picture which end on the singular fiber at $n_B'\pmod{L}$. Indeed, the lattice $\tilde L_i$ is precisely the lattice of points fixed by (at least) $Z_{q/i}$ (in the sense that $\kappa_q^{i} a=a\pmod{L}$), as one may verify by comparing with the positions of the singular fibers provided above. We therefore now denote the FI parameters by $\xi_{n_B' \lambda i \RR}$. Of course, they only depend on $n_B'$ via its equivalence class in $\tilde L_i/\parens{\frac{1}{\kappa_q^i-1}L_i}=\tilde L_i/L$ -- i.e., on the $Z_{q/i}$-fixed point $n_B'\pmod{L}$ -- and we also have the $Z_q$ identifications $\xi_{(\kappa_q n_B') (\chi \lambda) i \RR}=\xi_{n_B' \lambda i \RR}$. Note that a triplet $(i,n_B',\lambda)$ determines a point $x\in T^4$ stabilized by $\chi^{i}$ with coordinates $a=n_B', z=\frac{1}{\kappa_q^{-i}-1}\lambda$. We can therefore write \eqref{eq:D2mod} as
\be \sum_{\substack{i=1,\ldots,\floor{q/2}:\\ n_B',z\in \tilde L_i}} \frac{f_i\, \xi_{n_B' [(\kappa_q^{-i}-1)z] i \RR}}{1-\kappa_q^i} \sigma^i + {\rm h.c.} \ , \label{eq:D2reform} \ee
where $f_i$ is defined below \eqref{eq:halfSum}.

At this point, our discussion necessarily trifurcates, as the $L_i$-cosets over which we can Poisson resum depend on $q=3,4,6$.

\subsection{$q=3$} \label{sec:e6}

Since $\floor{q/2}=1$, $i$ is only allowed to be 1, and so we drop all $i$ subscripts. For example, $N^u\equiv N^u_{n,1,0}=n^u+\sqrt{3}\kappa_{12}^* u$, $N^v\equiv N^v_{n,1,0}=n^v+\sqrt{3}\kappa_{12} v$, $\tilde L\equiv \tilde L_1$, and $\xi_{n_B' \lambda \RR}\equiv \xi_{n_B' \lambda 1\RR}$.

We now Poisson resum over the fiber copy of $L'$. We recall that $L=\ZZ[\kappa_3]$, in which $n_F$ is valued, is isomorphic to $\ZZ^3/\avg{(1,1,1)}$, with the isomorphism given by $a+b\kappa_3+c\kappa_3^2\leftrightarrow (a,b,c)$. We can thus represent elements of $L$ by points $(a,b,0)\in \ZZ^3$. Comparing with \eqref{eq:NuNv0}, we identify $(a,b)$ with $(\tilde n^3,\tilde n^4)$. We then recall from \S\ref{sec:FI} that $L'$-cosets are labelled by the value of $a+b+c\pmod{3} = \tilde n^3+\tilde n^4 \pmod{3}$. We therefore define $\lambda\in \{0,1,2\}$ by $\lambda\equiv \tilde n^3+\tilde n^4 \pmod{3}$. Fixing $\lambda$ then means that we want to Poisson resum over those values of $\tilde n^3$ and $\tilde n^4$ such that $\tilde x=\tilde n^3$ and $\tilde y=\frac{\tilde n^3+\tilde n^4-\lambda}{3}$ are integral. That is, we Poisson resum over $(\tilde x,\tilde y)\in \ZZ^2$. We denote the conjugate variables by $(k_3,k_4)$.

After making the change of variables
\be
\gamma'_g = (P,Q) = (-2k_3-k_4, -k_3) \ , \quad 
k_3 = -Q \, ,\  k_4 = - P + 2Q \ ,
\ee
and performing the trivial sums over $t=\pm 1$, we obtain the following result:
\begin{align}
\varpi(\zeta) &= \varpi^{\rm orb}(\zeta)+\varpi^{\rm inst}(\zeta) \nonumber \\
\varpi^{\rm inst}(\zeta) &= -\frac{i}{2\zeta}\omega_+^{\rm inst} + \omega_K^{\rm inst} - \frac{i\zeta}{2}\omega_-^{\rm inst} \nonumber \\
&= \sum_{(\tilde n^1,\tilde n^2)\in \ZZ^2} \sum_{\gamma'_g\in\ZZ^2 \backslash \{(0,0)\}} \sum_{\lambda\in\{0,1,2\}} \varpi^{\rm inst}_{n_B'\gamma'_g\lambda} \nonumber \\
&= \sum_{(\tilde n^1,\tilde n^2)\in \ZZ^2} \sum_{\gamma'_g\in\ZZ^2 \backslash \{(0,0)\}} \sum_{\lambda\in\{0,1,2\}} \parens{-\frac{i}{2\zeta}\omega_{n_B'  \gamma'_g \lambda +} + \omega_{n_B'  \gamma'_g \lambda K} - \frac{i\zeta}{2}\omega_{n_B' \gamma'_g \lambda -}} \ , \nonumber \\
\omega^{\rm inst}_{n_B'\gamma'_g\lambda} &= - \frac{i}{8\pi^2}\cdot \parens{- \frac{4\pi^4 R^2}{3} |\xi_{n_B'\lambda \RR}|^2} \kappa_2^{P+Q} \kappa_3^{- \lambda (-P+2Q)} e^{i(P\theta_m+Q\theta_e)} \times \nonumber \\
&\quad  d\Y^{\rm sf}_{\gamma'_g}\wedge\parens{-|Z_{\gamma'}| K_1(2\pi R |Z_{\gamma'}|) d\log(Z_{\gamma'}/\bar Z_{\gamma'}) + K_0(2\pi R |Z_{\gamma'}|)\parens{\frac{1}{\zeta} dZ_{\gamma'} - \zeta d\bar Z_{\gamma'} } } \ .
\end{align}
We henceforth write $\kappa_3^{\lambda(P-2Q)}$ in the slightly more aesthestically pleasing form $\kappa_3^{\lambda(P+Q)}$. We also write $e^{i(P\theta_m+Q\theta_e)}$ as $(-1)^{P Q}e^{i\theta_{\gamma'_g}}$. Next, we write $\gamma'_g=n\gamma_g=n(p,q)$, where $n>0$ and $p$ and $q$ are relatively prime. Since, as in \cite{mz:K3HK}, the geometry of the string webs that we are detecting at second order in the FI parameters is quite simple, the string web winding charge is also divisible by $n$, and so we write $\gamma'_w=n\gamma_w$. Correspondingly, we define
\be
Z_{\gamma''} = \frac{1}{n} Z_{\gamma'} = (p\tau_F+q)\parens{a-n_B'} \ ,
\ee
where $\gamma''=\gamma_g+\gamma_w$; note that $\gamma''$ does not contain any flavor charge associated to the non-abelian global symmetries. Since $(-1)^{PQ}=(-1)^{n^2 pq}=(-1)^{npq}=(-1)^{n(1+p+q)}$ (where the last equality used the fact that $p$ and $q$ cannot both be even), we now have
\begin{align}
\varpi^{\rm inst}(\zeta) &= \sum_{\gamma_g} \varpi^{\rm eff}_{\gamma_g} \nonumber \\
\varpi^{\rm eff}_{\gamma_g} &= -\frac{i}{8\pi^2} d\Y^{\rm sf}_{\gamma_g}(\zeta)\wedge \sum_{(\tilde n^1,\tilde n^2)\in\ZZ^2} \sum_{\lambda\in\{0,1,2\}} \sum_{n>0} n^2 e^{in\theta_{\gamma_g}} (-1)^{n} \kappa_3^{n\lambda(p+q)}
\parens{-\frac{4\pi^4 R^2}{3} |\xi_{n_B'\lambda \RR}|^2} \times \nonumber \\
&\quad \parens{-|Z_{\gamma''}| K_1(2\pi R n |Z_{\gamma''}|) d\log(Z_{\gamma''}/\bar Z_{\gamma''}) + K_0(2\pi R n |Z_{\gamma''}|) \parens{\frac{1}{\zeta} dZ_{\gamma''} - \zeta d\bar Z_{\gamma''}}} \ . \label{eq:wEff}
\end{align}
Recalling that the contribution from a BPS state with gauge charge $m\gamma_g$, where $m>0$ and $\gamma_g$ is a primitive element of the gauge charge lattice, and flavor charge $\gamma_f$ is
\begin{align}
\Omega(\gamma) \varpi^{\rm inst}_\gamma(\zeta) &= - \frac{i \Omega(\gamma)}{8\pi^2} d\Y^{\rm sf}_{m\gamma_g}(\zeta) \wedge \sum_{n>0} e^{in(\theta_{m\gamma_g} + \theta_{\gamma_f})} \times \nonumber \\
&\qquad \parens{ -|Z_{m\gamma_g}+ Z_{\gamma_f}| K_1(2\pi R n |Z_{m\gamma_g}+ Z_{\gamma_f}|) d\log((Z_{m\gamma_g}+ Z_{\gamma_f})/(\bar Z_{m\gamma_g}+ \bar Z_{\gamma_f})) \right. \nonumber \\
&\left. \qquad\qquad+ K_0(2\pi R n |Z_{m\gamma_g}+ Z_{\gamma_f}|)\parens{\frac{1}{\zeta} dZ_{m\gamma_g} - \zeta d\bar Z_{m\gamma_g} } } \nonumber \\
&= - \frac{i m^2 \Omega(\gamma)}{8\pi^2} d\Y^{\rm sf}_{\gamma_g}(\zeta) \wedge \sum_{n>0} e^{inm(\theta_{\gamma_g} + \frac{1}{m} \theta_{\gamma_f})} \times \nonumber \\
&\qquad \parens{ -|Z_{\gamma_g}+\frac{1}{m} Z_{\gamma_f}| K_1(2\pi R n m |Z_{\gamma_g}+\frac{1}{m} Z_{\gamma_f}|) d\log((Z_{\gamma_g}+\frac{1}{m} Z_{\gamma_f})/(\bar Z_{\gamma_g}+\frac{1}{m} \bar Z_{\gamma_f})) \right. \nonumber \\
&\left. \qquad\qquad+ K_0(2\pi R n m |Z_{\gamma_g}+\frac{1}{m} Z_{\gamma_f}|)\parens{\frac{1}{\zeta} dZ_{\gamma_g} - \zeta d\bar Z_{\gamma_g} } } \nonumber \\
&= -\frac{im^2\Omega(\gamma)}{8\pi^2} d\Y^{\rm sf}_{\gamma_g}(\zeta)\wedge \sum_{n>0:\,m|n} e^{in(\theta_{\gamma_g}+\frac{1}{m} \theta_{\gamma_f})} \times \nonumber \\
&\qquad \parens{ -|Z_{\gamma_g} + \frac{1}{m} Z_{\gamma_f}| K_1(2\pi R n |Z_{\gamma_g}+\frac{1}{m} Z_{\gamma_f}|) d\log((Z_{\gamma_g}+\frac{1}{m} Z_{\gamma_f})/(\bar Z_{\gamma_g}+\frac{1}{m} \bar Z_{\gamma_f})) \right. \nonumber \\
&\left.\qquad\qquad + K_0(2\pi R n |Z_{\gamma_g}+\frac{1}{m} Z_{\gamma_f}|) \parens{\frac{1}{\zeta} dZ_{\gamma_g} - \zeta d\bar Z_{\gamma_g}} } \ , \label{eq:wInst}
\end{align}
we see that we are on the right track. For, summing the contributions \eqref{eq:wInst} over all charges of the form $\gamma=m\gamma''+\tilde\gamma_f$ with fixed $\gamma_g=(p,q)$, where $\tilde\gamma_f$ denotes the part of the flavor charge $\gamma_f=m\gamma_w+\tilde\gamma_f$ associated to the non-abelian global symmetries, yields
\begin{align}
\varpi^{\rm eff}_{\gamma_g} &= -\frac{i}{8\pi^2} d\Y^{\rm sf}_{\gamma_g}(\zeta) \wedge \sum_{n>0} e^{in\theta_{\gamma_g}} \sum_{m|n} m^2 \sum_{\gamma_w,\tilde\gamma_f} \Omega(m\gamma''+\tilde\gamma_f) e^{in\theta_{\gamma_f}/m} \times \nonumber \\
&\qquad \parens{ -|Z_{\gamma''}| K_1(2\pi R n |Z_{\gamma''}|) d\log(Z_{\gamma''}/\bar Z_{\gamma''}) + K_0(2\pi R n |Z_{\gamma''}|) \parens{\frac{1}{\zeta} dZ_{\gamma''} - \zeta d\bar Z_{\gamma''}} } \ , \label{eq:wEffGen}
\end{align}
and so all that remains is to undo the Taylor expansion in the real masses that takes us from \eqref{eq:wEffGen} to \eqref{eq:wEff}.

We note that
\be Z_{\gamma''} = Z_{\gamma_g} + \frac{1}{m} Z_{\gamma_f} = (p\tau_F+q)\parens{a-n_B'} \label{eq:Zcomb} \ee
is independent of $m$. Geometrically, this is because the string webs we are detecting are simply $m$ superposed strings. Therefore, while the factor of $e^{in\theta_{\gamma_g}}$ in \eqref{eq:wEff} allows us to extract those terms in $\varpi^{\rm inst}$ that contribute with a given value of $p,q,n$, and \eqref{eq:Zcomb} similarly allows us to read off $n_B'$, we cannot similarly grade by $m$. So, we should compare the terms in \eqref{eq:wEff} and \eqref{eq:wEffGen} which have the same values of $p,q,n,n_B'$:
\be \sum_{m|n} \sum_{\tilde\gamma_f\in P(E_6)} m^2 \Omega(m\gamma''+\tilde\gamma_f) e^{in\theta_{\tilde\gamma_f}/m} = n^2 (-1)^n \sum_{\lambda\in\{0,1,2\}} \kappa_3^{n\lambda(p+q)} \parens{- \frac{4\pi^4 R^2}{3} |\xi_{n_B'\lambda\RR}|^2} \ . \label{eq:info} \ee
Here, $\tilde\gamma_f$ runs over the weight lattice $P(E_6)$ of the $E_6$ Lie algebra associated to the singular fiber at $n_B' \pmod{L}$. We recall that $\gamma''=\gamma_g+\gamma_w$, where $\gamma_w$ (which depends on $p,q,n_B'$) labels a charge associated to an abelian global symmetry.

We briefly comment further on this global symmetry. The flavor contribution to \eqref{eq:Zcomb} -- i.e., the term proportional to $n_B'$ -- is the central charge associated to a $U(1)^2$ winding symmetry (in the string web picture). This is in contrast with the $q=2$ case studied in \cite{mz:K3HK}, where there was a $U(1)^4$ winding symmetry, associated to the charges $mp\tilde n^1$, $mp\tilde n^2$, $mq\tilde n^1$, and $mq\tilde n^2$. In the present case, only certain combinations of these charges are conserved. This may be understood by noting that $\frac{1}{\kappa_3-1}=\frac{\kappa_{12}^7}{\sqrt{3}}$ and
\be -m(p\tau_F+q)\cdot \frac{\kappa_{12}^7}{\sqrt{3}} (\tilde n^1+\tau_B\tilde n^2) = \frac{\kappa_{12}}{\sqrt{3}} m\parens{(q\tilde n^1-p\tilde n^2)+\kappa_3(p\tilde n^1+q\tilde n^2-p\tilde n^2)} \ , \label{eq:windingZ} \ee
so all winding central charges are integral linear combinations of $\frac{\kappa_{12}}{\sqrt{3}}$ and $\frac{\kappa_{12}^{5}}{\sqrt{3}}$, where the coefficients (i.e., the conserved charges) are, respectively, $mq\tilde n^1-mp\tilde n^2$ and $mp\tilde n^1+mq\tilde n^2-mp\tilde n^2$. That is, the winding central charge lattice is $\tilde L$. Similar observations hold for $q=4,6$: for $q=4$, this lattice is $\tilde L_2$, while for $q=6$ it is the lattice $\tilde L_2+\tilde L_3$ comprised of points of the form $\ell_2+\ell_3$, where $\ell_2\in \tilde L_2$ and $\ell_3\in \tilde L_3$.

Since we are clearly only finding field theory spectra, as in \cite{mz:K3HK} -- that is, the rest of the little string theory spectrum is not contributing at this order -- the degeneracies $\Omega$ on the left side of \eqref{eq:info} are independent of $n_B'$, and so the only way that $n_B'$ enters on the left side is via the fact that we can choose different mass parameters $\theta_{\tilde\gamma_f}$ at different singular fibers. Of course, all of the weights of an $E_6$ representation contribute together, and so these $\Omega$'s are all determined by the indices associated to $E_6$ irreducible representations. That is,
\be \Omega(m\gamma_g+\tilde\gamma_f) = \sum_{\R} d_{\tilde\gamma_f,\R} \Omega(m,p,q,\R) \ , \label{eq:repWt} \ee
where $d_{\tilde\gamma_f,\R}$ is the multiplicity of the weight $\tilde\gamma_f$ in the weight system associated to $\R$. We can thus rewrite \eqref{eq:info} as
\be \sum_{m|n} \sum_{\R} m^2 \Omega(m,p,q,\R) \phi_\R(n\theta/m) = n^2 (-1)^n \sum_{\lambda\in\{0,1,2\}} \kappa_3^{n\lambda(p+q)}\parens{- \frac{4\pi^4 R^2}{3} |\xi_{n_B'\lambda\RR}|^2} \ , \label{eq:repInfo} \ee
where
\be \phi_\R(n\theta/m) = \sum_{\tilde\gamma_f\in P(E_6)} d_{\tilde\gamma_f,\R} e^{in\theta_{\tilde\gamma_f}/m} \ee
is the character of $\R$ evaluated at the point $n\theta/m$ in the maximal torus of $E_6$. (Here, by $\theta$ we mean the restriction of the homomorphism $\theta:\Gamma_{\rm flavor}\to \RR/2\pi \ZZ$ to the relevant $E_6$ weight lattice.)

We henceforth denote the combination of characters appearing on the left side of \eqref{eq:repInfo} by
\be F_{n,p,q}(\theta) = \sum_{m|n} \sum_{\R} m^2 \Omega(m,p,q,\R) \phi_\R(n\theta/m) \ . \label{eq:Fdef} \ee
At quadratic order in the FI parameters, we have found that for all $n$, this function agrees with the simple expression on the right hand side of \eqref{eq:repInfo}. This turns out to be a rather weak constraint on the BPS spectrum, but it is nevertheless quite interesting that we have learned something about $\Omega(m,p,q,\R)$ for all values of $m,p,q$! The way in which the MN theory satisfies \eqref{eq:repInfo} is far more interesting than the analogous mechanism in the $SU(2)$ $N_f=4$ theory studied in \cite{mz:K3HK}. For, the spectrum has arbitrarily large irreducible representations and `imprimitivity,' $m$, and the values of $\Omega$ and representations that appear at each $p,q,m$ are quite different. (In contrast, recall that the spectrum for $SU(2)$ $N_f=4$ simply has hypers $(\Omega=1)$ in 8-dimensional $\Spin(8)$ representations for $m=1$ and vectors $(\Omega=-2)$ in the trivial representation for $m=2$. So, in this theory one simply has
\be F_{n,p,q}(\theta) = \piecewise{\phi_{\R_{p,q}}(n\theta)-8}{2|n}{\phi_{\R_{p,q}}(n\theta)}{2\nmid n} \ , \quad \R_{p,q}=\left\{\begin{array}{rl} \mathbf{8_v} & : 2|p\wedge 2\nmid q \\ \mathbf{8_s} & : 2\nmid p\wedge 2\nmid q \\ \mathbf{8_c} & : 2\nmid p \wedge 2|q \end{array}\right. \ .) \label{eq:SU2fns0} \ee Indeed, the right hand side of \eqref{eq:repInfo} might lead one to believe that the BPS spectrum depends on $p,q$ only via the combination $p+q\pmod{3}$, but even this is not true: \cite{neitzke:e6} demonstrated that $(m,p,q)=(1,1,0)$ and $(1,1,3)$ have quite different BPS spectra. Nevertheless, $F_{1,1,0}$ and $F_{1,1,3}$ both satisfy \eqref{eq:repInfo}, as do all other functions $F_{n,p,q}$ that can be determined from the results of \cite{neitzke:e6,neitzke:e7}. (Of course, in stating this we are getting ahead of ourselves, as one needs to know the change of variables from $\theta$ to $\xi$. We will determine this shortly.)

We expect that at higher orders in the FI parameters, we will continue to find simple expressions from Poisson resumming the output of the hyper-K\"ahler quotient, and that this will allow for the complete determination of the BPS index. However, we note that while one might hope that this would imply that there are patterns in the coefficients of the Taylor series of the functions $F_{n,p,q}$ about the orbifold point that would allow one to guess the entire spectrum, it is possible that any patterns will only be easy to discern when one includes all contributions to $\varpi(\zeta)$ at a given order in the FI parameters, as opposed to solely focusing on the approximation $\varpi^{\rm inst}(\zeta)$. We leave the comparison of the Higgs and Coulomb branch formalisms at higher orders in the FI parameters to \cite{mz:K3HK2}.

We now work out the relationship between $\theta$ and $\xi$. Since we have already found the relationships between $\theta_{n_B',i}$ and $\delta\theta_{n_B',i}=\theta_{n_B',i}-\theta^{(0)}_{n_B',i}$ and between $\delta\theta_{n_B',i}$ and $\eta_{n_B',\alpha}$, it remains only to relate $\eta_{n_B',\alpha}$ to the FI parameters $\xi_{n_B'\lambda\RR}$. We know that the $\eta$s within a singular fiber above $n_B'$ are paired up as $\{\eta_{n_B',1},\eta_{n_B',2}\},\{\eta_{n_B',4},\eta_{n_B',5}\},\{\eta_{n_B',3},\eta_{n_B',6}\}$, with each pair associated to an $A_2$ singularity within the singular fiber. Our first task is to determine this association. To do so, we employ an observation from \cite{zwiebach:webs2}: there is a linear map from the gauge plus flavor lattice to a vector space over $\QQ$ with an embedded lattice, and a string web can only exist if its charges map into this lattice. Expanding the image of a charge under this map in a basis for this lattice, we call the resulting coefficients the `invariant charges' of the charge, and then the rule can be restated as the requirement that the invariant charges be integrally quantized. In practice, \cite{zwiebach:webs2} found that this yields a restriction on the flavor representations that can appear with a certain gauge charge in terms of the `congruence class'\footnote{For mathematicians, this is essentially a synonym for the central character.} of the representation, defined in \cite{lemire:congruence}, which we denote by $C(\R)$. For $E_6$, these classes correspond to the elements of $Z_3$; furthermore, classes 1 and 2 are exchanged by the outer automorphism of $E_6$ which conjugations all representations, while class 0 is invariant. With our basis for the gauge charge lattice, the integrality constraint takes the form $C(\R)=-m(p+q) \pmod{3}$. Experimentally, we find that the second order Taylor expansions about $\theta^{(0)}$ of characters of representations $\R$ with $C(\R)=1$ all take the form of a real constant times
\be \kappa_3 (\eta_{n_B',1}^2-\eta_{n_B',1}\eta_{n_B',2}+\eta_{n_B',2}^2) + \kappa_3^{-1}(\eta_{n_B',4}^2-\eta_{n_B',4}\eta_{n_B',5}+\eta_{n_B',5}^2)+(\eta_{n_B',3}^2-\eta_{n_B',3}\eta_{n_B',6}+\eta_{n_B',6}^2) \ , \ee
and comparing with \eqref{eq:repInfo} with $n=1$ lets us match $\lambda=0$ with $\{\eta_{n_B',3},\eta_{n_B',6}\}$, $\lambda=1$ with $\{\eta_{n_B',4},\eta_{n_B',5}\}$, and $\lambda=2$ with $\{\eta_{n_B',1},\eta_{n_B',2}\}$. (The same conclusion obtains if we instead examine characters of representations with $C(\R)=2$.)

Within each $A_2$ singularity, associated to a fixed point $x$ of $T^2_F$ in the singular fiber over $n_B'$, or equivalently to a choice of $\lambda$, the order of the roots is reversed by the non-trivial outer automorphism (which conjugates all representations). This is Weyl-equivalent to negating the $\eta$s, which is part of the phase freedom mentioned below \eqref{eq:D2mod}. So, at higher orders in the FI parameters the order of the $\eta$s will matter, but at present we arbitrarily fix an ordering of each of the three pairs of $\eta$s. Having done so, we can label these parameters as $\eta_{n_B'\lambda A}$, where $A=1,2$, and they parametrize a traceless diagonal matrix (an element of the $A_2$ Cartan subalgebra) via
\be {\rm diag}(\eta_{n_B'\lambda 1},\eta_{n_B'\lambda 2}-\eta_{n_B'\lambda 1},-\eta_{n_B'\lambda 2}) \ . \label{eq:etaMat} \ee
Comparing with \eqref{eq:D2reform}, which here takes the form
\be \frac{\kappa_{12}}{\sqrt{3}} \xi_{n_B' \lambda \RR}\sigma + {\rm h.c.} \ , \label{eq:xiMat} \ee
we find that
\be \xi_{n_B'\lambda\RR} \propto \eta_{n_B'\lambda 1}-\kappa_6 \eta_{n_B'\lambda 2} \ . \label{eq:xiEta0} \ee
Specifically, we require \eqref{eq:etaMat} and \eqref{eq:xiMat} to be equal, up to a real non-zero constant of proportionality. Thanks to the aforementioned phase freedom, we can require this constant of proportionality to be positive. This implies that the constant of proportionality in \eqref{eq:xiEta0} should also be positive. Explicitly, we take the change of variables to be
\begin{align}
\xi_{n_B'0\RR}&=\frac{3}{2\pi^2 R}(\eta_{n_B',6}-\kappa_6 \eta_{n_B',3}) \nonumber \\
\xi_{n_B'1\RR}&=\frac{3}{2\pi^2 R}(\eta_{n_B',5}-\kappa_6 \eta_{n_B',4}) \nonumber \\
\xi_{n_B'2\RR}&=\frac{3}{2\pi^2 R}(\eta_{n_B',1}-\kappa_6 \eta_{n_B',2}) \ . \label{eq:xiEta}
\end{align}
Composing \eqref{eq:xiEta} and \eqref{eq:etaDTheta} gives the change of variables from $\xi$s to $\delta\theta$s. With this in hand, one may verify that all of the data on the $E_6$ MN theory from \cite{neitzke:e6,neitzke:e7} satisfies \eqref{eq:repInfo}; this comparison with \eqref{eq:repInfo} is how we fixed the proportionality constant in \eqref{eq:xiEta0}.

\subsection{$q=4$} \label{sec:E7}

Now, we have both $i=1$ and $i=2$. In the former case, $L_1$ is of index 2 in $L$, and we label $L_1$-cosets by $\lambda\in\{0,1\}$ such that $\lambda\equiv \tilde n^3+\tilde n^4 \pmod{2}$. That is, after fixing $\lambda$, we define $\tilde x=\tilde n^3$ and $\tilde y=\frac{\tilde n^3+\tilde n^4-\lambda}{2}$ and Poisson resum over $(\tilde x,\tilde y)\in \ZZ^2$. When $i=2$, we instead have $L_2=2L$, and we label $L_2$-cosets by $\lambda^3,\lambda^4\in\{0,1\}$ such that $\lambda^3\equiv \tilde n^3\pmod{2}$ and $\lambda^4\equiv \tilde n^4\pmod{2}$. So, after fixing $\lambda$ we define $\tilde x=\frac{\tilde n^3-\lambda^3}{2}$ and $\tilde y=\frac{\tilde n^4-\lambda^4}{2}$ and Poisson resum over $(\tilde x,\tilde y)\in \ZZ^2$. For $i=2$, we also have the identifications $\xi_{(\kappa_4 n_B')(0,0)2 \RR}=\xi_{n_B'(0,0)2\RR}$, $\xi_{(\kappa_4 n_B')(1,1)2\RR}=\xi_{n_B'(1,1)2\RR}$, $\xi_{(\kappa_4 n_B')(0,1)2\RR}=\xi_{n_B'(1,0)2\RR}$, and $\xi_{(\kappa_4 n_B')(1,0)2\RR}=\xi_{n_B' (0,1)2\RR}$.

For $i=1$, we make the change of variables 
\be
\gamma'_g = (P,Q) = (-k_3-k_4,-k_3) \ , \quad 
k_3 = -Q \, ,\  k_4 = - P + Q \ , \label{eq:q4i1Defs}
\ee
while for $i=2$ we define
\be
\gamma'_g = (P,Q) = (-k_3,k_4) \ , \quad 
k_3 = -P \, ,\  k_4 = Q  \ . \label{eq:q4i2Defs}
\ee
Remembering the factor of $f_i$ from \eqref{eq:halfSum} and that $\xi_{n_B' \lambda (q/2)\RR}$ is real, we obtain
\begin{align}
\omega^{\rm inst}_{n_B' \gamma'_g\lambda 1} &= - \frac{i}{8\pi^2}\cdot \parens{- 2\pi^4 R^2 |\xi_{n_B' \lambda 1 \RR}|^2} \kappa_2^{- \lambda (-P+Q)} \kappa_2^{P+Q} e^{i(P\theta_m+Q\theta_e)} \times \nonumber \\
&\quad  d\Y^{\rm sf}_{\gamma'_g}\wedge\parens{-|Z_{\gamma'}| K_1(2\pi R |Z_{\gamma'}|) d\log(Z_{\gamma'}/\bar Z_{\gamma'}) + K_0(2\pi R |Z_{\gamma'}|)\parens{\frac{1}{\zeta} dZ_{\gamma'} - \zeta d\bar Z_{\gamma'} } } \nonumber \\
\omega^{\rm inst}_{n_B' \gamma'_g\lambda 2} &= - \frac{i}{8\pi^2}\cdot \parens{- \half \pi^4 R^2 \xi_{n_B' \lambda 2 \RR}^2} \kappa_2^{\lambda^3 P - \lambda^4 Q} \kappa_2^{P+Q} e^{i(P\theta_m+Q\theta_e)} \times \nonumber \\
&\quad  d\Y^{\rm sf}_{\gamma'_g}\wedge\parens{-|Z_{\gamma'}| K_1(2\pi R |Z_{\gamma'}|) d\log(Z_{\gamma'}/\bar Z_{\gamma'}) + K_0(2\pi R |Z_{\gamma'}|)\parens{\frac{1}{\zeta} dZ_{\gamma'} - \zeta d\bar Z_{\gamma'} } } \ .
\end{align}
As in \S\ref{sec:e6}, we now define $(P,Q)=n(p,q)$. Finally, recalling that the $i=1$ terms contribute only to $E_7$ singular fibers, whereas $i=2$ terms contribute to the $E_7$ singular fibers and the $\Spin(8)$ singular fiber, we find the following analogues of \eqref{eq:repInfo}:
\be
F_{n,p,q}(\theta) = n^2 (-1)^{n} \brackets{ \sum_{\lambda\in Z_2} (-2\pi^4 R^2 |\xi_{n_B' \lambda 1\RR}|^2) (-1)^{n\lambda(p+q)} + \sum_{\lambda\in Z_2^2} (-\half \pi^4 R^2 \xi_{n_B' \lambda 2\RR}^2) (-1)^{n(\lambda^3 p+\lambda^4 q)}} \label{eq:repInfoQ4Z4}
\ee
when $n_B'\pmod{L}$ is a $Z_4$-fixed point and
\be F_{n,p,q}(\theta) = n^2 (-1)^{n}  \sum_{\lambda\in Z_2^2} (-\half \pi^4 R^2 \xi_{n_B' \lambda 2\RR}^2) (-1)^{n(\lambda^3 p+\lambda^4 q)} \label{eq:repInfoQ4Z2} \ee
when $n_B'\pmod{L}$ is a $Z_2$-fixed point. The latter equality is the familiar result from \cite{mz:K3HK} for the $SU(2)$ $N_f=4$ field theory, whereas \eqref{eq:repInfoQ4Z4} corresponds to the $E_7$ MN theory. We note that, as expected, the $\Spin(8)$ theory has 4 real mass parameters whereas the $E_7$ theory has 7: in the latter case, $\xi_{n_B'(1,0)2\RR}$ and $\xi_{n_B'(0,1)2\RR}$ are identified.

We now note a fascinating aspect of these formulae: \eqref{eq:repInfoQ4Z4} and \eqref{eq:repInfoQ4Z2} correspond to two completely different SCFTs, and yet they share an ingredient -- namely, the $i=2$ term. The interpretation of this term in the two formulae is slightly different, since in \eqref{eq:repInfoQ4Z4} two of the $i=2$ FI parameters are identified, but nevertheless it means that the $E_7$ MN and $SU(2)$ $N_f=4$ functions $F_{n,p,q}$ coincide if we specialize certain parameters appropriately:
\be F^{E_7 \, {\rm MN}}_{n,p,q}(\xi_{n_B'\lambda1\RR}=0) = F^{SU(2) \, N_f=4}_{n,p,q}(\xi_{n_B' (1,0) 2 \RR}=\xi_{n_B' (0,1) 2 \RR}) \ . \label{eq:q4conjOld} \ee
Of course, this is all only at quadratic order in the FI parameters, but the way in which this correspondence is produced by Poisson resummation seems somewhat robust, in the sense that one might expect that at all orders, some terms contribute only to the spectra of the CFTs associated to $Z_4$-stabilized singular fibers whereas others contribute to both $Z_2$- and $Z_4$-stabilized singular fibers, but no terms contribute only to $Z_2$-stabilized singular fibers. (Others still will involve multiple singular fibers, but these will correspond to BPS states in the little string theory which do not correspond to BPS states in either the $SU(2)$ $N_f=4$ or $E_7$ MN SCFTs.) This leads us to conjecture that \eqref{eq:q4conjOld} holds at \emph{all} orders in the real FI parameters, not just at leading order. For the sake of clarity, and because we really like this equation, we restate it without the extraneous subscripts which are vestiges of the little string theory:
\be \boxed{F^{E_7 \, {\rm MN}}_{n,p,q}(\xi_{\lambda1\RR}=0) = F^{SU(2) \, N_f=4}_{n,p,q}(\xi_{(1,0) \RR}=\xi_{(0,1) \RR}) \ . } \label{eq:q4conj} \ee
Remarkably, despite the complexity of the $E_7$ MN spectrum and the simplicity of the $SU(2)$ $N_f=4$ spectrum, this identity is indeed satisfied by all of the data from \cite{neitzke:e7}!\footnote{Our leading order results actually suggest a stronger conjecture, namely that
\be F^{E_7 \, {\rm MN}}_{n,p,q} = F^{SU(2)\, N_f=4}_{n,p,q}(\xi_{n_B' (1,0) 2 \RR}=\xi_{n_B' (0,1) 2 \RR}) + f_{n,p,q}(\xi_{n_B'\lambda 1\RR}) \ , \ee
where $f_{n,p,q}$ depends only on the FI parameters $\xi_{n_B'\lambda 1\RR}$ and is determined at leading order by \eqref{eq:repInfoQ4Z4}. However, we have experimentally found that this conjecture fails at higher orders. This is not surprising: a term that involves two sums, one over $\tilde L_1$ and one over $\tilde L_2$, might reasonably be expected to make contributions to $F^{E_7\, {\rm MN}}_{n,p,q}$ that involve both $i=1$ and $i=2$ FI parameters (and to make no contributions to $F^{SU(2)\, N_f=4}_{n,p,q}$).} We will find evidence that there are similar relations between the BPS spectra of the SCFTs which coexist in the $q=6$ F-theory orbifold configuration. We thus expect that this principle generalizes, e.g. to the higher rank theories obtained from multiple D3-brane probes. It would be interesting to understand these relationships from additional points of view.

Here, we provide one such explanation of this phenomenon, which makes no reference to little string theories or hyper-K\"ahler quotients. We do so by observing that $(\CC\times T^2)/Z_2$ is a degree 2 topological cover (or finite \'{e}tale cover, if one prefers) of $(\CC\times T^2)/Z_4$ if we restrict to $\tau_F=i$ in the former quotient, and this fact persists in the resolved spaces as long as we only turn on certain combinations of mass parameters. In other words, there exists a degree 2 map from the $SU(2)$ $N_f=4$ Coulomb branch (on $S^1_R$) down to that of the $E_7$ MN theory such that the former is locally isomorphic to the latter -- and in particular carries the same local metric. Since the metrics of these spaces are related to the respective BPS spectra of these theories, we find that there must be relations between these spectra. It seems likely that there is a similar explanation of this phenomenon that involves relationships between string webs in the two theories. This same reasoning relates resolutions of the $(\CC\times T^2)/Z_q$ orbifolds with $q=2,3$ with resolutions of the $q=6$ orbifold and supports the conjectures \eqref{eq:q6conj} that we will make in the next subsection. More generally, similar reasoning likely relates instanton corrections to the Coulomb branches of various theories in diverse dimensions.

In order to verify \eqref{eq:q4conj}, as well as to check that \eqref{eq:repInfoQ4Z4} holds at second order in the FI parameters, one needs the change of variables from $\eta$s to $\xi$s. We determine this as in the last subsection. The $\eta$s within an $E_7$ singular fiber above $n_B'$ are grouped as $\{\eta_{n_B',1},\eta_{n_B',2},\eta_{n_B',3}\},\{\eta_{n_B',4},\eta_{n_B',5},\eta_{n_B',6}\},\{\eta_{n_B',7}\}$, where the first two groups correspond to $A_3$ singularities and the singlet corresponds to an $A_1$ singularity. We employ the integrality constraint $C(\R)=m(p+q)\pmod{2}$ ($C(\R)$ for $E_7$ is valued in $Z_2$) in order to learn that we should study charges with $p+q\equiv 1\pmod{2}$ in order to match these groups with singularities, as in the previous subsection. Specifically, \eqref{eq:repInfoQ4Z4}, combined with a couple results from \cite{neitzke:e7}, lets us match $\{\xi_{n_B'01\RR},\xi_{n_B'(0,0)2\RR}\}$ and the fixed point $z=0$ with $\{\eta_{n_B',1},\eta_{n_B',2},\eta_{n_B',3}\}$, which we thus also denote by $\{\eta_{n_B'01},\eta_{n_B'02},\eta_{n_B'03}\}$; $\{\xi_{n_B'11\RR},\xi_{n_B'(1,1)2\RR}\}$ and the fixed point $\frac{1+i}{2}$ with $\{\eta_{n_B',4},\eta_{n_B',5},\eta_{n_B',6}\}\equiv \{\eta_{n_B' \frac{1+i}{2} 1},\eta_{n_B' \frac{1+i}{2} 2},\eta_{n_B' \frac{1+i}{2} 3}\}$; and $\xi_{n_B'(1,0)2\RR}=\xi_{n_B'(0,1)2\RR}$ and the fixed points $\half$ and $\frac{i}{2}$ (which are identified) with $\eta_{n_B',7}$.

For a $Z_4$ fixed point, the $\eta$s parametrize the $A_3$ Cartan subalgebra via
\be {\rm diag}(\eta_{n_B'z 1}, \eta_{n_B' z 2} - \eta_{n_B' z 1}, \eta_{n_B' z 3} - \eta_{n_B' z 2}, - \eta_{n_B' z 3}) \ . \label{eq:etaMatQ4} \ee
Meanwhile, for the fixed point at $z=0$, \eqref{eq:D2reform} takes the form
\be \frac{1+i}{2} \xi_{n_B' 0 1 \RR}\sigma + {\rm h.c.} + \half \xi_{n_B' (0,0) 2 \RR} \sigma^2 \ , \label{eq:FIZ0} \ee
while for the fixed point at $z=\frac{1+i}{2}$, \eqref{eq:D2reform} takes the form
\be \frac{1+i}{2} \xi_{n_B' 1 1 \RR}\sigma + {\rm h.c.} + \half \xi_{n_B' (1,1) 2 \RR} \sigma^2 \ . \label{eq:FIZother} \ee
Comparing these expressions yields the change of variables
\begin{align}
\xi_{n_B' 01 \RR} &= \frac{1}{\pi^2 R}(\eta_{n_B' 1}-i\eta_{n_B' 2}-\eta_{n_B' 3}) \nonumber \\
\xi_{n_B' 11 \RR} &= \frac{1}{\pi^2 R}(\eta_{n_B' 4}-i\eta_{n_B' 5}-\eta_{n_B' 6}) \nonumber \\
\xi_{n_B' (0,0) 2 \RR} &= \frac{2}{\pi^2R}(\eta_{n_B' 1}-\eta_{n_B' 2}+\eta_{n_B' 3}) \nonumber \\
\xi_{n_B' (1,1) 2 \RR} &= \frac{2}{\pi^2R}(\eta_{n_B' 4}-\eta_{n_B' 5}+\eta_{n_B' 6}) \nonumber \\
\xi_{n_B' (1,0) 2 \RR} &= \frac{2}{\pi^2R} \eta_{n_B' 7} \ ,
\end{align}
when $n_B'$ hosts an $E_7$ singular fiber. Lastly, modulo our small changes in conventions from \cite{mz:K3HK}, the change of variables for a $\Spin(8)$ singular fiber is as in \cite{mz:K3HK}.

Lastly, in order to ease the verification of \eqref{eq:q4conj} we provide the functions $F^{SU(2)\, N_f=4}_{n,p,q}$, which follow straightforwardly from \eqref{eq:SU2fns0}:\footnote{0 and 1 subscripts are swapped relative to \cite{mz:K3HK}, thanks to the shift from $\tilde z$ to $z$.}
\begin{align}
F_{n,p,q}(\theta)
&= -4(1+(-1)^n) + 4 \left\{\begin{array}{rl}
\cos(n x_{11})\cos(n x_{01})+(-1)^n \cos(n x_{00}) \cos(n x_{10}) & : 2|p\wedge 2\nmid q \\
(-1)^n \cos(n x_{11}) \cos(n x_{00}) + \cos(n x_{01}) \cos(n x_{10}) & : 2\nmid p\wedge 2\nmid q \\
\cos(n x_{11}) \cos(n x_{10}) + (-1)^n \cos(n x_{01}) \cos(n x_{00}) & : 2\nmid p \wedge 2|q
\end{array} \right. \nonumber \\
&= \sum_{r=1}^{\infty} \frac{2(in)^{2r}}{(2r)!} \times \nonumber \\
&\left\{\begin{array}{rl} (x_{11}+x_{01})^{2r} + (x_{11}-x_{01})^{2r} + (-1)^{n} (x_{00}+x_{10})^{2r} + (-1)^{n} (x_{00}-x_{10})^{2r} & : 2|p \wedge 2\nmid q \\
(-1)^n (x_{11}+x_{00})^{2r} + (-1)^n(x_{11}-x_{00})^{2r} + (x_{01}+x_{10})^{2r} + (x_{01}-x_{10})^{2r} & : 2\nmid p \wedge 2\nmid q \\
(x_{11}+x_{10})^{2r} + (x_{11}-x_{10})^{2r} + (-1)^n(x_{01}+x_{00})^{2r} + (-1)^n (x_{01}-x_{00})^{2r} & : 2\nmid p \wedge 2|q
\end{array} \right. \ . \label{eq:SU2fns}
\end{align}
To lighten the notation in these equations, we have defined $x_{ab}\equiv \frac{\pi^2 R}{2} \xi_{n_B' (a,b) 2 \RR}$.

\subsection{$q=6$} \label{sec:E8}

Now, we have $i=1,2,3$. For $i=1$, we Poisson resum over all of $L$. For $i=2$, we have $L_2=L'$, and we label $L'$-cosets by $\lambda\in\{0,1,2\}$ such that $\lambda\equiv \tilde n^3+\tilde n^4\pmod{3}$. So, we Poisson resum over $\tilde x=\tilde n^3$ and $\tilde y=\frac{\tilde n^3+\tilde n^4-\lambda}{3}$ in $\ZZ^2$. We note the identifications $\xi_{(\kappa_6 n_B')02\RR}=\xi_{n_B' 02\RR}$, $\xi_{(\kappa_6 n_B')12\RR}=\xi_{n_B' 22\RR}$, and $\xi_{(\kappa_6 n_B')22\RR}=\xi_{n_B' 12\RR}$. Finally, we have $L_3=2L$, and so we label $L_3$-cosets by $\lambda^3,\lambda^4\in\{0,1\}$ such that $\lambda^3\equiv \tilde n^3 \pmod{2}$ and $\lambda^4\equiv \tilde n^4\pmod{2}$. We thus Poisson resum over $\tilde x=\frac{\tilde n^3-\lambda^3}{2}$ and $\tilde y=\frac{\tilde n^4-\lambda^4}{2}$ in $\ZZ^2$. The identifications amongst the FI parameters take the form $\xi_{(\kappa_6 n_B')(0,0)3\RR}=\xi_{n_B'(0,0)3\RR}$, $\xi_{(\kappa_6^2 n_B')(0,1)3\RR}=\xi_{(\kappa_6 n_B') (1,0)3\RR}=\xi_{n_B' (1,1)3\RR}$.

The definitions of $(P,Q)$ in terms of $(k_3,k_4)$ now take the form
\begin{align}
\gamma_g'&=(P,Q)=(-k_3-k_4,-k_3) \ , & k_3&=-Q\ , \  & k_4&=-P+Q & (i&=1) \nonumber \\
\gamma_g'&=(P,Q)=(-2k_3-k_4,-k_3) \ , & k_3&=-Q\ , \ & k_4&=-P+2Q & (i&=2) \nonumber \\
\gamma_g'&=(P,Q)=(-k_3, k_4) \ , & k_3&=-P\ , \ & k_4&=Q & (i&=3) \ .
\end{align}
In terms of these, we have
\begin{align}
\omega^{\rm inst}_{n_B'\gamma'_g 1} &= - \frac{i}{8\pi^2}\cdot \parens{- 4\pi^4 R^2 |\xi_{n_B' 1 \RR}|^2} \kappa_2^{P+Q} e^{i(P\theta_m+Q\theta_e)} \times \nonumber \\
&\quad  d\Y^{\rm sf}_{\gamma'_g}\wedge\parens{-|Z_{\gamma'}| K_1(2\pi R |Z_{\gamma'}|) d\log(Z_{\gamma'}/\bar Z_{\gamma'}) + K_0(2\pi R |Z_{\gamma'}|)\parens{\frac{1}{\zeta} dZ_{\gamma'} - \zeta d\bar Z_{\gamma'} } } \nonumber \\
\omega^{\rm inst}_{n_B'\gamma'_g\lambda 2} &= - \frac{i}{8\pi^2}\cdot \parens{- \frac{4}{3} \pi^4 R^2 |\xi_{n_B'\lambda 2 \RR}|^2} \kappa_3^{-\lambda(-P+2Q)} \kappa_2^{P+Q} e^{i(P\theta_m+Q\theta_e)} \times \nonumber \\
&\quad  d\Y^{\rm sf}_{\gamma'_g}\wedge\parens{-|Z_{\gamma'}| K_1(2\pi R |Z_{\gamma'}|) d\log(Z_{\gamma'}/\bar Z_{\gamma'}) + K_0(2\pi R |Z_{\gamma'}|)\parens{\frac{1}{\zeta} dZ_{\gamma'} - \zeta d\bar Z_{\gamma'} } } \nonumber \\
\omega^{\rm inst}_{n_B'\gamma'_g\lambda 3} &= - \frac{i}{8\pi^2}\cdot \parens{- \half \pi^4 R^2 \xi_{n_B'\lambda 3 \RR}^2} \kappa_2^{\lambda^3 P - \lambda^4 Q} \kappa_2^{P+Q} e^{i(P\theta_m+Q\theta_e)} \times \nonumber \\
&\quad  d\Y^{\rm sf}_{\gamma'_g}\wedge\parens{-|Z_{\gamma'}| K_1(2\pi R |Z_{\gamma'}|) d\log(Z_{\gamma'}/\bar Z_{\gamma'}) + K_0(2\pi R |Z_{\gamma'}|)\parens{\frac{1}{\zeta} dZ_{\gamma'} - \zeta d\bar Z_{\gamma'} } } \ .
\end{align}
There are now 3 singular fibers with different associated global symmetries: $E_8$, $E_6$, and $\Spin(8)$. Terms with all three values of $i$ contribute to the field theory associated to the $E_8$ singular fiber, terms with $i=2$ contribute to the $E_6$ singular fiber, and terms with $i=3$ contribute to the $\Spin(8)$ singular fiber. We thus have
\begin{align}
F_{n,p,q}(\theta) &= n^2 (-1)^{n} \brackets{ (-4 \pi^4 R^2 |\xi_{n_B' 1\RR}|^2) + \sum_{\lambda\in Z_3} (-\frac{4}{3}\pi^4 R^2 |\xi_{n_B' \lambda 2\RR}|^2) \kappa_3^{n\lambda(p+q)} \right. \nonumber \\
&\left. \qquad + \sum_{\lambda\in Z_2^2} (-\half \pi^4 R^2 \xi_{n_B' \lambda 3\RR}^2) (-1)^{n(\lambda^3 p+\lambda^4 q)}} \label{eq:e6Info}
\end{align}
for the $E_8$ singular fiber,
\be F_{n,p,q}(\theta) = n^2 (-1)^{n} \sum_{\lambda\in Z_3} (-\frac{4}{3}\pi^4 R^2 |\xi_{n_B' \lambda 2\RR}|^2) \kappa_3^{n\lambda(p+q)} \ee
for the $E_6$ singular fiber, and
\be F_{n,p,q}(\theta) = n^2 (-1)^{n} \sum_{\lambda\in Z_2^2} (-\half \pi^4 R^2 \xi_{n_B' \lambda 3\RR}^2) (-1)^{n(\lambda^3 p+\lambda^4 q)} \ee
for the $\Spin(8)$ singular fiber. There are, respectively, 8, 6, and 4 real mass parameters associated to these SCFTs, as expected: for the $E_8$ singular fiber, we have $\xi_{n_B' 12\RR} = \xi_{n_B' 22\RR}$ and $\xi_{n_B' (1,0)3\RR}=\xi_{n_B' (0,1) 3\RR}=\xi_{n_B' (1,1) 3\RR}$.

Working out the change of variables for an $E_8$ singular fiber is easier than it was in the previous subsections, since at the orbifold point $E_8$ breaks to $A_1\oplus A_2\oplus A_5$, and all three of these $A_n$ algebras are distinct. We thus associate $\{\xi_{n_B'1\RR},\xi_{n_B'02\RR},\xi_{n_B'(0,0)3\RR}\}$ to $A_5$, $\xi_{n_B'12\RR}$ to $A_2$, and $\xi_{n_B'(1,1)3\RR}$ to $A_1$. Following the strategy described in the previous subsections, we then find
\begin{align}
\frac{\xi_{n_B'1\RR}}{1-\kappa_6}\sigma+\frac{\xi_{n_B'02\RR}}{1-\kappa_3}\sigma^2&+{\rm h.c.}+\frac{\xi_{n_B'(0,0)3\RR}}{2}\sigma^3 \nonumber \\
&\propto {\rm diag}(\eta_{n_B',7},\eta_{n_B',6}-\eta_{n_B',7},\eta_{n_B',5}-\eta_{n_B',6},\eta_{n_B',4}-\eta_{n_B',5},\eta_{n_B',3}-\eta_{n_B',4},-\eta_{n_B',3}) \nonumber \\
\frac{\xi_{n_B' 12\RR}}{1-\kappa_3}\sigma^2 + {\rm h.c.} &\propto {\rm diag}(\eta_{n_B',1},\eta_{n_B',2}-\eta_{n_B',1},-\eta_{n_B',2},\eta_{n_B',1},\eta_{n_B',2}-\eta_{n_B',1},-\eta_{n_B',2}) \nonumber \\
\half \xi_{n_B'(1,1)3\RR} \sigma^3 &\propto \eta_{n_B',8}\sigma^3 \label{eq:e8Change}
\end{align}
and
\begin{align}
\xi_{n_B' 1 \RR} &= \frac{1}{2\pi^2 R}(\eta_{n_B',7}+\kappa_6^{-1}\eta_{n_B',6}+\kappa_3^{-1}\eta_{n_B',5}-\eta_{n_B',4}+\kappa_3\eta_{n_B',3}) \nonumber \\
\xi_{n_B' 02 \RR} &= \frac{3}{2\pi^2R}(\eta_{n_B',7}+\kappa_3^{-1}\eta_{n_B',6}+\kappa_3\eta_{n_B',5}+\eta_{n_B',4}+\kappa_3^{-1}\eta_{n_B',3}) \nonumber \\
\xi_{n_B' (0,0) 3 \RR} &= \frac{2}{\pi^2 R}(\eta_{n_B',7}-\eta_{n_B',6}+\eta_{n_B',5}-\eta_{n_B',4}+\eta_{n_B',3}) \nonumber \\
\xi_{n_B' 1 2 \RR} &= \frac{3}{2\pi^2 R}(\eta_{n_B',1}+\kappa_3^{-1}\eta_{n_B',2}) \nonumber \\
\xi_{n_B' (1,1) 3 \RR} &= \frac{2}{\pi^2 R} \eta_{n_B',8} \ .
\end{align}
The proportionality constants in \eqref{eq:e8Change} were determined using data from \S\ref{sec:experiment}. The changes of variables for $\Spin(8)$ and $E_6$ singular fibers are similar to those in the $q=2,3,4$ cases.

As in the last subsection, we are led to conjecture relationships between the $E_8$ MN theory and the $E_6$ MN and $SU(2)$ $N_f=4$ theories:
\begin{align}
F^{E_8 \, {\rm MN}}_{n,p,q}(\xi_{n_B' 1 \RR}=\xi_{n_B' \lambda 2 \RR}=0) &= F^{SU(2) \, N_f=4}_{n,p,q}(\xi_{n_B' (1,0) 3 \RR}=\xi_{n_B' (0,1) 3 \RR}=\xi_{n_B' (1,1) 3 \RR}) \nonumber \\
F^{E_8 \, {\rm MN}}_{n,p,q}(\xi_{n_B' 1 \RR}=\xi_{n_B' \lambda 3 \RR}=0) &= F^{E_6 \, {\rm MN}}_{n,p,q}(\xi_{n_B' 1 2 \RR}=\xi_{n_B' 2 2 \RR}) \ . \label{eq:q6conjOld}
\end{align}
Or, stripped of the extraneous subscripts, we have
\begin{empheq}[box=\fbox]{align}
F^{E_8 \, {\rm MN}}_{n,p,q}(\xi_{1 \RR}=\xi_{\lambda 2 \RR}=0) &= F^{SU(2) \, N_f=4}_{n,p,q}(\xi_{(1,0) \RR}=\xi_{(0,1) \RR}=\xi_{(1,1) \RR}) \nonumber \\
F^{E_8 \, {\rm MN}}_{n,p,q}(\xi_{1 \RR}=\xi_{\lambda 3 \RR}=0) &= F^{E_6 \, {\rm MN}}_{n,p,q}(\xi_{1 \RR}=\xi_{2 \RR}) \ . \label{eq:q6conj}
\end{empheq}
We have far less data available to support these conjectures, but we will find some in \S\ref{sec:experiment}. While the number of tests we are able to perform with this data is small, it turns out that each of these is quite strong, as the functions involved are quite complicated, even after specializing their arguments as in \eqref{eq:q6conj}. We note that while in general we cannot fix the sign of the proportionality constant in the relationship between the $\eta$s and $\xi$s without working at higher orders in the FI parameters -- or, equivalently, we cannot determine whether the $\eta$s should be ordered as in \eqref{eq:e8Change} or whether we should reverse the $\eta$s associated to one or more $A_n$ subalgebras of $E_8$ -- there are relative signs between the $E_6$ and $E_8$ MN theories that are fixed if we want \eqref{eq:q6conj} to hold.

\subsection{Discussion} \label{sec:discuss}

We conclude with a discussion of some related topics.

\subsubsection{Relationship between different constructions of MN and $SU(2)$ $N_f=4$ SCFTs}

We begin by sketching the relationship between the F-theory picture of the $SU(2)$ $N_f=4$ and MN theories that was reviewed in the introduction and the class S ones of \cite{w:MGaugeSol,gaiotto:classS,GMN:classS,tachikawa:e8}, which are are obtained by wrapping a number of M5-branes on a sphere with four punctures for $SU(2)$ $N_f=4$ and three punctures for the MN theories. We denote this punctured sphere by $C$. That one should suspect the existence of such a relationship is suggested by the fact that the class S construction of these theories involves $q$ M5-branes -- that is, these constructions involve twisted compactification of the $A_{q-1}$ 6d $\N=(2,0)$ SCFT.

Our starting point is F-theory on $(\CC\times T^2)/Z_q$ with a probe D3-brane. We first compactify a dimension transverse to both the D3-brane and the orbifold into a large circle. T-dualizing this and lifting to M-theory yields an M5-brane wrapping the $T^2$ in $(\CC\times T^2)/Z_q$. We now claim that the low energy limit of this configuration defines the same 4d CFT as that of $q$ M5-branes wrapping the zero section of $T^* C$. To see this, we change our perspective: instead of regarding $(\CC\times T^2)/Z_q$ as an elliptic fibration over $\CC/Z_q$, we regard it as a holomorphic vector bundle -- specifically, the holomorphic cotangent bundle -- over $T^2/Z_q$. This change of perspective is particularly amusing because in the former picture, once we move onto the Coulomb branch -- i.e., move the M5-brane away from the origin of $\CC/Z_q$ -- it is natural to say that there is one M5-brane wrapping $T^2$, whereas in the latter it is natural to say that there are $q$ wrapping $T^2/Z_q$. Since the complement of the fixed points in $T^2/Z_q$, which we denote by $C'$, and $C$ coincide as punctured Riemann surfaces, we learn that $T^* C'$ and $T^* C$ are the same holomorphic vector bundles. But, embedding M5-branes into the cotangent bundle implements the twist of the class S construction \cite{GMN:classS}, and so the low energy limits of M5-branes wrapping the zero sections of these bundles are only sensitive to this complex geometry. They therefore define the same CFTs.

One gap in this argument is the fact that in order to complete the class S construction, we must associate certain data -- namely partitions of $q$ (often described via Young tableaux with $q$ boxes) -- to the punctures \cite{gaiotto:classS}. These specify the orders of the poles that the degree-$d$ differentials $\phi_d$ appearing in the Seiberg-Witten curve
\be \lambda^q = \sum_{d=2}^q \phi_d \lambda^{q-d} \ee
may have at the punctures, as well as the mass parameters of the theory. Rather than study this in detail, we simply make some observations here that explain the number and associated flavor symmetries of the punctures. The entire flavor groups are not visible in the class S construction; instead, only certain subgroups associated to the punctures are manifest and one argues by comparing with weakly coupled limits that the flavor symmetry enhances in the infrared limit that defines the 4d SCFT. The groups that are visible precisely coincide with those in \eqref{eq:G3d}, which was concerned with the 3d flavor groups at the orbifold point. The reason that these two different problems have the same answer is that they are both equivalent to asking for the gauge symmetry of M-theory on $(\CC\times T^2)/Z_q$: for the 3d flavor group, this gauge symmetry appears as a global symmetry on a probe M2-brane, while in the present context this gauge symmetry appears as a global symmetry on wrapped M5-branes. We thus see that the puncture types correspond to the stabilizer subgroups of the fixed points of $T^2$, since the M-theory geometry near a fixed point with stabilizer $Z_Q$ will locally look like $\CC^2/Z_Q$ and introduce the corresponding $SU(Q)$ gauge symmetry. For $q=2$, all four punctures of $C$ in the class S picture are of the same type, with associated $SU(2)$ symmetries, and all four fixed points of $T^2$ have $Q=2$; for $q=3$, all three punctures are of the same type, with associated $SU(3)$ symmetries, and all three fixed points have $Q=3$; for $q=4$, there is one puncture with an associated $SU(2)$ symmetry and two of another type with an associated $SU(4)$ symmetry, and there is one fixed point with $Q=2$ and two with $Q=4$; and for $q=6$ there are three different types of punctures with associated $SU(2)$, $SU(3)$, and $SU(6)$ symmetries and fixed points with $Q=2,3,6$, respectively.

This discussion implies amusing relationships between Borel-de Siebenthal theory and the class S construction: the extended Dynkin diagram of the full global symmetry is obtained by taking the Dynkin diagrams of the visible global symmetries and attaching them all to a single new node, whose valence coincides with the number of punctures. Indeed, Borel-de Siebenthal theory is of quite general applicability. For example, the theory applies whenever one considers adjoint (or group-valued, in the case of Wilson-`t Hooft lines) Higgsing, of either gauge or global symmetries, as this pattern of spontaneous symmetry breaking always preserves the rank of the symmetry. In particular, this is always the case when the field breaking the symmetry is in a vector multiplet, as is the case for 3d $\N=4$ mass parameters. As another simple application of this theory, we note that $SU(N)^3$ global symmetries in class S constructions cannot enhance to a single simple group in the infrared for $N>3$; this follows from studying the Dynkin diagrams that obtain by deleting trivalent nodes from extended Dynkin diagrams.

Interestingly, the F-theory and M-theory constructions give slightly different answers to the question of why the MN SCFTs are isolated, while the $SU(2)$ $N_f=4$ theory has a marginal coupling constant, $\tau_F$. In F-theory, it is simply the case that the $(\CC\times T^2)/Z_q$ orbifolds only exist for special values of $\tau_F$ for $q=3,4,6$, as elliptic curves with other values of $\tau_F$ do not have $Z_q$ automorphisms, whereas any value of $\tau_F$ suffices for $q=2$. In contrast, in M-theory we have performed the change of perspective described above, i.e. we have quotiented $T^2$ by $Z_q$. Therefore, the space of marginal parameters for the MN SCFTs is the moduli space $\M_{0,3}$ of genus 0 Riemann surfaces with three marked points, which is a point, whereas the space of marginal parameters for $SU(2)$ $N_f=4$ is $\M_{0,4}\cong \M_{1,1}$.\footnote{We should be a bit more precise about whether the marked points are ordered or not. For $q=2$, the points should be unordered, since the punctures are all of the same type. Since the sets of $S_4$ orbits and $S_3$ orbits in the moduli space $\PP^1\backslash\{0,1,\infty\}=\HH/\Gamma(2)$ of genus 0 Riemann surfaces with four distinct ordered marked points turn out to coincide -- where $S_4$ permutes all four marked points and $S_3$ permutes only three of them -- one finds that the (coarse) moduli space of genus 0 Riemann surfaces with four distinct unordered marked points is $\HH/SL(2,\ZZ)\cong \M_{1,1}$. When there are three marked points, the moduli space with the marked points ordered is already a point, and so forgetting the order between some or all of them still leaves one with a point.}

It now follows that the relationship between the Higgs and Coulomb pictures that was explained in \cite{mz:K3HK} implies that the moduli space of $Z_q$-equivariant $SU(q)$ Higgs bundles on $T^2$ (Higgs) coincides with the moduli space of $SU(q)$ Higgs bundles on $C$ (Coulomb). We denote the latter by Coulomb, even though this moduli space also has a hyper-K\"ahler quotient construction via a D4-brane picture (obtained by compactifying the M5-branes on a circle), because this D4-brane picture does not have the discrete B-field that breaks the global symmetry of the theory to its maximal torus, and so it is more analogous to the Coulomb side of 3d mirror symmetry. Correspondingly, the associated hyper-K\"ahler quotient requires input from non-perturbative physics in the form of singular boundary conditions at the punctures. In particular, when all mass parameters vanish, we learn that a moduli space of \emph{smooth} equivariant Higgs bundles on $T^2$ coincides with a moduli space of \emph{singular} Higgs bundles on $C$. It would be interesting to exhibit this isomorphism explicitly, for general mass parameters.

This discussion should generalize to cover many other theories, such as those studied in \cite{kapustin:nonLag}. This likely provides a useful means of studying many class S field theories, and perhaps also Higgs bundle moduli spaces. For example, the global symmetries are more readily apparent in the D3-brane picture than in the class S one.

\subsubsection{Experimental observations} \label{sec:experiment}

In this subsection, we demonstrate that the mere knowledge that the Taylor expansions of the functions $F_{n,p,q}$ (defined in \eqref{eq:Fdef}) about the orbifold point $\theta^{(0)}$ (specified in \eqref{eq:orbPt}) ought to involve simple expressions -- thanks to the existence of the hyper-K\"ahler quotient formalism -- can allow one to guess some new values of the BPS index in the MN theories. It is even conceivable that this approach could allow one to determine the entire spectra, but we defer this determination to \cite{mz:K3HK2}, where we will proceed systematically by comparing with the Higgs branch results. The main input that we require consists of a number of constraints on the functions $F_{n,p,q}$ (most of which are of the form of constraints on the BPS spectra that appear in the definition of these functions). We therefore begin by introducing these constraints.

As we reviewed in \cite{mz:K3HK}, the F- and M-theory constructions of 4d $\N=2$ field theories described in the previous subsection yield geometric pictures of BPS states as, respectively, string webs terminating on D3- and 7-branes or M2-branes wrapping holomorphic curves and terminating on an M5-brane. Using this picture, \cite{zwiebach:webs} derived a constraint on the gauge and flavor charges of BPS states that can appear in these theories. This makes use of the fact that the genera of the holomorphic curves associated to BPS states are given by
\be (\gamma,\gamma)=2g_{\gamma}-2+m \ , \label{eq:genusDef} \ee
where the left hand side denotes a symmetric relative intersection pairing on the charge lattice which is defined axiomatically in \cite{zwiebach:webs2} (see also \cite{hutchings:relInt1,hutchings:relInt2,halverson:strings}). For each field theory, this ends up taking the form
\be (\gamma,\gamma) = -\tilde\gamma_f\cdot \tilde\gamma_f+m^2 f(p,q) \ , \label{eq:intPair} \ee
where the product in the first term employs the Cartan-Killing form and $f(p,q)$ is a positive-definite quadratic form that depends on the theory. Here, we have employed our usual notation $\gamma=m\gamma_g+\tilde\gamma_f$, where $m\gamma_g=m(p,q)$ is the gauge charge, with $\gcd(p,q)=1$ and $m>0$, and $\tilde\gamma_f$ is the flavor charge. The constraint is then simply that
\be (\gamma,\gamma)+2-m\ge 0 \ , \label{eq:posConstr} \ee
so that the formula defining $g_\gamma$ is non-negative. It was found in \cite{zwiebach:webs} that the charges of the $SU(2)$ $N_f\le 4$ theory that satisfied this constraint, as well as the integrality constraint on the invariant charges from \cite{zwiebach:webs2} that was employed above, were precisely those that appear in the BPS spectrum in the weak coupling chamber. The same conclusion was reached in \cite{halverson:ad} for the Argyres-Douglas theories that arise on a D3-brane probing a $II$, $III$, or $IV$ singular fiber.

An analogous study was performed in \cite{neitzke:flavorString} for the MN theories. There, it was found that the constraints of \cite{zwiebach:webs2,zwiebach:webs} were insufficient on their own for identifying the charges that appear in the MN theories, but that an additional constraint from \cite{sethi:FWebs} -- again motivated by this geometric picture -- exactly identified the charges that were known from \cite{neitzke:e6,neitzke:e7} to appear in the $E_6$ and $E_7$ MN theories' spectra. Specifically, this takes the form
\be (\gamma,\gamma')\ge 0 \label{eq:intConstr} \ee
for all $\gamma,\gamma'$ that appear in the spectrum and have gauge charges which differ by a positive rational multiple. The ansatz of \cite{neitzke:flavorString} was then that any charge with integral invariant charges and $g_\gamma=0,m=1$ appears in the spectrum and that the rest of the spectrum can be bootstrapped from this starting point by combining the above three constraints. We will soon find strong evidence that this ansatz does not, in general, correctly identify those charges that appear in the spectrum.

While we are reviewing results of \cite{neitzke:flavorString}, we take this opportunity to mention some experimental observations in this reference concerning relationships between this geometric picture of BPS states as holomorphic curves with boundary and the data from \cite{neitzke:e6,neitzke:e7} on the values of $\Omega(m,p,q,\R)$. To state this, we first define $\Omega(m,p,q,\R)=(-1)^{m+1} m \, \Omega_{\rm red}(m,p,q,\R)$; the motivation for working with this reduced index is the experimental observation in \cite{neitzke:e6,neitzke:e7} that $\Omega_{\rm red}$ is always a non-negative integer. The experimental finding of \cite{neitzke:flavorString} regarding the BPS index (for charges for which it is non-vanishing) was then a strong dependence of $\Omega_{\rm red}(m,p,q,\R)$ on the genus of a holomorphic curve associated to the highest weight $\tilde\gamma_\R$ of the irreducible representation $\R$. In particular, it was found that $\Omega_{\rm red}=1$ when $g_{m,p,q,\tilde\gamma_\R}=0$, $\Omega_{\rm red}\in\{2,3\}$ when $g_{m,p,q,\tilde\gamma_\R}=1$, and $\Omega_{\rm red}$ tends to grow rapidly with genus from there. Toward the end of this subsection, we will build on these results.

Another constraint on the BPS spectrum comes from S-duality \cite{zwiebach:webs}. For the $E_6$ MN theory, this group is $Z_6\cong Z_2\times Z_3$, and both of these ways of thinking about this group are useful. The action of the $Z_3$ generator is $p\tau_F+q\mapsto \kappa_3(p\tau_F+q)$, or equivalently $(p,q)\mapsto (-p+q,-p)$, while the action of the $Z_2$ generator is given by the combination of $(p,q)\mapsto -(p,q)$ with $\tilde\gamma_f\mapsto -\tilde\gamma_f$ -- that is, we act by the outer automorphism of $E_6$ which conjugates all representations. Alternatively, we can describe this group action in terms of the action of the $Z_6$ generator: $p\tau_F+q\mapsto \kappa_6(p\tau_F+q)$, or equivalently $(p,q)\mapsto (q,-p+q)$, combined with the outer automorphism. The S-duality group of the $E_7$ MN theory is $Z_4$, whose generator acts by $p\tau_F+q\mapsto \kappa_4(p\tau_F+q)$, or equivalently $(p,q)\mapsto (q,-p)$. Finally, the S-duality group of the $E_8$ MN theory is again $Z_6$, which acts in the same way as in the $E_6$ MN theory, except it acts trivially on flavor charges. Note that these duality constraints subsume the CPT requirement $\Omega(-\gamma)=\Omega(\gamma)$; this is obvious for the $E_6$ theory, while for the $E_7$ and $E_8$ theories this follows from the fact that all $E_7$ and $E_8$ representations are self-conjugate (i.e., either real or pseudo-real), and so it is automatic that $\Omega(m\gamma_g+\tilde\gamma_f)=\Omega(m\gamma_g-\tilde\gamma_f)$. We stress that all of these S-duality actions leave $m$ invariant, as is clear from the more general fact that $SL(2,\ZZ)$ elements map a vector of relatively prime integers to another such vector. Therefore, S-duality acts naturally on the functions $F_{n,p,q}$: we act on $p,q$ as above, and for the $E_6$ MN theory we also sometimes negate their argument, $\theta$, or equivalently conjugate them. S-duality invariance implies that these actions leave the functions invariant. We note by contrast that the fact that all $E_7$ and $E_8$ representations are self-conjugate implies that the functions $F_{n,p,q}$ in the $E_7$ and $E_8$ MN theories are all real.

The final constraint that we employ is that $F_{n,p,q}$ must be invariant under the Weyl group $W_{3d}$ of the global symmetry group $G_{3d}$ at the orbifold point (see \eqref{eq:G3d}). That is, the $r$-th order part $F^{(r)}_{n,p,q}$ of $F_{n,p,q}$ is a symmetric (under $W_{3d}$) $r^{\text{th}}$-order symmetric\footnote{Note that symmetric is being used in two different senses here. In the latter usage, we refer to the fact that as an $r^{\text{th}}$-order term in a Taylor expansion, we naturally get an element of $\Sym^r \mf{h}_{3d}^{\vee}$ by the symmetry of partial derivatives. In the former usage, we refer to further invariance under the natural $W_{3d}$-action. We shall henceforth mean both meanings simultaneously when we refer to the space of symmetric tensors.} tensor on the Cartan subalgebra of $G_{3d}$, i.e. an element of $(\Sym^r \mf{h}_{3d}^{\vee})^{W_{3d}}$, which has a nice interpretation as $H^{2r}(BG_{3d}; \mb{C})$. As a useful mnemonic, this ring $H^{2*}(BG_{3d};\CC)$ of symmetric tensors is exactly the ring of characteristic classes. ($H^*(BG_{3d};\CC)$ is already concentrated in even degree, so it differs from $H^{2*}(BG_{3d};\CC)$ only in that degrees in the latter are half of degrees in the former.) For example, $SU(3)$ has ring of symmetric tensors freely generated by ``$c_2$'' and ``$c_3$.'' Hence, there are unique (up to scaling) quadratic and cubic (symmetric -- an adjective which we henceforth implicitly assume) tensors and there is a two-dimensional space of sextic tensors. Renaming the two generators of $H^{2*}(BSU(3);\CC)$ as $a,b$, where $a$ has degree 2 and $b$ has degree 3, we thus learn that for $q = 3$, where $G_{3d} = SU(3)^3$ (up to a central isogeny, i.e. a quotient, which we presently ignore), the ring of symmetric tensors on $\mf{h}_{3d}$ is generated as a polynomial ring by $a_0, b_0, a_1, b_1, a_2, b_2$, where the subscript refers to one of the three copies of $SU(3)$.

Explicitly, the quadratic and cubic tensors on the Cartan are given in terms of the usual $\eta$ basis associated to fundamental weights (see \eqref{eq:etaDTheta}) by
\begin{eqnarray}
a_0 &=& \eta_6^2 - \eta_6 \eta_3 + \eta_3^2 \nonumber \\
b_0 &=& 3\sqrt{3} \eta_6 \eta_3 (\eta_6 - \eta_3) \nonumber \\
a_1 &=& \eta_5^2 - \eta_5 \eta_4 + \eta_4^2 \nonumber \\
b_1 &=& 3\sqrt{3} \eta_5 \eta_4 (\eta_5 - \eta_4) \nonumber \\
a_2 &=& \eta_1^2 - \eta_1 \eta_2 + \eta_2^2 \nonumber \\
b_2 &=& 3\sqrt{3} \eta_1 \eta_2 (\eta_1 - \eta_2) \ . \label{eq:e6tensors}
\end{eqnarray}
These tensors are only well-defined up to scaling; we chose the above normalizations for aesthetic effect. As an aside, we also note that in terms of the $\xi$ parametrization of the $A_2$ Cartans (see \eqref{eq:xiEta}), the quadratic and cubic tensors are, respectively, proportional to $|\xi|^2$ and $\xi^3 - \overline{\xi}^3$; of course, this may change slightly if one finds (pursuant to the discussion after \eqref{eq:D2mod}) that \eqref{eq:xiEta} requires small modifications, but in any case these tensors will have simple expressions in terms of the FI parameters. We stress that nothing in this section depends on the change of variables from $\eta$ to $\xi$, since any computations of terms $F^{(r)}_{n,p,q}$ that we perform in this section will consist solely of Taylor expanding functions of the form \eqref{eq:Fdef}, which are naturally expressed in terms of the $\delta\theta$ variables, and therefore (via \eqref{eq:etaDTheta}) the $\eta$ variables.

\bigskip

We now take a brief detour to give an abstract way by which one might intuit the appearance of cohomology in this context. (This paragraph and the following one may safely be skipped.) We first note that $F_{n,p,q}$ can be regarded as a sum of characters of virtual representations of the 4d flavor group $G_{4d}$ evaluated at different points in its Cartan subalgebra, $\mf{h}$. (The factor $m^2 \Omega(m,p,q,\R)$ can simply be interpreted as taking $m^2 \Omega(m,p,q,\R)$ copies of $\R$.) For the moment, we neglect this translation to different points in $\mf{h}$ and simply consider the character of a single (possibly virtual) representation of $G_{4d}$. This can be understood as the image of the Atiyah-Segal completion map followed by the Chern character map
\be \Rep(G_{4d})\cong K_{G_{4d}}(\pt) \to K(BG_{4d}) \stackrel{\text{ch}}{\to} H^*(BG_{4d};\CC) \ , \ee
where $\Rep(G_{4d})$ is the ring of virtual representations of $G_{4d}$. Note that the Atiyah-Segal completion map (the first map above) is close to an isomorphism; more precisely, it is a completion at the augmentation ideal $I \subset \mathrm{Rep}(G_{4d})$ consisting of representations of virtual dimension zero. The (complex) Chern character, on the other hand, is genuinely an isomorphism once one applies $(-) \otimes_{\mb{Z}} \mb{C}$, i.e. once one tensors $K(BG_{4d})$ with the complex numbers. And in terms of this composite map, the $r^{\text{th}}$-order information of the character is precisely the $2r^{\text{th}}$-order information of the Chern character. This presentation makes it easy to read off various properties: for example, absent any other information, knowing the character to any finite order is insufficient to reconstruct the full character, since arbitrary powers of the ideal $I$ are nontrivial and hence yield examples of representations whose characters vanish to arbitrarily high orders.

Next, we consider the expansion of this character about the integral multiple $n/m$ of the orbifold point. Since the centralizer of $e^{in\theta^{(0)}/m}$ contains $G_{3d}=C_{G_{4d}}(e^{i\theta^{(0)}})$, the virtual $G_{4d}$ representation decomposes into virtual $G_{3d}$ representations. The discussion of the previous paragraph then applies, with $G_{4d}$ replaced by $G_{3d}$. Abstractly, we are now considering the stalk about $n\theta^{(0)}/m$ of the delocalized equivariant Chern character \cite{baum:equivCoh}. Finally, $F_{n,p,q}$, as the sum of a number of $G_{4d}$ characters evaluated at multiples of $\theta^{(0)}$, can now instead be regarded as the sum of a number of elements of $H^{2*}(BG_{3d};\CC)$, which is itself in $H^{2*}(BG_{3d};\CC)$. 

\bigskip

We now write $F^{(2)}_{n,p,q}$ in the $E_6$ MN theory in terms of the tensors \eqref{eq:e6tensors}. We also employ the notation $(P,Q)=n(p,q)$ from earlier subsections; this is motivated by the fact that $p,q$ will likely always appear multiplied by $n$ in expressions that are obtained by Poisson resumming results from the hyper-K\"ahler quotient formalism. Translating \eqref{eq:repInfo} into this new notation, we have
\be F_{n,p,q}^{(2)} = 3 (-1)^{n+1} n^2 \sum_{\lambda \in Z_3} \kappa_3^{\lambda(P+Q)} a_{\lambda} \ . \ee
This is easily seen to behave properly under S-duality, since the $Z_6$ generator negates $p+q\pmod{3}$ and this is undone by conjugation of the function.

We know that the third-order character must similarly be a sum of the $b_{\lambda}$, and comparing to data from \cite{neitzke:e6,neitzke:e7}, we guess the general expression
\be F_{n,p,q}^{(3)} = \frac{1}{2} (-1)^{n+1} n^2 I \sum_{\lambda \in Z_3} \kappa_3^{\lambda(P+Q)} b_{\lambda} \ , \ee
where
\be I = P^2 - PQ + Q^2 \ee
is $n^2$ times the duality-invariant quadratic form denoted $f(p,q)$ in \eqref{eq:intPair}.

We now pause to specify generators of the algebra $\Ss$ over $\RR$ of all S-duality-invariant polynomials in $P,Q$ with complex coefficients. Since the action of S-duality on the gauge charges does not introduce complex numbers, we can take these generators to have purely imaginary or real coefficients. Similarly, since this action maps homogeneous polynomials to homogeneous polynomials of the same degree, we can take these generators to be homogeneous. Since the $Z_2$ factor in the S-duality group acts on these polynomials by the combination of $(P,Q)\mapsto (-P,-Q)$ with conjugation of the polynomial, we can further take these generators to either be homogeneous of odd degree and have purely imaginary coefficients or to be homogeneous of even degree and have real coefficients. So, the problem is now to find homogeneous generators of the algebra $\Ss'$ of polynomials which are invariant under the action of the $Z_3$ factor of the S-duality group. We consider this algebra over $\CC$ in order to make complex changes of variables. Making such a change from $P,Q$ to $\Q,\bar\Q$, where $\Q=P\kappa_3+Q$, this is then the same as asking for homogeneous polynomials in $\Q,\bar\Q$ which are invariant under $\Q\mapsto \kappa_3\Q$. We can now take our generators to be of the form $\Q^a \bar\Q^b+\kappa_3^{a-b}\Q^a\bar\Q^b+\kappa_3^{b-a}\Q^a\bar\Q^b=(1+\kappa_3^{a-b}+\kappa_3^{b-a})\Q^a\bar\Q^b$. But, this is non-vanishing iff $3|(a-b)$, and in this case we simply have the invariant monomial $\Q^a \bar\Q^b$. Taking $I=\Q\bar\Q$ to be a generator, we can now assume that either $a=0$ or $b=0$. We thus see that $\Ss'$ is generated by $I$, $A=\Q^3$, and $\bar A = \bar\Q^3$, and the only relation is $I^3=A\bar A$. Again making a change of basis, we finally take our generators for $\Ss$ to be $I$, $A_1$, and $A_2$, where
\be A_1 = i(P^3-3P^2 Q + Q^3) \ , \quad A_2 = i PQ(P-Q) \ . \ee
$I^3=A\bar A$ translates to the relation $-I^3=A_1^2+3 A_1 A_2 + 9 A_2^2$. This paragraph makes it clear that S-duality is quite restrictive.

At this point, we have enough data to form a reasonable guess for $\Omega_{\rm red}(1,1,4,\R)$ for all irreducible representations $\R$. (However, we will soon argue that this guess is wrong.) The constraints from \cite{zwiebach:webs2,zwiebach:webs} imply that the allowed pairs $(\R,g_{\gamma_{1,1,4,\tilde\gamma_\R}})$ are $(\mathbf{1728},0), (\mathbf{351'},0),(\mathbf{351},1),(\mathbf{27},2)$. (The extra constraint from \cite{neitzke:flavorString} does not rule any of these out.) Supposing, per the observations of \cite{neitzke:flavorString}, that $\Omega_{\rm red}=1$ for the representations with $g_{\gamma_{1,1,4,\tilde\gamma_\R}}=0$, we are left with two integers that need to be determined. The second- and third-order expressions $F^{(r)}_{1,1,4}$ give us the following constraints:
\begin{eqnarray}
-2 + 1 -2 \Omega_{\rm red}(1,1,4,\mathbf{351})+ \Omega_{\rm red}(1,1,4,\mathbf{27}) &=& 1 \nonumber \\
-8 -5 + 4 \Omega_{\rm red}(1,1,4,\mathbf{351}) + \Omega_{\rm red}(1,1,4,\mathbf{27}) &=& 13 \ ,
\end{eqnarray}
yielding $\Omega_{\rm red}(1,1,4,\mathbf{351}) = 4$, $\Omega_{\rm red}(1,1,4,\mathbf{27}) = 10$. We note that this value of $\Omega_{\rm red}(1,1,4,\mathbf{351})$ violates the trend from \cite{neitzke:flavorString} that $\Omega(m,p,q,\R)\in\{2,3\}$ when $g_{m,p,q,\tilde\gamma_\R}=1$; this is a first hint that this guess may be faulty.

Pressing on, $F_{n,p,q}^{(4)}$ must be a quadratic expression in the $a$ tensors. For the purely magnetic spectrum, we conjecture the following:
\be F_{n,1,0}^{(4)} = \frac{1}{4} (-1)^{n+1} n^4 \frac{n^2 - 3}{2} \sum_{\lambda \in Z_3} \kappa_3^{\lambda P} a_{\lambda}^2 + \mathbf{1}_{3 \nmid n} (-1)^{\lfloor n/3 \rfloor} n^3 \sum_{\lambda < \mu \in Z_3} \kappa_3^{-(\lambda + \mu)P} a_{\lambda} a_{\mu} \ . \ee
Here $\mathbf{1}_{3 \nmid n}$ is an indicator function, i.e. it is $1$ when $3 \nmid n$ and $0$ when $3 \mid n$. In addition, we conjecture that for more general $p,q$, the first term in this expression generalizes to
\be F_{n,p,q}^{(4)} = \frac{1}{4} (-1)^{n+1} n^2 \frac{I^2 - 3n^2}{2} \sum_{\lambda \in Z_3} \kappa_3^{\lambda (P+Q)} a_{\lambda}^2 + \cdots \ . \label{eq:gen4E6} \ee
We note that S-duality is quite useful here: we only really have enough data to guess this expression for $p=1$, but if we assume that certain complicated high-order expressions in $\Ss$ that vanish for $p=1$, such as $A_2^2 + n^2 (I-n^2)^2$, are not present then we obtain a plausible conjectural constraint on the entire spectrum.

This last conjecture yields stronger evidence that the proposed spectrum with $(m,p,q)=(1,1,4)$ is wrong, as $F^{(4)}_{1,1,4}$, when computed using this spectrum, disagrees with \eqref{eq:gen4E6}. We therefore now relax the assumption that the prescription of \cite{neitzke:flavorString} yields the precise list of global symmetry representations that appear in the BPS spectrum. If we suppose that at least one of the genus $0$ representations is missing and solve the second- through fourth-order constraints, we find the possibilities $\{(\mathbf{1728},0,1),(\mathbf{351},1,3),(\mathbf{27},2,9)\}$ and $\{(\mathbf{351'},0,1),(\mathbf{351},1,3),(\mathbf{27},2,6)\}$, where here each triple is of the form $(\R,g_{\gamma_{1,1,4,\tilde\gamma_\R}},\Omega_{\rm red}(1,1,4,\R))$. Both of these are consistent with \eqref{eq:gen4E6}. In addition, they both have $\Omega(1,1,4,\R)\in\{2,3\}$ when $g_{1,1,4,\tilde\gamma_\R}=1$.

Carrying on, we next hypothesize that
\be F_{n,1,q}^{(5)} = \Big((-1)^{n+1}n^2\Big) \frac{1}{24} \frac{I^3 - I^2 - 6In^2 - 13I - 3n^4 + 12n^2}{10} a_0b_0 + \cdots \ . \label{eq:r5e6} \ee
This is consistent with the first of the possible $(m,p,q)=(1,1,4)$ BPS spectra that we have just found, but not the second. We take this agreement as strong evidence that $\{(\mathbf{1728},0,1),(\mathbf{351},1,3),(\mathbf{27},2,9)\}$ is, indeed, the correct BPS spectrum. To reiterate the point, let us suppose our ans{\"a}tze for the second- through fifth-order expansions are correct and rederive the $(1,1,4)$ spectrum making no assumptions whatsoever as to any of the multiplicities \emph{a priori} other than that only $\mathbf{1728}, \mathbf{351'}, \mathbf{351}$, and $\mathbf{27}$ may appear. We find the following system of equations:
\begin{align}
-2\Omega_{\rm red}(1,1,4,\mathbf{1728}) + \Omega_{\rm red}(1,1,4,\mathbf{351'}) - 2\Omega_{\rm red}(1,1,4,\mathbf{351}) + \Omega_{\rm red}(1,1,4,\mathbf{27}) &= 1 \nonumber \\
-8\Omega_{\rm red}(1,1,4,\mathbf{1728}) - 5\Omega_{\rm red}(1,1,4,\mathbf{351'}) + 4\Omega_{\rm red}(1,1,4,\mathbf{351}) + \Omega_{\rm red}(1,1,4,\mathbf{27}) &= 13 \nonumber \\
14\Omega_{\rm red}(1,1,4,\mathbf{1728}) + 11\Omega_{\rm red}(1,1,4,\mathbf{351'}) + 26\Omega_{\rm red}(1,1,4,\mathbf{351}) - \Omega_{\rm red}(1,1,4,\mathbf{27}) &= 83 \nonumber \\
128\Omega_{\rm red}(1,1,4,\mathbf{1728}) + 53\Omega_{\rm red}(1,1,4,\mathbf{351'}) + 20\Omega_{\rm red}(1,1,4,\mathbf{351}) - \Omega_{\rm red}(1,1,4,\mathbf{27}) &= 179 \ ,
\end{align}
which indeed has the unique solution
\begin{eqnarray}
\Omega_{\rm red}(1,1,4,\mathbf{1728}) &=& 1 \nonumber \\
\Omega_{\rm red}(1,1,4,\mathbf{351'}) &=& 0 \nonumber \\
\Omega_{\rm red}(1,1,4,\mathbf{351}) &=& 3 \nonumber \\
\Omega_{\rm red}(1,1,4,\mathbf{27}) &=& 9 \ .
\end{eqnarray}
Summing up, we take this finding for the $(m,p,q)=(1,1,4)$ BPS spectrum as strong evidence that the prescription of \cite{neitzke:flavorString} does not yield the precise list of global symmetry representations that appear in the BPS spectrum.

We now demonstrate that integrality can serve as an interesting constraint. We focus on \eqref{eq:r5e6}. Suppose (toward a contradiction) that it is correct for general $p$. For $(n,p,q)=(1,2,5)$, this constraint would give the following equation:
\begin{align}
-75\Omega_{\rm red}(1,2,5,\mathbf{7371})&+128\Omega_{\rm red}(1,2,5,\mathbf{1728})+53\Omega_{\rm red}(1,2,5,\mathbf{351'}) \nonumber \\
&+20\Omega_{\rm red}(1,2,5,\mathbf{351})-\Omega_{\rm red}(1,2,5,\mathbf{27}) = \frac{3073}{5} \ . \label{eq:fracConstr}
\end{align}
The non-integrality on the right-hand side is patently ridiculous as $\Omega_{\rm red}(1,2,5,\R)=\Omega(1,2,5,\R)$ should be integral for all $\R$. Hence, \eqref{eq:r5e6} cannot not hold for arbitrary $p$. In contrast, for $p = 1$ and $q=0$ the necessary integrality condition (to avoid fractions on the right-hand sides of equations as in \eqref{eq:fracConstr}) is that
\be \frac{n^6 - 10n^4 - n^2}{10} \in \mb{Z} \ . \ee
Indeed, this is true, although perhaps not immediately obvious, as $n^6 \equiv n^2 \pmod{5}$ by Fermat's little theorem. Next, we consider the right hand side of \eqref{eq:r5e6} for arbitrary $p$ but now, for simplicity, specialize to $n = 1$. The necessary integrality constraint would then be that
\be \frac{I^3 - I^2 - 19I - 9}{10} \in \mb{Z} \ , \ee
i.e. that $(I - 1)(I^2 + 1) \equiv 0 \pmod{10}$. This constraint fails for, say, $I = 19$, as is the case for $(n,p,q) = (1,2,5)$. However, one may check that this integrality constraint does in fact hold if one additionally assumes that $p = 1$. So, integrality is a rather useful constraint that can effectively differentiate between clearly false ans{\"a}tze and those with a possibility of being true.

Continuing onwards, we next find that
\be F_{n,1,0}^{(6)} = (-1)^{n+1} \frac{n^4 (5n^6 - 126n^4 + 273n^2 + 16)}{20160} a_0^3 + \cdots \ee
and
\begin{align}
F_{n,1,0}^{(8)} &= (-1)^{n+1} n^2 \times \nonumber \\
& \parens{ -\frac{17n^{12}-1265n^{10}+40161n^8-534875n^6+2229722n^4-3420720n^2+1710720}{159667200} } a_0^4 + \cdots \ .
\end{align}
We unfortunately do not yet have enough data to posit the extension to more general gauge charges. 

One could imagine continuing this procedure, alternating between using available data on the BPS spectrum to guess general expressions for some $F^{(r)}_{n,p,q}$ and using the available expressions for $F^{(r)}_{n,p,q}$ to guess new values of the BPS index. However, at present we are content with stopping here, as by now this method should be clear; we hope that it is now evident, from these relatively minimal calculations, how additional data, from whatever source, can produce further constraints on the entire spectrum of the $E_6$ MN theory.

\bigskip
 
We now move on to the $E_7$ MN theory; we will be far more brief here. We proceed analogously to the $E_6$ case, first setting up notation for the basic tensors. Namely, for an $A_3$ factor, we will use $a, b, c$ to denote the unique quadratic, cubic, and quartic tensors (up to scaling). We also denote the quadratic tensor for an $A_1$ factor by $a'$.

We showed in \eqref{eq:repInfoQ4Z4} that
\be F_{n,p,q}^{(2)} = 4(-1)^{n+1}n^2 \Big(a_0+(-1)^{P+Q}a_1+(-1)^P\mathbf{1}_{2|(P+Q)}a' \Big) \ . \ee
 Next, we surmise from the data in \cite{neitzke:e7} that
\be F_{n,p,q}^{(3)} = 8(-1)^{n+1}n^2I \Big(b_0-(-1)^{P+Q}b_1\Big) \ , \ee
where now $I = P^2 + Q^2$ is $n^2$ times the $f(p,q)$ of this gauge lattice. Note that in the above expressions, $a_0, a_1, a'$ refer to the quadratic tensors for the $A_3, A_3, A_1$ factors of $G_{3d}$. Specifically, for the $A_3$ factors, we use the $i = 1$ values of $\lambda$ as our subscripts.

These results are already enough to calculate, for example, the BPS spectrum with $(m,p,q)=(1,1,2)$. \cite{neitzke:flavorString} tells us that the representations which can contribute are $\mathbf{912}$ and $\mathbf{56}$, so the second- and third-order constraints yield
\begin{eqnarray}
-2\Omega_{\rm red}(1,1,2,\mathbf{912}) + \Omega_{\rm red}(1,1,2,\mathbf{56}) &=& 1 \nonumber \\
2\Omega_{\rm red}(1,1,2,\mathbf{912}) + \Omega_{\rm red}(1,1,2,\mathbf{56}) &=& 5 \ ,
\end{eqnarray}
whose solution is $\Omega_{\rm red}(1,1,2,\mathbf{912}) = 1$, $\Omega_{\rm red}(1,1,2,\mathbf{56})=3$. We note that this spectrum satisfies the relationship \eqref{eq:q4conj} with the BPS spectrum of the $SU(2)$ $N_f=4$ theory. If one wished, one could instead \emph{assume} \eqref{eq:q4conj}; this would then likely prove to be an extremely powerful constraint on the $E_7$ MN spectrum.

\bigskip

Moving on to $q = 6$, if we are willing to assume that $\Omega_{\rm red}(1,1,0,\mathbf{248})=1$, since $\R=\mathbf{248}$ has $g_{1,1,0,\tilde\gamma_\R}=0$, then the zeroth-order constraint already yields $\Omega_{\rm red}(1,1,0,\mathbf{1}) = 3$ as the reduced index for the only other allowed representation with these gauge charges. We pause for a moment to appreciate the implications of this statement: the knowledge of the set of representations that can appear, together with the knowledge that instanton corrections to the flat metric vanish at the orbifold point and this weak assumption about $\Omega_{\rm red}(1,1,0,\mathbf{248})$, is enough information to exactly pin down the BPS index for this gauge charge. If we include the second order constraint, \eqref{eq:e6Info}, then we can drop the assumption about $\Omega_{\rm red}(1,1,0,\mathbf{248})$.

If we continue to make this assumption about $g = 0$ contributions, then $F^{(0)}_{2,1,0}$ and $F^{(2)}_{2,1,0}$, together with the information about the $(m,p,q)=(1,1,0)$ spectrum from the last paragraph, suffice to determine the $(m,p,q)=(2,1,0)$ spectrum. Indeed, the BPS index a priori may have contributions from $\mathbf{3875}, \mathbf{248}, \mathbf{1}$, and by assuming that $\Omega_{\rm red}(2,1,0,\mathbf{3875})=1$ we may calculate that $\Omega_{\rm red}(2,1,0,\mathbf{248})=2$ and $\Omega_{\rm red}(2,1,0,\mathbf{1})=4$. Exactly the same representations can contribute to the $(m,p,q)=(1,1,2)$ spectrum, and by assuming that $\Omega_{\rm red}(1,1,2,\mathbf{3875})=1$ we may similarly calculate that $\Omega_{\rm red}(1,1,2,\mathbf{248})=3$ and $\Omega_{\rm red}(1,1,2,\mathbf{1})=6$. We note that all results from this paragraph and its predecessor are consistent with the relationships \eqref{eq:q6conj} between the $E_8$ MN theory and the $E_6$ MN and $SU(2)$ $N_f=4$ theories.

\bigskip

We conclude this subsection by discussing another class of experimental observations. Specifying the BPS index in terms of multiplicities of irreducible representations is perfectly reasonable, but one might expect weight multiplicities to be better behaved. This is because in order to differentiate holomorphic curves with different geometries, and thus obtain a concrete geometric definition of the BPS degeneracies, it is necessary to turn on generic small complex masses. (This may destroy the `weakly coupled' chamber in moduli space whose BPS spectrum we are investigating, but we expect that for any fixed charge if one goes far enough away from the origin then the BPS index will stabilize at its `weakly coupled' value; see the analogous discussion in the context of the $SU(2)$ $N_f=4$ SCFT on page 137 of \cite{GMN:classS}.) But, upon doing so, the flavor symmetry is Higgsed to its maximal torus, whose irreducible representations simply correspond to weights. So, the topological constraints on a string web depend only on the gauge charge and weight. Mathematically, this reflects the fact that the weight multiplicities are the relevant open string Gromov-Witten invariants, and the weight specifies the relative homology class of the curve. One might therefore expect patterns to emerge if one works with BPS indices of weights, rather than representations.

We therefore now re-analyze the data on the BPS index from \cite{neitzke:e6,neitzke:e7}, as well as our new data, studying $\Omega_{\rm red}$ at the level of weights, rather than representations. That is, we have $\Omega(m\gamma_g+\tilde\gamma_f)=(-1)^{m+1}m \, \Omega_{\rm red}(m,p,q,\tilde\gamma_f)$, where as usual $\gamma_g=(p,q)$ and $\Omega(m\gamma_g+\tilde\gamma_f)$ and $\Omega(m,p,q,\R)$ are related as in \eqref{eq:repWt}. It suffices to restrict to dominant weights $\tilde\gamma_f$ (those in the fundamental Weyl chamber), as weight multiplicities are automatically constant on Weyl orbits. Note some straightforward consequences: if $\tilde\gamma_1$ and $\tilde\gamma_2$ are dominant weights of a representation $\R$ and $\tilde\gamma_1$ is a higher weight than $\tilde\gamma_2$, i.e. $\tilde\gamma_1-\tilde\gamma_2$ is positive, then we have
\be (\tilde\gamma_1-\tilde\gamma_2)\cdot \tilde\gamma_1 \ge 0 \ , \quad (\tilde\gamma_1-\tilde\gamma_2)\cdot \tilde\gamma_2 \ge 0 \quad \Rightarrow \quad \tilde\gamma_1\cdot \tilde\gamma_1 \ge \tilde\gamma_1 \cdot \tilde\gamma_2 \ge \tilde\gamma_2 \cdot \tilde\gamma_2 \ . \ee
Furthermore, the first of these last two inequalities must be strict, since
\be (\tilde\gamma_1-\tilde\gamma_2)\cdot(\tilde\gamma_1-\tilde\gamma_2) = \tilde\gamma_1\cdot\tilde\gamma_1+\tilde\gamma_2\cdot\tilde\gamma_2 - 2\tilde\gamma_1\cdot\tilde\gamma_2 > 0 \ . \ee
Therefore, if $\R$ appears in the BPS spectrum with gauge charge $m,p,q$, then \eqref{eq:intPair} and \eqref{eq:genusDef} together imply that $g_{m,p,q,\tilde\gamma_1} < g_{m,p,q,\tilde\gamma_2}$. In particular, genus 0 curves only appear in the Weyl orbits of highest weights.

We now show that the weight multiplicities $\Omega_{\rm red}(m,p,q,\tilde\gamma_f)$ display quite interesting regularity. First, if $\tilde\gamma_f$ corresponds to a genus $0$ curve, it is certainly experimentally true that $\Omega_{\rm red}(m,p,q,\tilde\gamma_f)$ is always one (or zero -- as noted earlier, it may not appear in the BPS spectrum at all). Indeed, per the comments of the last paragraph, this observation is equivalent to the observation of \cite{neitzke:flavorString} that the irreducible representation $\R$ with highest weight $\tilde\gamma_f$ has $\Omega_{\rm red}=1$.

Next, we experimentally find that, in both the $E_6$ and $E_7$ MN theories, it is always the case that $\Omega_{\rm red}(m,1,0,\tilde\gamma_f)=11-m$ when $g_{m,1,0,\tilde\gamma_f}=1$. (Our data in the $E_8$ theory differs slightly from this trend, and we will shortly comment on several seemingly related irregularities.) We also find in the $E_6$ theory that $\Omega_{\rm red}(1,1,q,\tilde\gamma_f)=11-q$ for $q \ge 1$ -- but, notably, not $q = 0$ -- when $g_{1,1,q,\tilde\gamma_f}=1$. (Note that S-duality implies that $\Omega_{\rm red}(1,1,1,\tilde\gamma_f)=\Omega_{\rm red}(1,1,0,-\tilde\gamma_f)$.) Note that both of these formulas predict that for $m$ or $q$ (respectively) sufficiently large, there no longer exist any genus $1$ contributions. Indeed, this does seem to experimentally be true: any contributions are either genus $0$ or genus $2$ or higher.

Returning to gauge charges with $(p, q) = (1, 0)$ in the $E_6$ theory, where we have the most data, we find that genus $2$ reduced weight multiplicities are similarly now determined by a quadratic function of $m$. Namely, when $g_{m,1,0,\tilde\gamma_f}=2$, we have $$\Omega_{\rm red}(m,1,0,\tilde\gamma_f) = \binom{m}{2}-12m+82.$$ (We will use the $\binom{m}{k}$ basis for writing polynomials in $m$ as they have the convenient property of spanning the lattice of polynomials which yield integer output on integer input.) Similarly for $g_{m,1,0,\tilde\gamma_f} = 3$, one finds that all reduced multiplicities are either of the form
\be \Omega_{\rm red}(m,1,0,\tilde\gamma_f) = -\binom{m}{3}+13\binom{m}{2}-94m+497 \ee
or
\be \Omega_{\rm red}(m,1,0,\tilde\gamma_f) = -\binom{m}{3}+13\binom{m}{2}-95m+515 \ . \ee
Once again, $g = 3$ contributions seem to be disallowed for sufficiently large gauge charge; in general, it appears that any given nonzero genus may only appear within some annulus in gauge space. Moreover, we note the regularity in the coefficients of these polynomials for $g=1,2,3$. Similar trends seem to continue at higher genera.

It would be quite interesting to obtain a better understanding of these intriguing regularities and irregularities among the weight multiplicity data. However, we leave this exploration for later. It will benefit from having far more data on which to experiment.

\section*{Acknowledgements}
We thank S. Kachru for comments on a draft of this paper and A. Neitzke for helpful correspondence.
The research of A.T. was supported by the National Science Foundation under NSF MSPRF grant number 
1705008. This publication is funded in part by the Gordon and Betty Moore Foundation through grant GBMF8273 to Harvard University to support the work of the Black Hole Initiative. This publication was also made possible through the support of a grant from the John Templeton Foundation. The research of M.Z. was also supported by DOE grant DE-SC0007870.

\newpage
\appendix

\bibliography{Refs}

\end{document}